\documentclass[preprint,showpacs,preprintnumbers,amsmath,amssymb]{revtex4}

\usepackage[dvips]{graphicx}
\usepackage{graphicx}
\usepackage{dcolumn}
\usepackage{bm}
\usepackage{amsmath}
\usepackage{amsfonts}
\usepackage{amssymb}
\usepackage{color}
\usepackage{pstcol}
\usepackage[all]{xy}

\begin{document}

\title{Test of some Fundamental Principles in Physics via Quantum Interference with Neutrons and Photons}

\author{Abel Camacho}
\email{acq@xanum.uam.mx} \affiliation{Departamento de F\'{\i}sica,
Universidad Aut\'onoma Metropolitana--Iztapalapa\\
Apartado Postal 55--534, C.P. 09340, M\'exico, D.F., M\'exico.}

\author{ A. Camacho--Galv\'an}
\email{abel@servidor.unam.mx}
\affiliation{ DEP--Facultad de Ingenier{\'\i}a\\
Universidad Nacional Aut\'onoma de M\'exico.}

\date{\today}

\begin{abstract}
The limitations and possibilities that the concept of quantum
interference offers as a tool for testing fundamental physics are
explored here. The use of neutron interference as an instrument to
confront against measurement readouts some of the principles
behind metric theories of gravity will be analyzed, as well as
some discrepancies between theory and experiment. The main
restrictions that this model embodies for the study of some of the
features of the structure of space--time will be explicitly
pointed out. For instance, the conditions imposed by the necessary
use of the semiclassical approximation. Additionally, the role
that photon interference could play as an element in this context
is also considered. In this realm we explore the differences
between first--order and second--order coherence experiments, and
underline the fact that the Hanbury--Brown--Twiss effect could
open up some interesting experimental possibilities in the
analysis of the structure of space--time. The void, in connection
with the description of wave phenomena, implicit in the principles
of metric theories is analyzed. The conceptual difficulties, that
this void entails, are commented.
\end{abstract}


\maketitle

\tableofcontents
\section{Introduction}

\subsection{What is Fundamental Physics?}

The work that you are about to read bears the title {\it Test of
some Fundamental Principles in Physics via Quantum Interference
with Neutrons and Photons}, so it would be a good idea if we from
the very beginning had a precise idea of the meaning of the words
appearing in the title, just for the sake of completeness. Already
from its etymology the word physics is related to the concept of
experiment. Indeed, the origin of this word is Greek, and its
meaning is {\it nature} \cite{OED1}. In other words, it is a
science which attempts to provide a description and explanation to
the natural phenomena. Nowadays this remark may be considered
superfluous, but there was a time in which the explanation of the
phenomena in nature did not entail the necessity of confronting
the corresponding models against any experiment. Many times the
disobedience of this unspoken rule was accompanied with personal
problems. In this context we only need remember that Galileo was
forbidden to hold Copernican views \cite{Gaga1}, or the tragic
death of Miguel Servet \cite{Servet1}. Fortunately now the
possibility of questioning any scientific model does not always
imply a perilous situation.

Now that we have mentioned the fact that Physics is an
experimental science, we face a new question, what is the meaning
of the phrase Fundamental Principles in Physics? We could argue
that Fundamental Principles in Physics denotes those assumptions
that are considered the bedrock of our description of nature.
Though our answer is a correct one it also seems to be not very
illustrative. Let us provide a more profound definition resorting
to an example. Consider the Newtonian description of dynamics
\cite{Newton1}. The core part of this theory is contained in the
famous three laws of motion, and no problem in the realm of
classical mechanics can be solved without resorting, in one way or
in another, to these laws. In other words, the Newtonian laws of
motion, or any of its equivalent formulations, Poisson brackets,
Hamilton formulation, etc., \cite{Jose1} constitute an example of
Fundamental Principles in Physics, since they provide the
possibility of making predictions about the behavior of nature.
Clearly, these predictions can be confronted against measurement
readouts. Fortunately the story does not finish here. Indeed, the
Newtonian view of the universe assumes also some other concepts,
which are taken for granted as a conceptual background. For
instance, in this perspective of the universe the ideas of time,
space, and simultaneity are absolute, i.e., they do not depend on
the observer \cite{Newton1}. The possibility of testing these
premises remained for several centuries outside the possibilities
of mankind. The technology that allowed the option of checking
them appeared at the end of the nineteenth century. Among the
experiments in this context, we may mention the Michelson and
Morley proposal \cite{Mandel1}, which is in essence an
interference experiment and the Trouton--Noble experiment
\cite{Trouton1, Chase1, Teukolsky1}. In this last case a suspended
parallel-plate capacitor is held by a fine torsion fiber and is
charged. If the aether theory were correct, the change in
Maxwell's equations due to the Earth's motion through the aether
would lead to a torque causing the plates to align perpendicular
to the motion.

The options that light interferometry offers us do not end with
Michelson--Morley, some of the most important proposals to test
gravity do involve laser interferometry \cite{Turyshev1}.

We may at this point answer our original question stating that the
concept of Fundamental Principles in Physics will embody, not
only, some fundamental equations, as could be the Newtonian motion
laws, or the Schr\"odinger equation, but also basic assumptions
concerning the structure of time and space. Additionally, as will
be shown below, some physical theories, as quantum mechanics, have
more than one interpretation. Therefore, another issue to be
addressed in the present work comprises the use of interference
devices in order to discard some models concerning quantum theory,
for instance the hidden variables theory \cite{Bohm1, Nelson1}.

Now that we have clearly stated the goals in this work let us
answer a second question that at this point can be posed: What can
be obtained from the application of interference techniques in
connection with fundamental principles in physics? The answer to
this question can be contemplated from two different points of
view: (i) The unification between general relativity and quantum
theory is currently one of the most challenging theoretical tasks
in modern physics. The situation in this issue requires now the
help of phenomenology as an important element which shall mitigate
some of the deficiencies that the models show. Indeed,
phenomenology can help to fix bounds upon some parameters that
appear in connection with these theories \cite{Amelino3}, and
which cannot be obtained from the corresponding approaches
\cite{Kiefer1, Rovelli1}; (ii) Looking at the experimental side of
this issue we notice that the requirements involved in this kind
of experiments have spurred the development of new technology, as
the case of the detection of gravitational waves clearly shows
\cite{Kudoh1}. In other words, theoretical work and the
development of technology live in a state that could be denoted a
symbiosis, since each of them benefits from the other one.

\subsection{What is Classical Interference?}

The idea of interference takes us to the notion of wave motion, a
concept that at this point allows us to remain completely within
the realm of classical physics. One of the most important physical
phenomenon involving the concept of interference is the
propagation of light. Robert Boyle and Robert Hooke made the first
observation of interference, though the wave behavior of light was
recognized until Fresnel confirmed Young's experiments. The delay
in this aspect is explained by the fact that this hypothesis
disagreed with the corpuscular model introduced by Newton
\cite{Guenther1}. By the way, this last remark also shows that
sometimes the weight of a scientific figure becomes more important
than experimental evidence.

The existence of interference (either of light, of quantum matter,
etc.) requires the fulfillment of several conditions. The first
one is the superposition principle, the one exists in connection
with motion equations which are linear differential equations. We
may define this principle stating that the sum of solutions to the
corresponding motion equation is also a solution to equation
\cite{Mandel1}. If this principle does not hold, then the
phenomenon of interference, in general, disappears.

Let us illustrate this last assertion with an example. Consider a
string, for instance, it can be the string of a violin, when the
performer plays a staccato note (a bowed string behaves in a very
different way \cite{Acustica1}), such that any pulse generated by
the performer travels with speed $v$. Let us denote the amplitude
of this pulse by $y $. It can be shown that the dynamics of any
pulse, in a very idealized scheme, is given by \cite{Guenther1}

\begin{eqnarray}
\frac{\partial^2 y}{\partial x^2} = \frac {1}{v^2}\frac{\partial^2
y}{\partial t^2}\label{Interference1} .
\end{eqnarray}

Notice that if $f$ and $g$ are twice differentiable, then
$f(vt-x)$ and $g(vt+x)$ do satisfy, each one of them,
(\ref{Interference1}). Define $h_{(\pm)}(x, t) = f(vt-x) \pm
g(vt+x)$. The linearity of the motion equation implies that
$h_{(\pm)}(x, t)$ satisfies (\ref{Interference1}).

Let us now analyze the consequences of a non--linear term in the
equation, and, in order to do this, consider a generalization of
(\ref{Interference1}) in the following form

\begin{eqnarray}
\frac{\partial^2 y}{\partial x^2} + \frac{\partial (y)^2}{\partial
x}= \frac {1}{v^2}\frac{\partial^2 y}{\partial
t^2}\label{Interference3}.
\end{eqnarray}

Assume that $f(vt-x)$ and $g(vt+x)$ do satisfy it, and introduce
$h_{(+)}(x, t)$ into (\ref{Interference3})

\begin{eqnarray}
\frac{\partial^2 f}{\partial x^2} + \frac{\partial^2 g}{\partial
x^2}+\frac{\partial (f)^2}{\partial x}+ \frac{\partial
(g)^2}{\partial x} + 2\frac{\partial (fg)}{\partial x}= \frac
{1}{v^2}\Bigl(\frac{\partial^2 f}{\partial t^2} + \frac{\partial^2
g}{\partial t^2}\Bigr)\label{Interference4}.
\end{eqnarray}

It is readily seen that $h_{(+)}(x, t)$ is not a solution to our
motion equation. In other words, the sum of two solutions of the
motion equation is not anymore a solution. This remark has several
consequences. For instance, as will be shown below, the effect
called interference is related to the feature of summing solutions
and ending up with another solution. Therefore, phenomena ruled by
non--linear motion equations will not, in general, have
interference as one of its characteristics.

Now that we fathom better the implications of a non--linear motion
equation let us analyze the possibilities that its counterpart
offers us. Harking back to (\ref{Interference1}) assume now that
at a certain point $\tilde{x}$ and at a certain time $\tilde{t}$
we have that $f(v\tilde{t}-\tilde{x}) = -a$ and
$g(v\tilde{t}+\tilde{x})= +a$, where $a>0$. The physical meaning
of $a$ can be easily understood with the example of a violin. We
may suppose that positive values of $y$ denote displacements along
the positive direction of the $y$--axis. This last remark implies
that the right propagating pulse $f(vt-x)$ has, at the
aforementioned position and time, a negative displacement along
the $y$--axis, whereas $g(vt-x)$ has a positive displacement.
Clearly, we have that $h_{(+)}(\tilde{x}, \tilde{t}) =0$. The
displacement of the string is neither $a$  nor $-a$, as a matter
of fact there is a vanishing displacement of the string. This is a
simple example of destructive interference.

This case of a staccato note on a violin string is an example of
the superposition principle, and clearly it does not exhaust the
possibilities in this direction. Indeed, physical situations
satisfying the superposition principle teem in physics. For
instance, in an adiabatic process the propagation of sound can be
contemplated as a first--order linear differential equation for
the pressure as a function of the density of the air
\cite{Colin1}. The propagation of light, a case analyzed below, is
also related to the superposition principle. Additionally, we may
mention the motion of water, under certain circumstances. These
simple examples show that every day we are in contact with the
phenomenon of interference, though we may not notice it.

At this point we may give an answer to the question: What is
interference? It is a direct consequence of the superposition
principle, and it is the fact that at any point in which two
perturbations are superposed the resulting perturbation is the
algebraic sum of each one of them.

Preparing the material for the topics that we will here discuss
let us mention that a very important phenomenon associated with a
linear differential equation of motion is the propagation of
light. Indeed, the behavior of the electromagnetic field is
governed by Maxwell's equations \cite{Scully1}. The debate between
the corpuscular behavior of light and the possibility of a wave
description held for many years. One of the reasons behind this
longevity can be tracked down to the fact that the observation of
interference requires very restrictive conditions
\cite{Guenther1}. In connection with light interference appears as
a set of dark and bright bands denoted by fringes. The bright
regions appear when a number of waves add together to produce an
intensity maximum of the resultant wave, a case called
constructive interference. On the other hand, destructive
interference happens if the involved waves add together to produce
an intensity minimum of the resultant wave. In our oil--dominated
world we may find these fringes on oil films on a wet roadway, a
situation that, unfortunately, nowadays is quite common.

But, what are the conditions that allow the detection of
interference? As a part of the answer we may say that interference
involves always small dimensions. For instance, two slits
separated by only a couple of millimeters, or in the case of the
aforementioned oil films, they have a thickness of very few
millimeters. These dimensions are not imposed by the wavelength of
light, but by a fundamental property called coherence
\cite{Mandel2}. This last statement could sound puzzling, so let
us explain the concept of coherence a little bit better. This idea
will be handled in section III, though here we provide a brief
introduction to the main concepts.

Consider two electric fields, then they are said to be coherent if
the interference term

\begin{eqnarray}
<\vec{E}_1\cdot\vec{E}_2> = \sqrt{I_1I_2}\cos\big(\delta\big)
\label{Inter1},
\end{eqnarray}
is nonzero within the region occupied by both electric fields,
here $I_i = <\vec{E}_i\cdot\vec{E}_i>$, and $\delta$ is the phase
between these two fields. We may rephrase this last remark saying
that two waves are coherent if the associated electric fields have
a constant phase relation $\delta$. In other words, the property
of coherence is related to the comparison of the relative phase
between these fields. This comparison can be done in two different
ways, temporal and spatial \cite{Mandel2}.

We may even assert that the simplest manifestations of
correlations in light fields are interference effects that arise
when two light beams are superposed. A good comprehension of this
property is due, since any experiment intending to prove some
physical model, and embodied in the context of interference, must
take into account the fundamental features defining this property.

From the last paragraph a question assails us, namely, in the case
of light, what are the conditions that entail the possibility of
observing interference?  At this point it is noteworthy to comment
that in optics the corresponding {\it elongations} to be specified
are the electric and magnetic fields strengths. Obviously, in this
process we must take into consideration the phases of the involved
waves. The situation is simplest when the phases do not vary
noticeable in time. Of course, this is a very rough idealization,
and, as shown below, we only require them to be constant over time
intervals of the length of the observation time. When this last
condition is fulfilled an interference pattern will emerge. Let us
now make this last statement more precise. For the sake of
simplicity take the case of two plane waves with equal frequencies
and linearly polarized. In addition, the amplitudes will be
considered constant and real, $E_{(1)}$ and $E_{(2)}$, where
$\vec{k}_{(j)}$, $\omega_{(j)}$, and $\theta_{(j)}$ denote the
wave vector, the frequency, and the phase, respectively.

\begin{eqnarray}
E_{(j)}(\vec{r}, t) =
E_{(j)}\exp\Bigl\{i\Bigl(\vec{k}_{(j)}\cdot\vec{r}-\omega_{(j)}t-\theta_{(j)}\Bigr)\Bigr\},~~~j=1,2
\label{Interference5}.
\end{eqnarray}

The usual definition of intensity is one--half of the square of
the electric field strength averaged over several oscillation
periods \cite{Mandel1}. The superposition principle states that
the total electric field strength is the sum of $E_{(1)}(\vec{r},
t)$ and $E_{(2)}(\vec{r}, t)$. Then we obtain for the intensity

\begin{eqnarray}
I(\vec{r}, t) = E^2_{(1)}+E^2_{(2)}
+2E_{(1)}E_{(2)}\cos\Bigl\{\Bigl(\vec{k}_{(2)}-\vec{k}_{(1)}\Bigr)\cdot\vec{r}
-\Bigl(\omega_{(2)}-\omega_{(1)}\Bigr)t
-\Bigl(\theta_{(2)}-\theta_{(1)}\Bigr)\Bigr\}
\label{Interference6}.
\end{eqnarray}

The definition of monochromatic wave also implies that
$\theta_{(j)}$ must be constants. If the frequencies are the same,
then we have a standing interference pattern, the one hinges
critically upon the difference $\theta_{(2)}-\theta_{(1)}$. On the
other hand, if the waves differ in the frequency, then we obtain a
sinusoidal dependence on time of the interference pattern. This
analysis is quite simple, nevertheless, the experimental
requirement of monochromaticity was at reach only after the
appearance of lasers \cite{Paul1}.

The problem with light emitted by conventional sources is that it
exhibits fast fluctuations, in both amplitude and phase of the
electric field. This entails that two independents beams, i.e.,
two beams emitted from different sources or from two different
parts of the same source, cannot produce a detectable interference
pattern. Indeed, a fleeting glance at (\ref{Interference6}) shows
that the pattern formed at a certain time and point will be
displaced by a random fraction of the fringe spacing each time
that the phases, $\theta_{(2)}$ and $\theta_{(1)}$, change their
value.

Up to this point everything seems to be in its place.
Nevertheless, from the last arguments we cannot understand how the
Michelson--Morley experiment provided, by means of an interference
experiment, the data that led, at least partially, Einstein to the
formulation of the Special Theory of Relativity. The answer is
very simple. The interference term appearing on the right--hand
side of (\ref{Interference6}) depends only upon the parameter
$\theta_{(2)}-\theta_{(1)}$, in other words, the only condition
that we need is $\theta_{(2)}-\theta_{(1)} = const.$ The phases
may fluctuate, but not independently, a definite correlation
between them must exist. Experimentally, this can be achieved
making the interfering beams replicas of only a primary beam. The
use of a beam splitter suffices. This is the core of the idea in
the Michelson--Morley experiment \cite{Scully1}.

How far can we walk testing fundamental principles in physics if
we resort only to classical interference experiments? We should
not belittle the possibilities that classical experiments could
provide us. Indeed, remember that one of the most important set of
data that Einstein had at his disposal (in the formulation of
Special Theory of Relativity \cite{Einstein1}) was the null
experiment, no existence of aether \cite{Whittaker1}, that the
Michelson--Morley experiment implied. This experiment contains the
empirical evidence behind one of the two premises of special
relativity, namely, the speed of light is a constant, i.e.,
independent from the motion of the observer \cite{Ohanian1}.
Moreover, the possibilities of light interferometry also embrace
the test of the validity of Lorentz symmetry \cite{Camacho1,
Camacho2}. At least if we consider the introduction of some kind
of deformed dispersion relation, a fact related to some quantum
gravity theories \cite{Amelino1, Amelino2}. This point will be
explained in section II.

Up to now everything remains in the realm of classical physics. If
the reader has been patience enough and followed our arguments up
to this point she/he could have the impression that classical and
quantum interference processes are events clearly distinctive.
This is not the case. For instance, in the detection process of
light the corpuscular behavior becomes dominant. Indeed, an
interference pattern will be formed only if there is a large
number of photons impinging upon the detecting screen
\cite{Paul1}.

\subsection{What is Quantum Interference?}

Let us now address the issue of quantum interference. The advent
of quantum theory opened up an ocean of possibilities in
connection with interference experiments. Indeed, the use of
particles in relation with interference devices is completely
excluded in the context of classical mechanics \cite{Jose1}, since
in classical mechanics particles are not endowed with wave--like
behavior. This may sound redundant, but as mentioned below,
particles in quantum mechanics may show wave--like features.
Quantum interference can be considered as the interference of the
de Broglie wave associated with an ensemble of particles with
itself. This wave--like and particle--like features ascribed to
one physical entity, as light, or neutrons, is a very common
characteristic in modern physics \cite{Omnes1}. Additionally, it
is inexorably linked to the concept of measurement in quantum
theory \cite{Lamb1}. This dual behavior has been the reason for
many debates, and has spurred several modifications to the
interpretation of the theory, for instance, the statistical
interpretation \cite{Penha1}, in which the quantum description of
any system is from the very beginning assumed as incomplete. In
other words, quantum mechanics is considered not a fundamental
theory, but a phenomenological model.

The emergence of quantum theory cannot be ascribed to the labor of
one researcher, rather it is the confluence of the ideas of many
people, beginning with Planck and his work on black body
radiation, and continuing with Schr\"odinger, Dirac, Pauli, etc.
In the non--relativistic limit, and for particles with no spin,
the fundamental motion equation is \cite{Sakurai1}

\begin{eqnarray}
i\hbar\frac{\partial \psi}{\partial t} =
-\frac{\hbar^2}{2m}\nabla^2\psi +V\psi\label{Schroedinger1}.
\end{eqnarray}

It is a linear equation, and, according to the Copenhagen
interpretation \cite{Omnes1}, the variable $\vert\psi(\vec{r},
t)\vert^2$ is related to the probability of detecting the
corresponding system in a neighborhood of point $\vec{r}$ at time
$t$. Additionally, $\psi$ is easily expressed as a vector of a
Hilbert space. Since the description of a quantum system is given
by elements of a Hilbert space, then the sum of two of them is an
allowed physical state. This brings into quantum theory the
phenomenon of interference. This has being, since the very
inception of the theory, a controversial and controverted point,
and will be one of the core points in the present work.

Let us explain this fact a little bit better. Consider a cat whose
state will be described by quantum mechanics, here we will make
use of the argument of Schr\"odinger's cat \cite{Omnes1}. Our cat
remains always inside a closed box containing a radioactive
substance, the one is also a quantum system. The state of the
radioactive substance can be expressed as the superposition of two
orthogonal states, a non--decayed state and a decayed one. In
consequence, the state of the cat will be expressed in terms of
two orthogonal states, $\psi_{(d)}$ and $\psi_{(a)}$ the cat is
dead and the cat is alive, respectively. The physical state of the
cat, $\psi_{(c)}$, reads then $\psi_{(c)} = \alpha_{(d)}\psi_{(d)}
+ \alpha_{(a)}\psi_{(a)}$, where $\alpha_{(a)}$ and $\alpha_{(d)}$
are complex numbers whose modulus is strictly larger than 0 and
smaller than 1.

The first conceptual problem that we encounter is the meaning of
$\psi_{(c)} = \alpha_{(d)}\psi_{(d)} + \alpha_{(a)}\psi_{(a)}$.
According to our classical understanding of the world the cat is
dead or alive, no place for a superposition of these two mutually
excluding cases can be considered within the classical perspective
of the universe. The superposition principle related to
(\ref{Schroedinger1}) entails a new possibility, which was absent
in the classical version. Clearly, a second problem appears in
connection with (\ref{Schroedinger1}). Indeed, we may pose the
following question: How do we recover a classical world from a
quantum theory? This is a very complicated problem, the one has
several approaches, some of them even deny the existence of the
measurement problem in quantum mechanics \cite{Bub1}. On the other
hand, we may find approaches in which the solution is obtained as
a consequence of the interaction of the corresponding system with
its environment \cite{Giulini1}, an idea named decoherence model
\cite{Zeh1}. At this point we may ask ourselves: Is there any
relation between the last paragraph and the task of the present
work? This is a good question, and luckily the answer is
affirmative. Let us fathom better this last statement. Decoherence
model predicts the existence of non--classical states, which, due
to the interaction with the environment, have a very short life
\cite{Joos1}. These non--classical states emerge as a direct
consequence of the superposition principle, and therefore, in
principle, they define a way in which decoherence model can be,
experimentally, tested. In other words, quantum interference could
allow us to discard the decoherence model, or to take it as a
fundamental element of the theory.

Now that we know that interference experiments can shed some light
upon some thorny points of the foundations and interpretation of
quantum theory we may address some additional issues. For
instance, can quantum interference experiments tell us something
about the relation between the structure of space--time and
quantum mechanics? Once again, we receive a gleaming yes. Indeed,
we may find a series of experiments that exhibit how gravity
appears in the realm of quantum theory. This experiment, first
performed in 1974 \cite{Colella1} contains an interference pattern
of thermal neutrons induced by gravity. A brief description of
this experiment is the following one. A nearly monoenergetic beam
of neutrons is split into two parts by silicon crystals, they
follow paths with different gravitational potential, and
afterwards they are brought together. The difference in the
gravitational potential implies the emergence of a phase shift
between the two beams. This difference in phase goes like (here
$m$ denotes the neutron mass)

\begin{eqnarray}
\Delta \phi\sim \Bigl(\frac{m}{\hbar}\Bigr)^2
 \label{Colella1}.
\end{eqnarray}

This is a purely quantum mechanical process, since the limit
$\hbar\rightarrow 0$ implies that the interference pattern gets
washed out, this experiment will be explained in a more complete
way below. The importance of this experiment can not be reduced to
one aspect. For instance, the dependence of (\ref{Colella1}) on
the mass of the involved particles has originated a hot debate
about the possibility of a non--geometric element of gravity at
the quantum realm \cite{Ahluwalia1, Ahluwalia2}. In other words,
gravity--induced quantum interference can be used to discuss the
possible breakdown, at the quantum level, of the Equivalence
Principle \cite{Will1}. This statement enhances the relevance of
quantum interference as a tool for testing the structure of
space--time.

Finally, as another point in which fundamental physics does have a
fundamental role in the development of technology let us mention
that quantum interference plays an important role in the
implementation of entangling quantum gates between atomic qubits
\cite{Simon1}, a fact that lies right in the center of the present
efforts  of quantum computation \cite{Haeffner1}.

Summing up, interference is a fundamental element of quantum
theory and can be used to test the validity of the equivalence
principle, in the construction of quantum computation, etc.

\subsection{Geometric Phases in Quantum Theory}

The concept of geometric phases could be an important tool in
connection with proposals that try to test some aspect of
fundamental physics, and here we explain its meaning. As a matter
of fact, examples of geometric phases, classical and quantum, teem
in physics, and many ordinary situations that we do not usually
associate with geometric phases can be rephrased in terms of them.
As an illustrative example of this let us consider the precession
of a Foucalt pendulum . The usual analysis of its movement is done
in terms of Coriolis force \cite{Jose1}, nevertheless, we may
contemplate this situation from a different perspective. Indeed,
suppose that we have a point mass with its corresponding
gravitational field, and that our Foucalt pendulum is transported
along a closed curve, say $A$, in the aforementioned gravitational
field. At this point we introduce two conditions upon the period
and amplitude of the motion of the pendulum, i.e., they are
smaller than the typical time and distance of the transport
motion, respectively. If additionally the curve lies on the
surface of a sphere, then, when the pendulum returns to its
initial position, its invariant plane will have rotated by some
non--vanishing angle. In the case which in the pendulum has been
transported along constant latitude, say $\alpha$, then the
rotation angle reads $2\pi\cos(\alpha)$ \cite{Jose1}. This is a
simple example of a classical phase. In general we may state that
if an integrable classical Hamiltonian describes a bound motion
which depends on parameters that suffer a very slow change, then
the adiabatic theorem \cite{Jose1} states that the action
variables of the motion are conserved. For angle variables the
change does not merely contain the time integral of the
instantaneous frequency, it also shows an extra angle which
depends only upon the circuit in the parameter space
\cite{Hannay1}.

All the foregoing arguments have concerned classical systems. Let
us now address the issue of quantum phases. The first work in
which a quantum phase was explicitly derived and analyzed was done
by Berry \cite{Berry1}. The first derivation of this phase
involved the adiabatic theorem, the one states that if a certain
Hamiltonian $H$ changes gradually from some initial form $H^{(i)}$
to a final one $H^{(f)}$, and a particle was initially in the
$n$th eigenstate of $H^{(i)}$, then it will be carried (according
to Schr\"odinger equation) into the $n$th eigenstate of $H^{(f)}$
\cite{Griffiths1}.

Consider a Hamiltonian $H(t)$ which shows a non--trivial time
dependence, then the eigenfunctions and eigenvalues themselves are
also time dependent.

\begin{eqnarray}
H(t)\psi_n(x,t) = E_n(t)\psi_n(x,t)\label{Berryphase1}.
\end{eqnarray}

At this point the adiabatic theorem is introduced, and in
consequence, if $H(t)$ changes in a gradual way, then our particle
picks up at most a time--dependent phase factor. This means that
the wavefunction, written in terms of its eigenfunction, reads

\begin{eqnarray}
\Psi_n(x,t) =
\psi_n(x,t)\exp\Bigl\{-\frac{i}{\hbar}\int_0^tE_n(\tau)d\tau
\Bigr\}\exp\Bigl\{i\gamma_n(t)\Bigr\}\label{Berryphase2}.
\end{eqnarray}

This new phase $\gamma_n(t)$ is called the geometric phase.

Introducing (\ref{Berryphase2}) into the time--dependent
Schr\"odinger equation the expression for the geometric phase can
be deduced. For instance, let us assume that the corresponding
Hamiltonian embodies more than one parameter changing with time
(for the case of only one parameter the geometric phase is
trivial), say $R_1(t)$, ..., $R_N(t)$. Under these circumstances
it can be shown \cite{Berry1} that if the Hamiltonian returns to
its original form after a time $T$, then the geometric phase can
be expressed as a line integral around a closed loop ${\tilde{C}}
$ in the corresponding parameter space

\begin{eqnarray}
\gamma_n(T) = i\oint_{\tilde{C}}<\psi_n\vert\nabla_R\psi_n>\cdot
d\vec{R}\label{Berryphase3}.
\end{eqnarray}

It is deeply rooted in quantum mechanics the acceptance of the
argument that the phase of a wave function is arbitrary, i.e., it
contains no physical information. However, phase variations and
the so--called geometric phase are gauge invariant quantities,
i.e., they may embody relevant physical information. The
observability of this effect is related to variations of the
corresponding phase shifts, i.e., auxiliary interference patterns
are required \cite{Bayer1}.

The Aharanov--Bohm \cite{Aharonov1} effect can be thought as an
example of Berry's phase, and its first experimental verification
was done resorting to electron holography \cite{Tonomura1}. Its
existence has been verified in other domains, for instance, using
an optical fiber \cite{Tomita1}. Since the experimental
verification of geometrical phases usually involves some kind of
interference experiment, then we may use them to test some
possible effects related to metric theories.

The Aharonov--Bohm effect can be considered one of the first
examples of a geometric phase, and deserves an analysis of its
own, since it involves some very interesting features. Consider a
particle with electric charge $q$, and an ideal solenoid (this
condition is imposed just for the sake of clarity) with
cylindrical symmetry, such that inside the solenoid the magnetic
field is constant, and along the axis of symmetry, and zero
outside, see figure 1.


\setlength{\unitlength}{0.00083300in}
\begingroup\makeatletter\ifx\SetFigFont\undefined

\def\x#1#2#3#4#5#6#7\relax{\def\x{#1#2#3#4#5#6}}%
\expandafter\x\fmtname xxxxxx\relax \def\y{splain}%
\ifx\x\y
\gdef\SetFigFont#1#2#3{%
  \ifnum #1<17\tiny\else \ifnum #1<20\small\else
  \ifnum #1<24\normalsize\else \ifnum #1<29\large\else
  \ifnum #1<34\Large\else \ifnum #1<41\LARGE\else
     \huge\fi\fi\fi\fi\fi\fi
  \csname #3\endcsname}%
\else \gdef\SetFigFont#1#2#3{\begingroup
  \count@#1\relax \ifnum 25<\count@\count@25\fi
  \def\x{\endgroup\@setsize\SetFigFont{#2pt}}%
  \expandafter\x
    \csname \romannumeral\the\count@ pt\expandafter\endcsname
    \csname @\romannumeral\the\count@ pt\endcsname
  \csname #3\endcsname}%
\fi \fi\endgroup
\begin{center}
\setlength{\unitlength}{2mm}
\begin{picture}(100,50)(-5,20)
\put(40,50){\circle{70}}\put(40,50){\oval(39,40)[l]}
\put(41,50){\oval(40,40)[r]}\put(35,26.5){\vector(-1,0){10}}
\put(45,26.5){\vector(1,0){10}} \put(40,29.5){\makebox(0,0){$S$}}
\put(40.5,70){\makebox(0,0){$D$}}
\put(40.5,55){\makebox(0,0){$Solenoid$}}
\put(35,20){\makebox(0,0)[lb]{\smash{\SetFigFont{12}{14.4}{rm}{Figure
1}}}}
\end{picture}
\end{center}

A source point, $S$ emits particles, and they can follow two
different trajectories, to the {\it left} or to the {\it right} of
the solenoid, as shown by the arrows in figure 1. Finally, at
point $D$ they are brought together. There is a striking feature
that emerges in the interference pattern. Though the particles
never enter the region in which the magnetic field is
non--vanishing, a non--null phase difference between the two beams
emerges

\begin{eqnarray}
\Delta =\frac{q}{\hbar c}\oint\vec{A}\cdot d\vec{l}\label{Ahb1}.
\end{eqnarray}

This last expression could deceive us, i.e., we could state that,
since in electrodynamics the vector potential $\vec{A}$ is not
unique \cite{Jackson1}, this expression is gauge--dependent. This
is not the case, and can be easily proved resorting to Stokes'
theorem, i.e., this phase becomes

\begin{eqnarray}
\Delta\phi =\frac{q}{\hbar c}\int_S\nabla\times\vec{A}\cdot
d\vec{S}\label{Ahb2}.
\end{eqnarray}

But $\nabla\times\vec{A} = \vec{B}$, and therefore $\Delta$ is
gauge--invariant. The situation in the context of the
Aharonov--Bohm effect is very rich. For instance, as a quite
surprising case let us mention that, under certain circumstances,
neutral particles with magnetic moment can exhibit the
Aharonov--Bohm effect \cite{ACasher1}. A quantum effect of the
Aharonov--Bohm type for particles with an electric dipole appears,
as a consequence of the relation between the topological
properties of the phase shift and the linear and angular momentum
of the electromagnetic field \cite{Spavieri1}. This topic is
spiced with a hot debate about the force that a neutron
experiences in connection with the Aharonov--Casher effect.
Indeed, some authors state that there is an electric force on a
classical model for a neutron \cite{Boyer1}, whereas others deny
the existence of this force \cite{APearle1}.

At this point we must have a more careful treatment of the
experimental situation associated to the gauge--invariance of the
effects of the Aharonov--Bohm type. Let us start mentioning that
this kind of experiments require the observation of the relative
displacement of the interference patterns, i.e., a sole
interference pattern does not suffice. The experimental detection
of the phase shift variation, $\Delta\phi$, needs one value of
$\vec{A}$, and a second one, say $\vec{A}_0$, which plays the role
of a reference parameter, see, for instance, the reference beam
mentioned in \cite{Tonomura1}. Hence the fringes related to
$\phi(\vec{A})$ are compared against those emerging from
$\phi(\vec{A}_0)$, and, in consequence, $\phi(\vec{A})-
\phi(\vec{A}_0)$ can be obtained. Clearly, a gauge transformation
leaves this last quantity unaltered.

Finally, we may also mention that there are several types of
Aharonov--Bohm effects. Indeed, for instance, the so--called
molecular Aharonov--Bohm effect \cite{Sjoqvist1} does not share
some of the properties to which we are used in the standard
Aharonov--Bohm effect, i.e., the molecular version is neither
non---local or topological. Additionally, in the context of
gravity there is an analogue of this effect when particles are
constrained to move in a region where the Riemann curvature tensor
does not vanish \cite{Marques1}.

\section{Neutron Interference}

\subsection{Metric Theories of Gravity and Quantum Interference}

\subsubsection{The Weak Equivalence Principle}

Modern physics has its foundations in two theories, namely,
quantum mechanics \cite{Omnes1} and general relativity
\cite{Misner1}. Any experimental test of general relativity must
bear clearly in mind the postulates of the theory. In order to
have a clear idea of the bounds involved with quantum
interferences experiments we now establish the different premises
of this theory. The starting point is the Weak Equivalence
Principle (WEP), the one states: {\it If an uncharged test body is
placed at an initial event of space--time and given an initial
velocity, then its subsequent trajectory will be independent of
its internal structure and composition} \cite{Will1}.

This principle has its experimental foundation in the universality
of free fall, an experiment first performed by Galileo
\cite{Gaga1}. It entails a relation between two different concepts
of mass. Let us address this point. Consider the case of a
classical particle freely falling in a homogeneous gravitational
field. The motion equation is given by \cite{Jose1}

\begin{eqnarray}
m_{(i)}\frac{d^2 \vec{r}}{dt^2} = -m_{(p)}g\hat{z}\label{Atomic1}.
\end{eqnarray}

In this last expression $m_{(i)}$, $m_{(p)}$, and $\hat{z}$ denote
the inertial mass, the passive gravitational mass, and the unit
vector along the $z$--axis (we assume that the gravitational field
is along this last axis and has a direction contrary to the
positive direction of $z$). The inertial mass appears already in
Newton's second law of motion \cite{Jose1}, whereas the passive
gravitational mass is the response of the particle to the presence
of a gravitational field, in this case the field of the Earth.
There is an additional concept of mass, active gravitational mass
\cite{Will1}, though we will not consider it here. The equivalence
principle states that $m_{(i)} = m_{(p)}$. In other words, the
mass term drops out from the motion equation, and this simple fact
is the reason that allows us to state that gravity, in classical
mechanics, is a purely geometric theory. This point was
distinguished already by Newton since he pointed out that the
universality of free fall implies that $m_{(i)}/m_{(p)}$ is a
constant for all bodies. This premise has been subjected to many
experimental tests, from Galileo \cite{Gaga1} to the sophisticated
experiments of our time \cite{Dittus1, Haugan1, Adel1,
Lockerbie1}. In general relativity the universality of free fall
is a fundamental element in the formulation of the theory in
purely geometrical terms, it appears in WEP \cite{Adler1}.

\subsubsection{Einstein Equivalence Principle}

Nevertheless, WEP does not suffice to define general relativity.
The axiom that divides the theories of gravity into metric
theories (those who satisfy it) and non--metric theories (those
who do not) is Einstein Equivalence Principle (EEP): {\it (i) WEP
is valid; (ii) the outcome of any local non--gravitational test
experiment is independent of the velocity of the freely falling
measuring device (a condition usually known as Local Lorentz
Invariance), and (iii) the outcome of any local non--gravitational
test experiment is independent of the where and when in the
universe it is performed (denoted as Local Position Invariance)}
\cite{Will1}.

Since physics is an experimental science, we must state quite
clearly the status of general relativity in this aspect. The
experimental confirmation of general relativity ranges from the
classical tests (deflection of light, in which a light beam
suffers an optical bending induced by the gravitational field of a
body \cite{Richard1}) to, for instance, experiments that are
designed to detect a time variation of the Newtonian gravitational
constant \cite{Ritter1}. Any test of the predictions of general
relativity is, in some way, an indirect tests of its axioms. Of
course, we must be quite careful, since, for instance, any
experimental confirmation of the gravitational red--shift
prediction would not be a proof of the validity of general
relativity. It would rather be a proof of the validity of all
metric theories of gravity which also contain this effect, such as
Brans--Dicke theory \cite{Brans1}, or some theories with prior
geometry \cite{Rosen1}.

\subsubsection{Gravitomagnetism}

We have explained that general relativity is a generalization of
the Newtonian gravity theory \cite{Misner1}, and some experiments
in which this difference appears have been mentioned and
described. There is one effect present in most metric theories,
including general relativity, which has no Newtonian counterpart,
namely, the gravitomagnetic effect, also known as Lense--Thirring
effect \cite{Ciufolini1}. The name of this effect is justified,
since in post--Newtonian approximation it differs from Newtonian
gravity as a magnetic force differs from an electric one. This
effect emerges as a consequence of mass--energy currents. Let us
explain this last statement a little bit better, and in order to
do this we will resort to an analogy with electrodynamics. In
electrodynamics, in a frame in which an electrically charge is at
rest, we have electric field. If the sphere starts to rotate, then
a magnetic field appears, and the strength of this field hinges
upon the angular velocity. In a similar manner, in general
relativity a non--rotating massive sphere produces the
Schwarzschild field \cite{Misner1}. As soon as the sphere begins
to rotate the gravitomagnetic effect emerges as an additional
element that modifies the structure of space--time. In contrast to
this situation in Newtonian theory the only source of gravity is
mass, if the mass of a sphere rotates or not it is completely
irrelevant for the calculation of the gravitational field. Though
this field has already been detected \cite{Ciufolini2}, it has to
be clearly stated that this experiment was performed employing
classical systems. Nevertheless, the possible consequences on
quantum systems, particularly on the coupling
spin--gravitomagnetic field, require a much deeper analysis, i.e.,
it is almost always assumed that the coupling orbital angular
momentum--gravitomagnetism can be extended to explain the coupling
spin--gravitomagnetic field \cite{Mashhoon2}. Nevertheless, this
assumption must be subject to experimental scrutiny
\cite{Ahluwalia1}.

\subsubsection{Gravity--Induced Interference}

The behavior of the mass parameter is quite different in classical
mechanics and in quantum theory. How does the situation look like
in quantum mechanics? Is the extrapolation from the classical
realm to the quantum domain straightforward? The answer is no. The
situation is rather different in Schr\"odinger equation (once
again we assume a homogeneous gravitational field) as can be seen
from the structure of the corresponding motion equation
\cite{Fluegge1}

\begin{eqnarray}
-\frac{\hbar^2}{2m}\nabla^2\psi +mgz\psi =
i\hbar\frac{\partial\psi}{\partial t}\label{Atomic2}.
\end{eqnarray}

It is readily seen that now the mass does not cancel. Notice also
that (\ref{Atomic2}) entails that mass appears always in the
combination $m/\hbar$. In other words, the detection of quantum
effects of gravity will inexorably imply the emergence also of
mass, and in consequence, we may wonder if at quantum level the
equivalence principle remains valid, or if it has to be restricted
to the classical world? The first direct evidence of the presence
of a gravitational field as a non--trivial quantum effect was
obtained in 1974 by Colella, Overhauser, and
Werner\cite{Colella1}, a proposal known as COW. There is an
additional result that could be considered also a test of gravity
in the quantum domain, the red--shift experiment performed by
Pound and Rebka, in which the effects of gravity upon the
frequency of a photon are experimentally confirmed \cite{Rebka1}.
Nevertheless, a careful analysis of this case shows that $\hbar$
does not appear explicitly, and hence the interpretation of this
experiment as a test of gravity in the quantum domain seems to be
a little bit feeble. The COW proposal has been repeated in a
series of experiments which showed an outstanding sophistication
of the involved technology \cite{Colella2, Colella3, Colella4}.

The phenomenon dealt with in this series of experiments can be
denoted a gravity--induced quantum interference. The idea is to
use an almost monoenergetic beam of thermal neutrons. This last
statement means a kinetic energy of about 20 MeV, which is
tantamount to a speed of $2000ms^{-1}$. The primary beam is split
into two parts, such that each one of the new beams travels along
different paths. These paths define a parallelogram of sides $l_1$
and $l_2$. Let us denote the vertices of this parallelograms by
$A$, $B$, $C$, and $D$, see figure 2

\setlength{\unitlength}{0.00083300in}
\begingroup\makeatletter\ifx\SetFigFont\undefined

\def\x#1#2#3#4#5#6#7\relax{\def\x{#1#2#3#4#5#6}}%
\expandafter\x\fmtname xxxxxx\relax \def\y{splain}%
\ifx\x\y
\gdef\SetFigFont#1#2#3{%
  \ifnum #1<17\tiny\else \ifnum #1<20\small\else
  \ifnum #1<24\normalsize\else \ifnum #1<29\large\else
  \ifnum #1<34\Large\else \ifnum #1<41\LARGE\else
     \huge\fi\fi\fi\fi\fi\fi
  \csname #3\endcsname}%
\else \gdef\SetFigFont#1#2#3{\begingroup
  \count@#1\relax \ifnum 25<\count@\count@25\fi
  \def\x{\endgroup\@setsize\SetFigFont{#2pt}}%
  \expandafter\x
    \csname \romannumeral\the\count@ pt\expandafter\endcsname
    \csname @\romannumeral\the\count@ pt\endcsname
  \csname #3\endcsname}%
\fi \fi\endgroup
\begin{picture}(3500,6000)(3000,-6888)
\thicklines

\put(4801,-2161){\line( 1, 0){3639}} \put(4801,-2161){\line(
0,-1){2339}} \put(4962,-4665){\line( 1, 0){3639}}
\put(8600,-4665){\line( 0, 1){2339}}

\put(8600,-2161){\circle{300}} \put(4801,-4665){\circle{300}}

\put(8550,-2250){\makebox(0,0)[lb]{\smash{\SetFigFont{12}{14.4}{rm}D}}}
\put(4750,-4750){\makebox(0,0)[lb]{\smash{\SetFigFont{12}{14.4}{rm}A}}}

\put(4600,-2250){\makebox(0,0)[lb]{\smash{\SetFigFont{12}{14.4}{rm}B}}}
\put(8650,-4750){\makebox(0,0)[lb]{\smash{\SetFigFont{12}{14.4}{rm}C}}}

\put(6800,-4900){\makebox(0,0)[lb]{\smash{\SetFigFont{12}{14.4}{rm}$\ell_{1}$}}}
\put(4600,-3600){\makebox(0,0)[lb]{\smash{\SetFigFont{12}{14.4}{rm}$\ell_{2}$}}}

\put(6500,-6000){\makebox(0,0)[lb]{\smash{\SetFigFont{12}{14.4}{rm}{Figure
2}}}}

\end{picture}


The primary beam is split at vertex $A$, one of the secondary
beams follows the side defined by the line passing through
vertices $A$ and $C$, this side has a length of $l_1$, upon
arrival at vertex $C$ the beam is deflected by a silicon crystal
and moves along the side defined by the line passing through
vertices $C$ and $D$, whose size reads $l_2$. In a similar way the
remaining secondary beam travels along $A-B-D$. At $D$ we find the
detecting screen. If these two paths, $A-B-D$ and $A-C-D$ lie in a
horizontal plane, then, there is vanishing relative phase shift
between the two beams. If we now rotate the plane, say an angle
$\theta$ around the side $A-C$ of the parallelogram, then a
non--vanishing relative phase shift appears, due to the fact that
the paths, these secondary beams follow, are located at heights
associated with different gravitational potential. It can be shown
that the phase difference between the two secondary beams,
$\Delta\phi$, has the following form

\begin{eqnarray}
\Delta\phi =
\frac{m^2gl_1l_2\lambda\sin(\theta)}{\hbar^2}\label{Atomic3}.
\end{eqnarray}

In this last expression $m$ and $\lambda$ denote the mass of the
neutrons and the de Broglie wavelength of the neutrons,
respectively. The interference pattern has a purely quantum
mechanical origin, i.e., in the limit $\hbar\rightarrow 0$ the
interference pattern {\it gets washed out}. Notice also that
(\ref{Atomic3}) tells us that mass does indeed appear in the form
of a function of the combination $m/\hbar$, as expected from the
analysis of the corresponding Schr\"odinger equation
(\ref{Atomic2}). The possibility of resorting to heavier species
for this kind of experiments has also been considered
\cite{Borde2}, though the situation with heavier samples
introduces the excitation of the internal states as a new variable
to be considered, a fact that experimentally can be considered a
shortcoming \cite{Borde3}. The arguments proving that
(\ref{Atomic3}) is gauge--invariant lie along the same line of
reasoning that those used in connection with the Aharonov--Bohm
effect, see third paragraph on page 16.

\subsubsection{Postulates of Metric Theories and Gravity--Induced Experiments}

What are the consequences of this experiment? Here we arrive at a
controversial and controverted issue. Indeed, the appearance of
the mass parameter in the interference pattern has led some people
to accept the idea that, in the quantum domain, gravity is not
purely geometric \cite{Sakurai1, Ahluwalia1, Ahluwalia2}. Of
course, as we have mentioned before, general relativity is weaved
with several hypotheses, one of them, the equivalence principle.
This axiom is closely related to a geometrical interpretation of
gravity \cite{Misner1}. In other words, this interpretation
implies the breakdown, in the quantum world, of the equivalence
principle. The phrase {\it equivalence principle} here means WEP.
This principle states that the motion of a particle can be reduced
to purely geometrical parameters and nothing else. The appearance
of $m$ in this phase difference, and that is the claim
\cite{Sakurai1, Ahluwalia1, Ahluwalia2}, could mean the breakdown
of WEP in the quantum domain. On the other hand, this same
experiment has been the main ingredient to state that the COW
experiments prove the validity of WEP. The argument reads: {\it
The experiment proves that the Newtonian potential
$m\vec{g}\cdot\vec{r}$ has to be taken into account in
Schr\"odinger's equation, and that this potential impinges upon
the interference pattern, as any other potential} \cite{Colella2}.
Clearly, a careful analysis of this issue has to be done. For
instance, a fleeting glance at the first experimental verification
of this gravity--induced interference has been considered as a
sound experimental verification of the equivalence principle in
the quantum limit \cite{Colella2}.

Nevertheless, as pointed out \cite{Bonse1} the COW experiment does
not suffice to prove the validity in the quantum level of the
aforementioned principle. An experiment that validates this
statement, up to an accuracy of $4\%$, \cite{Bonse1} resorts to
neutron interferometry on an accelerated inertial coordinate
system. This experiment does allow us to consider the possible
validity in the quantum world of the equivalence between a
gravitational field and an accelerated system. Additionally, it
also provides an indirect test of the weak equivalence principle.
Let us explain this last assertion a little bit better. Notice
that behind the equivalence between a gravitational field and an
accelerated coordinate system we may find an additional
requirement, namely, inertial mass has to be equal to the
gravitational passive mass, see the argument after equation (4) in
\cite{Bonse1}. In other words, the measurement readouts of
\cite{Bonse1} provide an indirect test of the weak equivalence
principle.

We may also comment that COW may be formulated in terms that do
not include the particle's mass \cite{Laemmerzahl10}. This could
be considered as a proof of the fact that quantum theory fulfills
the weak equivalence principle, though we must also add that this
point has also some thorny aspects. For instance, any
reformulation of the phase shift in terms of the wavelength of the
particle masks the mass dependence since the de Broglie wavelength
establishes a relation between mass and wavelength. Further
experiments and theoretical predictions can be found in the
literature \cite{Raum1, Varju1}.

There is an additional result which at this point has a particular
relevance in connection with our discussion of the validity of the
equivalence principle in the quantum realm. Indeed, the classical
origin of the gravitational modification of the phase of a neutron
beam has already been shown \cite{Mannheim1}. A careful analysis
of this last reference allows us to have a clear picture of the
assumptions of a COW experiment. We may notice that from square
one \cite{Mannheim1} considers the semiclassical approximation for
the motion of neutrons and the eikonal limit for massless
particles. The crucial question is the following one: does
\cite{Mannheim1} prove that the weak equivalence principle is
fulfilled in the quantum realm, or does it only show that in the
semiclassical region of quantum mechanics this principle holds?
This question requires a thorough analysis since the fulfillment
of a certain property in a very particular limit (the
semiclassical one) does not guarantee the validity of the involved
property in the most general scheme.

The generalization of the weak equivalence principle has already
been put forward \cite{Onofrio2, Laemmerzahl10}, though also some
possible conceptual difficulties have been pointed out
\cite{Camacho10}. At this point it is noteworthy to stress the
fact that this debate is still alive, and that no conclusive
evidence, in one direction or in the opposite one, exists. This
means that more work is needed in this direction \cite{Borde1}.

The possibility of resorting to these gravity--induced experiments
and use them as a tool for testing the foundations of general
relativity requires some modifications in the experimental
proposal. For instance, to test EEP requires, if we wish to prove
(or disprove) Local Lorentz Invariance or Local Position
Invariance (see above), to perform the experiment in different
freely falling frames. A corroboration of the validity, in the
quantum domain, of these two invariance properties would be
obtained if the corresponding readouts are always the same,
independent of the freely falling reference frame and of where and
when the experiment has been carried out. Additionally, the
validity of EEP implies that gravitation has to be described by a
metric theory. This condition implies that: {\it In any locally
freely falling frame the non--gravitational laws of physics are
those of special relativity} \cite{Will2}. Therefore, if the
results coincide with the outcome (when in Schr\"odinger's
equation we impose the condition $g=0$), then we could state that
for a quantum system locally the effects of gravity can be
transformed ({\it gauged}) away. In other words, this experiment
would be a confirmation that gravitation can be formulated within
the context of a metric theory. This kind of experiments have, up
to now, not been performed \cite{Will2}.

The phrase {\it gauged away} demands a deeper explanation. The
free fall experiment in the form of Einstein elevator
\cite{Misner1} implies that for an observer in free fall there is
a, sufficiently small, neighborhood in space--time in which the
gravitational field can be considered null. In other words, in
this neighborhood everything takes place as if the observer were
in an inertial coordinate frame. This implies, that at least
locally, gravity can be {\it gauged away}. Mathematically this can
also be easily understood. Indeed, consider a point $P$ in our
manifold, the flatness theorem \cite{Misner1} tell us that there
is a neighborhood around $P$ in which the metric is given by the
Minkowskian one. Since (this point will be explained further
below) the gravitational potential is encoded in the metric, then
the gravitational force is contained in the derivatives of the
metric. But the derivatives of the Minkowskian metric all vanish,
and in consequence, in this neighborhood of $P$ there is no
gravitational field. This is the meaning of the phrase {\it gauged
away}, for any point in our manifold we may find a neighborhood in
which the gravitational field vanishes.

Nevertheless, there is an additional interpretation for the phrase
"gauge-dependent" in the context of gravitation. A gauge
transformation can be introduced in the case of the linearized
version of Einstein equations. Indeed, in this situation the
metric can be written in the following form

\begin{eqnarray}
g_{\mu\nu} = \eta_{\mu\nu} + h_{\mu\nu}\label{Linearmetric1}.
\end{eqnarray}

In this last expression $\eta_{\mu\nu}$ denotes the Minkowskian
metric and $\vert h_{\mu\nu}\vert <<1$. A gauge transformation is
defined by ($\vert \xi^{\nu}_{,\mu}\vert <<1$)

\begin{eqnarray}
x^{\nu'} = x^{\nu} + \xi^{\nu}\label{Linearmetric2}.
\end{eqnarray}

It can be shown (see page 438, exercise 18.1 in \cite{Misner1})
that a gauge transformation does not modify the components of the
Riemann tensor. O course, the contractions of the Riemann tensor
are also invariant under this kind of transformations. In this
sense, (\ref{Atomic3}) is also gauge--invariant.

For the sake of completeness let us mention a couple of quantum
experiments (though not involving interference) testing Local
Lorentz Invariance, which are known as Hughes--Drever experiments
\cite{Hughes1, Drever1}. In these experiments the idea is to find
a bound to any possible anisotropy in the inertial mass of quantum
systems looking at the spacing among the energy levels of a
particular energy state. Finally, the effects of rotation upon the
interference pattern of thermal neutrons have also been considered
\cite{Anandan1, Mashhoon1}.

The description of wave propagation, within the principles of any
metric theory, has conceptual difficulties. Indeed, the principles
of these models are based upon the motion of point--like objects.
This fact implies that for wave phenomena, for instance, quantum
theory, the semiclassical approach has to be, from square one,
imposed, otherwise we encounter a question, which in the context
of metric theories, has no answer \cite{Mashhoon3}. The same
problem emerges if the description of light is attempted outside
the eikonal limit. This last statement does not mean that wave
propagation can not occur in metric theories. It only implies that
the propagation of wave phenomena faces, in the case of
non--vanishing wavelength, conceptual difficulties in the context
of the postulates of these theories.

\subsubsection{Geometric Phases and Gravitomagnetism}

One of the problems in the detection of gravitomagnetism comprises
the fact that it involves tiny perturbations in the orbit of the
used satellites \cite{Truehaft1}. At this point we may pose the
following question: Could this field be detected without having to
measure very small changes, either in the trajectory or in other
physical observable?

Additionally, the detection of this effect has always been
contemplated in the realm of classical systems, and this last
remark takes us to another question related directly with the
present work: Could this field be detected resorting to a quantum
interference experiment? Additionally, could this interference
experiment be used to test some fundamental property of physics?

The joint answer to these three questions will be provided by the
following proposal. Consider a $1/2$--spin particle immersed in
the gravitomagnetic field of a rotating sphere (this field will be
described in the PPN formalism for any metric theory of gravity
\cite{Misner1}). Additionally, we assume that its rotation axis
also spins. It will be shown that the interaction between spin and
gravitomagnetism predicts a geometric phase for the wave function,
the one does not depend upon the strength of the interaction. This
last comment gives an affirmative answer to our first question. We
may wonder at what stage does the strength of the gravitomagnetic
field appear in our proposal. The present work will show that the
strength of the gravitomagnetic field defines the adiabatic regime
\cite{Griffiths1}.

Let us now proceed to explain the proposed experiment of this part
of the work. A beam of $1/2$--spin particles (all in the same
initial state) is split into two. One of the beams will not be
allowed to interact with $\vec {J}$ (the definition of this
parameter appears below), whereas the second one will have its
spin state pointing always in the direction of $\vec {J}$.
Clearly, this last condition is obtained, as a consequence of the
adiabatic theorem, when $\vec {J}$ spins, sufficiently slow,
around a certain axis. After this angular momentum vector
completes one cycle we proceed to recombine these two beams. We
will show that the final probability involves a geometric phase
factor, which shall be non-vanishing for the case of a
non--trivial coupling between spin and gravitomagnetism. This last
explanation answers our second question; yes, there is a
interference proposal which involves the possible detection of the
gravitomagnetic field in the quantum domain. This last statement,
additionally, provides an answer to the third question. As shown
below, the emergence of the aforementioned geometric phase will
appear only if the coupling between spin and gravitomagnetism
given in the literature \cite{Mashhoon2}, but never subject to
experimental confirmation, is correct.

Of course, we need a source of gravitomagnetism, therefore let us
consider a rotating uncharged, idealized spherical body with mass
$M$ and angular momentum $\vec {J}$. In the formalism of the weak
field and slow motion limit the gravitomagnetic field may be
written, using the PPN parameters $\Delta_1$ and $\Delta_2$
\cite{Misner1}, as

\begin{eqnarray} \vec {B} = \Bigl({7\Delta_1 + \Delta_2\over
4}\Bigr){G\over c^2}{\vec {J} - 3(\vec {J}\cdot\hat {x})\hat
{x}\over |\vec {x}|^3}\label{Gravito1}.
\end{eqnarray}
\bigskip

In this last expression our parameters allow us to recover
different metric theories. For instance, ${7\Delta_1 +
\Delta_2\over 4} = 2$ implies general relativity, while
Brans--Dicke \cite{Brans1} appears if ${7\Delta_1 + \Delta_2\over
4} = {12 + 8\omega\over 8 + 4\omega}$. An interesting point
emerges in Ni's theory \cite{Ni1}, where ${7\Delta_1 +
\Delta_2\over 4} = 0$, i.e., there is no gravitomagnetic field.
This last example teaches us that a metric theory not necessarily
embodies a non--trivial gravitomagnetic effect.

As an additional condition we will assume that $\vec {J}$ rotates
around a certain axis, $\vec {e}_3$, with angular velocity
$\omega$, and that the direction of this axis and that of the
angular momentum defines an angle $\theta$. This implies that in
our coordinate system

 {\setlength\arraycolsep{2pt}\begin{eqnarray}
\vec {J} = J\Bigl[\cos(\omega t)\sin(\theta)\vec {e}_1 +
\sin(\omega t)\sin(\theta)\vec {e}_2 +  \cos(\theta)\vec
{e}_3\Bigr].
\end{eqnarray}}
\bigskip

As mentioned before, there is a spin $1/2$--system immersed in the
gravitomagnetic field of $M$, and located on $\vec {e}_3$ at a
distance $r$ from the center of our rotating  sphere. Usually the
coupling between gravitomagnetism and orbital angular momentum is
a copy of the behavior of orbital angular momentum under the
presence of a magnetic field \cite{Ciufolini1}. This has a
profound physical explanation, since in the weak--field limit
Einstein equations resemble the motion equations of
electrodynamics. Here we assume that the expression describing the
precession of orbital angular momentum can be also used for the
description of the dynamics in the case of intrinsic spin, i.e.,
the behavior of orbital angular momentum is copied into the
behavior of spin, which is a physical quantity without classical
analogue. This seems to be a reasonable assumption, though we must
underline the fact that up to now there is no experimental
evidence supporting it \cite{Will1, Will2}. The Hamiltonian
becomes (here $\vec{S}$ denotes the corresponding spin--operator)

\begin{eqnarray} H = - \vec {S}\cdot\vec {B}\label{Gravito2}.
\end{eqnarray}
\bigskip

We now introduce the following definitions

\begin{eqnarray} \omega_1= {7\Delta_1 + \Delta_2\over 2}{GJ\over
c^2r^3}\label{Gravito3},
\end{eqnarray}
\bigskip

we may rewrite (\ref{Gravito2}) as

{\setlength\arraycolsep{2pt}\begin{eqnarray} H = -
{\hbar\omega_1\over 2} \left(\begin{array}{cc}
      -2\cos(\theta), & e^{-i\omega t}\sin(\theta) \\
       e^{i\omega t}\sin(\theta), & 2\cos(\theta)
       \end{array}\right) \label{Gravito4}.
\end{eqnarray}}

The associated energy values are

\begin{eqnarray} E_{(\pm)} = \pm{\hbar\omega_1\over 2}\sqrt{1 +
3\cos^2(\theta)}\label{Gravito5}.
\end{eqnarray}

The eigenvector related to $E_{(+)}$ reads

{\setlength\arraycolsep{2pt}\begin{eqnarray} \psi_{(+)}(t) =
{\sin(\theta)\over\sqrt{2 + 6\cos^2(\theta) -4\cos(\theta)\sqrt{1
+ 3\cos^2(\theta)}}} \left(\begin{array}{c}
      1 \\
      {2\cos(\theta)- \sqrt{1 + 3\cos^2(\theta)}\over\sin(\theta)}e^{i\omega t}
       \end{array}\right)\label{Gravito6}.
\end{eqnarray}}

According to Berry \cite{Berry1}, if $\omega_1 >>\omega$, with the
initial spin state $\psi_{(+)}(t = 0)$, then the spin state is
provided by

\begin{eqnarray} \Psi_{(+)}(t) =
e^{iE_{(+)}t/\hbar}e^{i\gamma_{(+)}(t)}\psi_{(+)}(t)\label{Gravito7},
\end{eqnarray}

where $\gamma_{(+)}(t)$ is Berry's phase, a geometric term given
by \cite{Berry1}

\begin{eqnarray} \gamma_{(+)}(t) = i\int_0^t<\psi_{(+)}(t')\vert
{\partial\psi_{(+)}(t')\over\partial t'}>dt'\label{Gravito8}.
\end{eqnarray}

We may now write

\begin{eqnarray} \gamma_{(+)}(t) = -\omega t\label{Gravito9}.
\end{eqnarray}

At this point it is noteworthy to comment that the geometric phase
is independent of the magnitude of the gravitomagnetic field. A
glance at the previous work in this context \cite{Ciufolini1,
Ciufolini2} shows that the quantity to be observed always depends
upon the magnitude of the gravitomagnetic field, a fact that
represents a drawback, in the experimental realm. In the present
proposal, this factor does not emerge.

If $t = {2\pi\over \omega}$, this condition means that $\vec {J}$
has completed one rotation around $\vec {e}_3$, then the geometric
phase turns out to be

\begin{eqnarray} \gamma_{(+)}(t) = -2\pi\label{Gravito10}.
\end{eqnarray}
\bigskip
\bigskip

The condition defining the adiabatic regime is given by $\omega_1
>>\omega$, and we may rephrase this condition, resorting to
(\ref{Gravito3}), as

\begin{eqnarray} {7\Delta_1 + \Delta_2\over 2}{GJ\over c^2r^3}
>> \omega\label{Gravito11}.
\end{eqnarray}

The magnitude of the gravitomagnetic field appears in this last
expression, i.e., it defines the adiabatic regime. In this sense,
this present proposal is quite different from the usual
experimental ideas, which must detect tiny changes in
corresponding physical parameter \cite{Ciufolini1, Ciufolini2}.

Moreover, it was mentioned before that the extant experimental
confirmations of the gravitomagnetic effect involve macroscopic
objects, satellites as a matter of fact \cite{Ciufolini1,
Ciufolini2}. A careful analysis of the theoretical background
behind this experiment shows that it embodies the coupling between
gravitomagnetism and orbital angular momentum, a physical quantity
of classical origin. The behavior of spin under the presence of
gravitomagnetism is always assumed to be a copy of the case of
orbital angular momentum. This is a reasonable assumption,
nevertheless, it has not experimental evidence supporting it. The
present proposal gives the possibility of testing an aspect of
fundamental physics, namely, the manner in which spin couples to
gravitomagnetism. At this point we may wonder if the use of
geometric phases could tell us something about the validity, at
quantum level, of the postulates of metric theories, for instance,
WEP. It is noteworthy to comment that the so--called Hyper project
attempts to detect the Lense--Thirring effect resorting to atomic
interferometers. In addition it plans to measure the fine
structure constant, as well as the quantum gravity induced foam
structure of space \cite{Neugebauer1}.

The existence of gravitomagnetism is linked to the concept of
metric theories \cite{Will1}. Then the experimental corroboration
of the existence of this effect is an indirect test of the
postulates of metric theories, but it is not a direct test, for
instance, of WEP. Indeed, the quantum properties appear in
connection with the spin space, and not with the behavior of the
system in the configuration space, this can be seen from the fact
that the Hamiltonian (\ref{Gravito2}) does not contain the mass of
the particle. In other words, any postulate of metric theories
related to the mass parameter cannot be tested directly with this
kind of proposals.

Let us now fathom better the conditions under which the adiabatic
regime appears. We assume, for the sake of simplicity, that our
sphere is a homogeneous one (which implies $J = 2MR^2\Omega/5$,
here $\Omega$ is the angular velocity of $M$), then the validity
of the adiabatic regime is guaranteed if

\begin{eqnarray} {MR^2\Omega\over\omega r^3}
>> {5c^2\over G(7\Delta_1 + \Delta_2)}\label{Gravito12}.
\end{eqnarray}
\bigskip

This last expression defines the region for the experimental
parameters (M, $R$, etc.) in which the proposal could be used,
assuming a certain metric theory given by $\Delta_1$ and
$\Delta_2$.

\subsection{Non--Metric Theories of Gravity and Quantum Interference}

\subsubsection{Torsion and Gravity}

The current experimental data do not contradict general relativity
\cite{Will1, Will2}, and in consequence, we could take for granted
its validity, in the classical realm and in the quantum domain.
Another possibility in this context is to analyze generalizations
of general relativity and understand the bounds that the extant
experiments \cite{Will1, Will2} impose upon the extra parameters
that these generalizations contain. In this part of the present
work we will take this point of view, and analyze the options that
COW experiments offer in the context of one of the generalizations
of general relativity. The question now is: What generalization of
general relativity? The answer to this question will be given in
terms of the voids that the current experiments have. A fleeting
glance at the available experimental information on general
relativity clearly shows that there are more classical than
quantum tests of this theory. This is the starting criterion for
our generalization of general relativity which gives a promising
possibility for testing its principles in the quantum domain. Now
that we have defined our criterion, we must face once again our
question: What generalization? The answer is given, partially by
the Casimir invariants of the Poincar\'e group, i.e., mass and
spin. Mass is connected with the translational part of this group,
whereas spin with the rotational one \cite{Carmeli1}. Poincar\'e
group lies very deep in the principles of special relativity, but
if we look at general relativity we notice at once that mass is
taken into account as a source of the gravitational field, but
spin not \cite{Misner1, Will1}. This situation contains an
asymmetry which will provide us with our generalization. Indeed, a
possible generalization of general relativity is the
Einstein--Cartan theory \cite{Hehl1}, which introduces into its
formalism the concept of torsion, and connects it with spin. Mass
is a source for the gravitational field in general relativity, and
in the same spirit, the other Casimir invariant of the Poincar\'e
group, spin, will in Einstein--Cartan theory be a source of a
gravitational field. In other words, spin will modify the geometry
of space--time. In metric theories of gravity there is a coupling
between the energy--momentum tensor \cite{Misner1} and the metric
of the corresponding manifold. In a similar way, spin will be
coupled to a geometric element of the manifold, i.e., the
so--called contorsion tensor, which is related to rotational
degrees of freedom of space--time \cite{Hehl1}. The introduction
of this new element into the structure of the corresponding
space--time is done at the level of the so--called affine
connection, in which a non--symmetric part is included. This new
element behaves as a tensor, in contrast to the symmetric part
\cite{Hehl1}.

\begin{eqnarray}
S^k_{ij} = \frac{1}{2}\Bigl(\Gamma^k_{ij} -
\Gamma^k_{ji}\Bigr)\label{Cartan1}.
\end{eqnarray}

This is denoted as Cartan's torsion tensor, the $\Gamma^k_{ij}$
are called the affine connections. Geometrically Cartan's tensor
entails that if we build infinitesimal parallelograms in our
space--time, then they, in general, do not close. This tensor is
used to define the contorsion tensor, here denoted by $K^k_{ij}$
($\tilde{\Gamma}^k_{ij}$ are the Christoffel symbols computed as
usual, with the corresponding metric \cite{Misner1}).

\begin{eqnarray}\Gamma^k_{ij} = \tilde{\Gamma}^k_{ij} - K^k_{ij}
\label{Cartan2}.
\end{eqnarray}

 The corresponding Einstein tensor has the usual form

\begin{eqnarray}G_{ij} = R_{ij} -
\frac{1}{2}g_{ij}R\label{Cartan3}.
\end{eqnarray}

Though the structure is the same, there is a great difference with
general relativity \cite{Hehl1}, namely, the Ricci tensor,
$R_{ij}$, is asymmetric. Additionally, there is no torsion outside
the spinning matter distribution. Torsion is bound to matter and
cannot propagate through the vacuum. The influence of spin outside
the matter distribution appears through its influence on the
metric tensor. Torsion cannot influence macroscopic bodies
\cite{Yaskin1, Stoeger1}, its effects appear only in the evolution
of spin. Clearly, the interaction between spin and torsion
modifies the trajectory of a microscopic particle, i.e.,
Einstein--Cartan theory predicts a motion that will differ from
that emerging in the context of a metric theory of gravity, as
general relativity \cite{Hehl1}.

\subsubsection{Non--Newtonian Gravity}

We have mentioned several experiments that do not contradict the
theoretical predictions of general relativity. Just for the sake
of completeness, let us mention some additional results that are
also along this line. With this brief recount of the experimental
information we will try to understand better those regions in
which general relativity has not been tested very profoundly. This
aspect is important since it allows us to justify soundly a series
of interference experiments, which takes into account a
generalization of gravity. This conceptual broadening of gravity
becomes important in a region in which general relativity has not
been subjected to a very stringent analysis.

Among these proofs we may find, for instance, the gravitational
time dilation measurement \cite{Vessot1, Vessot2}, gravitational
deflection of electromagnetic waves \cite{Truehaft1}, time delay
of electromagnetic waves in the field of the sun \cite{Krisher1},
or the geodetic effect \cite{Dickey1}. A fleeting glance at these
observational results confirms the fact that they are tests of
weak field corrections to the Galilei--Newton mechanics, i.e.,
they do not involve any time dependence of the gravitational
field.

High--precision timing observations of pulsars have also been used
in the context of diverse topics. For instance, timing observation
over many years provided tight upper limits on the energy density
associated to gravitational waves. \cite{Taylor1}. The results
between theory and experiment agree at a level of $10^{-3}$. The
possibilities that binary pulsars offer do not finish here, they
can also be used as laboratories for testing strong--field gravity
\cite{Damour1}. Concerning binary pulsars at this point it is
noteworthy to mention that they are a confirmation of general
relativity done at the classical level. Here we mean that the
observations and predictions comprise the orbital dynamics of a
binary pulsar, for instance, orbital period, eccentricity
\cite{Damour2}, etc.

 After more than a decade of experiments \cite{Long1}, there is no
compelling e\-vi\-den\-ce for any kind of deviations from the
predictions of Newtonian gravity. At this point it is noteworthy
to comment an argument put forward in this context. Gibbons and
Whiting's (GW) phenomenological analysis of gravity data
\cite{Gibbons1} has proved that the very precise agreement between
the predictions of Newtonian gravity and observation for planetary
motion does not preclude the existence of large non--Newtonian
effects over smaller distance scales, i.e., precise experiments
over one scale do not ne\-ce\-ssarily constrain gravity over
another scale.

This has an important consequence since GW results conclude that
the current ex\-pe\-ri\-men\-tal constraints over possible
deviations did not severely test Newtonian gravity over the
$10$--$1000$m distance scale, usually called ``geophysical
window''.

New constraints on the possible ranges of a Yukawa term have been
given in an experiment carried out in 2000 \cite{Smith1}. This new
experiment improves the current limit for ranges between 10km and
1000km. Nevertheless, in the short range it can say nothing about
distances smaller than 1cm. This experiment is performed on a
classical system, namely, a 3 ton $^{238}U$ attractor rotates
around a torsion balance, which contains Cu and Pb macroscopical
test bodies. In this experiment the differential acceleration of
the test bodies toward the attractor was measured.

The displacement induced by an oscillating mass acting as a source
of gravitational field on a micromechanical resonator must be
mentioned \cite{Carugno1}, since it could provide evidence about
scalar interactions in the short range below 1mm.

Among the models that in the direction of non--inverse--square
forces currently exist we have Fujii's proposal \cite{Fujii1}. In
this idea a ``fifth force'', coexisting simultaneously with
gravity, comprises a modified Newtonian potential. The
corresponding Yukawa term is given by $V(r) = -G_{\infty}{mM\over
r}\left(1 + \alpha e^{-{r\over \lambda}}\right)$, here
$G_{\infty}$ describes the interaction between $m$ and $M$ in the
limit case $r\rightarrow\infty$, i.e., $G = G_{\infty}(1 +
\alpha)$, where $G$ is the Newtonian gravitational constant. This
kind of deviation terms arises from the exchange of a single new
quantum of mass $m_5$, the Compton wavelength of the exchanged
field is $\lambda = {\hbar\over m_5c}$ \cite{Fishbach2}. This
field is usually called dilaton.

If we take a look at the experimental efforts that have been done
in order to test the inverse--square law we will find that they
can be separated into two large classes: (i) Those experiments
which involve the direct measurement of the magnitude $G(r)$, they
compare preexisting laboratory Cavendish measurements of $G$
\cite{Yu1}; and (ii) the direct measurement of $G(r)$ with $r$
\cite{Zumberge1}. A relevant characteristic of these efforts has
to be mentioned, i.e., they remain always at the classical level,
the action of the Yukawa term is always on classical systems,
namely, classical test masses (Cavendish case), or in the case of
mine and Borehole experiments, once again, classical test
particles are employed. One of the exceptions around this topic is
the use of the Casimir effect \cite{Mostepanenko1, Milonni1},
where Planck constant, $\hbar$, appears as a parameter in the
experiment. Another quantum analysis may be found in
\cite{Laemmerzahl3}.

The information contained in all these works makes us wonder why
should we need analyze some possible deviation of the Newtonian
inverse--square force law? The answer to this question is a
two--fold one.

The first point to be mentioned here stems from the fact that the
agreement between general relativity and experiment might be
compatible with the existence of a scalar contribution to gravity,
such as a dilaton field \cite{Damour3}. This dilaton field emerges
in several theoretical attempts that try to formulate a unified
theory of elementary particle physics. As one of their
consequences they predict the existence of new forces (which are
usually referred to as ``fifth force''), whose effects extend over
macroscopic distances \cite{Fishbach1}. In some way, these new
forces simulate the effects of gravity, but a crucial point is
that they are not described by an inverse--square law, and even
more, they, generally, violate the Weak Equivalence Principle
(WEP) \cite{Fishbach1}. In other words, the presence of this kind
of forces, coexisting with gravity, could be detected by apparent
deviations from the inverse--square law, or from the violation of
WEP. These last arguments give us a partial answer to the last
question, namely, a strong theoretical motivation for analyzing
possible deviations from Newtonian gravity is to probe for new
fundamental forces in nature.

Additionally, as mentioned at the beginning of this section most
of the tests to which general relativity has been subjected lie in
the classical domain. The number of quantum tests is smaller,
\cite{Will1, Will2}. Our proposal, see below, resorts to quantum
theory, and in this sense it tends to fill this gap.

\subsubsection{Torsion--Induced Interference}

The importance of the COW experiments of gravity--induced
interference, and the corresponding improvement in the precision
that they have reached, is not constrained to the aforementioned
debate concerning the validity at the quantum realm of the
postulates around metric theories of gravity. There is an
additional fact that we now face. Indeed, the precision associated
to them tells us that there is a discrepancy of one percent
between theory and experiment \cite{Werner1}, which requires an
explanation. Several models can be proposed in this direction, and
among them we may find some which take into account variables
usually neglected. For instance, this discrepancy could be a
direct consequence of the way in which dynamical diffraction
interacts with bending and strains in the interferometer
\cite{Colella4}.

In this part of our work we analyze the possible role that torsion
could play as an element explaining part of this discrepancy
between experiment and theory. There is some previous work
concerning the detection of torsion as a fundamental geometrical
element of space--time, nevertheless, this analysis
\cite{Laemmerzahl1} cannot explain the discrepancy just mentioned,
since it is not a neutron or atomic interferometric experiment, it
is a Hughes--Drever type proposal and the presence of torsion is
tested through shifts in Zeeman lines.

At this point we pose the following question: Could the coupling
between the spin of the neutrons and the torsion of space--time be
held responsible for part of the discrepancy?

The case of a $1/2$--spin particle immersed in a Riemann--Cartan
space--time \cite{Hehl1} is now considered, and now we pose the
following question: How does the contribution to the interference
pattern, stemming from the coupling spin--torsion, look like? More
precisely, let us suppose that the spin part of the neutron beam's
wave function is the coherent linear superposition of two
contributions, one with z-component of the spin $1/2$, and the
other one with $-1/2$. It can be proved that the presence of
torsion could be detected, in principle, heeding the changes that
appear as a function of the way in which the superposition is done
\cite{Camacho3}. At this point it has to be mentioned that the
possible effects of torsion upon a neutron interference experiment
have already an old story \cite{Laemmerzahl2}. The presence of
torsion can be held responsible for the appearance of some
peculiarities. For instance, the particle's orbit is non--geodesic
\cite{Audretsch1}. At this point it is noteworthy to comment that
in these last works the analysis of the aforementioned discrepancy
was not done.

The results obtained from our analysis are quite interesting,
since they share some traits that emerged already in the
gravity--induced interference pattern. For instance, the quantum
mechanical trait of this effect depends on powers of $m/\hbar$,
and hence has a striking similarity with the conclusions of \cite
{Colella1}.

Let us consider a neutron interferometer, as in the COW
experiment, and assume that there is a coupling between the
torsion of space--time and the spin of our neutrons. The quantum
mechanical description of the neutrons requires now a Hilbert
space which has to be the tensor product of two contributions, to
wit, spin state space, $\mathcal{E}$$_s$, and the orbital state
space, $\mathcal{E}$$_r$ \cite{Cohen1}. The dynamics of the state
vector related to the neutron beam is given in the
non--relativistic limit of the Dirac equation, in a Newtonian
approximation of Riemann--Cartan space--time, by the Pauli
equation \cite{Laemmerzahl2}.

\begin{eqnarray} i\hbar{\partial\vert\psi>\over\partial t} =
-{\hbar^2\over 2m}\nabla^2\vert\psi> - i{\hbar^2\over
m}\kappa_{(0)}\sigma^l\partial_l\vert\psi> - mV\vert\psi> - \hbar
c\kappa_l\sigma^l\vert\psi> \label{Torsion1}.
\end{eqnarray}

In this last expression the following terms have been introduced,
$c$ is the speed of light, $V$ the Newtonian gravitational
potential, $\sigma^l$ Pauli matrices, and $\kappa_{\mu}$ the axial
part of the space--time torsion. Additionally, in (\ref{Torsion1})
we assume that $\kappa_{(0)} = 0$. This simplification will allow
us to fathom, in a clear manner, the consequences, upon the
interference pattern, of the space part of the axial part of the
torsion. The motion equation becomes now

\begin{eqnarray} i\hbar{\partial\vert\psi>\over\partial t} =
-{\hbar^2\over 2m}\nabla^2\vert\psi> - mV\vert\psi> - \hbar
c\kappa_l\sigma^l\vert\psi>\label{Torsion2}.
\end{eqnarray}

If $\phi$ describes the spin state vector, i.e.,
$\phi\in\mathcal{E}$$_s$, then its dynamics is governed by the
following equation

\begin{eqnarray} i\hbar{\partial\phi\over\partial t} = -\hbar
c\kappa_n\sigma^n\phi\label{Torsion3}.
\end{eqnarray}

Obviously a solution reads

\begin{eqnarray} \phi(t) =
\exp\Bigl\{ic\int_0^t\kappa_n\sigma^ndt'\Bigr\}\phi(t=
0)\label{Torsion4}.
\end{eqnarray}

Let us now consider an experiment similar to COW, figure 3. In
other words, two particles, starting at point $(O)$, move along
two different trajectories, say $C$ and $\tilde{C}$. Afterwards
these beams are detected at a certain point $S$. Additionally, we
take for granted the validity of the semiclassical approach.

Let us now specify, explicitly, our trajectories. $C$ is made up
of two contributions, namely, (O)--(A) which is horizontal, whose
length reads $l$, and (A)--(S), vertical, and with length equal to
$L$. $\tilde{C}$ comprises also two parts, (O)--(B) vertical, with
length $L$, and (B)--(S) horizontal, and size $l$. The horizontal
axis is $x$, and $y$ points upwards, such that the Newtonian
potential reads $V = gy$.

\setlength{\unitlength}{0.00083300in}
\begingroup\makeatletter\ifx\SetFigFont\undefined

\def\x#1#2#3#4#5#6#7\relax{\def\x{#1#2#3#4#5#6}}%
\expandafter\x\fmtname xxxxxx\relax \def\y{splain}%
\ifx\x\y
\gdef\SetFigFont#1#2#3{%
  \ifnum #1<17\tiny\else \ifnum #1<20\small\else
  \ifnum #1<24\normalsize\else \ifnum #1<29\large\else
  \ifnum #1<34\Large\else \ifnum #1<41\LARGE\else
     \huge\fi\fi\fi\fi\fi\fi
  \csname #3\endcsname}%
\else \gdef\SetFigFont#1#2#3{\begingroup
  \count@#1\relax \ifnum 25<\count@\count@25\fi
  \def\x{\endgroup\@setsize\SetFigFont{#2pt}}%
  \expandafter\x
    \csname \romannumeral\the\count@ pt\expandafter\endcsname
    \csname @\romannumeral\the\count@ pt\endcsname
  \csname #3\endcsname}%
\fi \fi\endgroup
\begin{picture}(3500,6000)(3000,-6888)
\thicklines

\put(4801,-2161){\line( 1, 0){3639}} \put(4801,-2161){\line(
0,-1){2339}} \put(4962,-4665){\line( 1, 0){3639}}
\put(8600,-4665){\line( 0, 1){2339}}

\put(8600,-2161){\circle{300}} \put(4801,-4665){\circle{300}}

\put(8550,-2250){\makebox(0,0)[lb]{\smash{\SetFigFont{12}{14.4}{rm}S}}}
\put(4750,-4750){\makebox(0,0)[lb]{\smash{\SetFigFont{12}{14.4}{rm}O}}}

\put(4600,-2250){\makebox(0,0)[lb]{\smash{\SetFigFont{12}{14.4}{rm}B}}}
\put(8650,-4750){\makebox(0,0)[lb]{\smash{\SetFigFont{12}{14.4}{rm}A}}}

\put(6800,-4900){\makebox(0,0)[lb]{\smash{\SetFigFont{12}{14.4}{rm}$l$}}}
\put(4600,-3600){\makebox(0,0)[lb]{\smash{\SetFigFont{12}{14.4}{rm}$L$}}}

\put(6300,-5500){\makebox(0,0)[lb]{\smash{\SetFigFont{12}{14.4}{rm}{Figure
3}}}}

\end{picture}


Then,

\begin{eqnarray} \kappa_n(A) = \kappa_n(0) +
{\partial\kappa_n\over\partial x}_{(0)}l\label{Torsion5},
\end{eqnarray}

\begin{eqnarray} \kappa_n(B) = \kappa_n(0) +
{\partial\kappa_n\over\partial y}_{(0)}L.\label{Torsion6}
\end{eqnarray}

This entails that at the screen, (S), (for the spin wave function
that passes through (A), $\phi_A(S)$, and for that passing through
(B), $\phi_B(S)$) we have

\begin{eqnarray} \phi_A(S) =
\exp\Bigl\{ic\sigma^n[\alpha_A\kappa_n(0) +
\beta_A{\partial\kappa_n\over\partial x}_{(0)} +
\gamma_A{\partial\kappa_n\over\partial y}_{(A)}]\Bigr\}\phi(t=
0)\label{Torsion7},
\end{eqnarray}

\begin{eqnarray} \phi_B(S) =
\exp\Bigl\{ic\sigma^n[\alpha_B\kappa_n(0) +
\beta_B{\partial\kappa_n\over\partial x}_{(B)} +
\gamma_B{\partial\kappa_n\over\partial y}_{(0)} ]\Bigr\}\phi(t=
0)\label{Torsion8}.
\end{eqnarray}

In these last two expressions we have (approximately)

\begin{eqnarray} \alpha_A = {m\tilde{\lambda}\over\hbar}\Bigl\{l
+ L/2 -
({m\tilde{\lambda}\over\hbar})^2gL^2/8\Bigr\}\label{Torsion9},
\end{eqnarray}

\begin{eqnarray} \beta_A = {m\tilde{\lambda}\over\hbar}l\Bigl\{(l
+ L)/2 -
({m\tilde{\lambda}\over\hbar})^2gL^2/8\Bigr\}\label{Torsion10},
\end{eqnarray}

\begin{eqnarray} \gamma_A =
{m\tilde{\lambda}\over\hbar}\Bigl\{L^2/2\Bigl[1/4 -
({m\tilde{\lambda}\over\hbar})^2gL/4\Bigr] + lL\Bigl[1/2 -
({m\tilde{\lambda}\over\hbar})^2g(2L +
3l)/4\Bigr]\Bigr\}\label{Torsion11},
\end{eqnarray}

\begin{eqnarray} \alpha_B = {m\tilde{\lambda}\over\hbar}\Bigl\{l
+ L/2 + ({m\tilde{\lambda}\over\hbar})^2gL\Bigl[l - L/8\Bigr]
\Bigr\}\label{Torsion12},
\end{eqnarray}

\begin{eqnarray} \beta_B =
3L^2{m\tilde{\lambda}\over\hbar}\Bigl\{1/4 -
({m\tilde{\lambda}\over\hbar})^2gL/8\Bigr\}\label{Torsion13},
\end{eqnarray}

\begin{eqnarray} \gamma_B =
{m\tilde{\lambda}\over\hbar}L^2\Bigl\{3/4\ -
({m\tilde{\lambda}\over\hbar})^213gL/(48)\Bigr\}\label{Torsion14}.
\end{eqnarray}

We must add that in our equations $\tilde{\lambda} =
\lambda/(2\pi)$, and $\lambda$ denotes the initial wavelength of
the neutron beam.

These wave functions may be cast in terms of a rotation of the
initial state

\begin{eqnarray} \phi_n(S) = \exp\Bigl\{-{i\over
2}\theta_v\vec{n}_v\cdot\vec{\sigma}\Bigr\}\phi(t=
0)\label{Torsion15}.
\end{eqnarray}

Here $v = A, B$. The definition of the components of the unit
vectors and the rotation angles are given by

\begin{eqnarray} \tau_n^{(A)} = \Bigl\{\alpha_A \kappa_n(0) +
\beta_A{\partial\kappa_n\over\partial x}_{(0)}l +
\gamma_A{\partial\kappa_n\over\partial
y}_{(A)}\Bigr\}\label{Torsion16},
\end{eqnarray}

\begin{eqnarray} (\vec{n}_{A})_n =
{\tau_n^{(A)}\over\sqrt{(\tau_x^{(A)})^2 + (\tau_y^{(A)})^2 +
(\tau_z^{(A)})^2}}\label{Torsion17},
\end{eqnarray}

\begin{eqnarray} \theta_A = -2c\sqrt{(\tau_x^{(A)})^2 +
(\tau_y^{(A)})^2 + (\tau_z^{(A)})^2}\label{Torsion55}.
\end{eqnarray}

In a similar manner for (B).

These results allow us to distinguish two different situations:
\bigskip

(i) $\vert l{\partial\kappa_n\over\partial y}\vert, \vert
l{\partial\kappa_n\over\partial x}\vert << \vert\kappa_n\vert$.
Therefore $\vec{n}_A = \vec{n}_B$, the axis of rotation of the
beams is the same, and, in consequence, they differ only in the
angle of rotation, $\theta_A \not= \theta_B$.
\bigskip

(ii) If the foregoing condition does not hold, then not only
$\theta_A \not= \theta_B$, also additionally, $\vec{n}_A \not=
\vec{n}_B$.

Let us now assume that $\phi(t= 0)$ is the linear coherent
superposition of states $\chi_{(+)}$ and $\chi_{(-)}$, where
$\sigma_z\chi_{(\pm)} = \pm\chi_{(\pm)}$, namely

\begin{eqnarray} \phi(t= 0) = c_{(+)}\chi_{(+)} +
c_{(-)}\chi_{(-)}\label{Torsion19}.
\end{eqnarray}

The interference pattern at $S$ is a function of the complete
state vector, i.e., $\vert\psi>$, whose dynamics evolves according
to (\ref{Torsion1}). This last argument may be rephrased stating
$I = \vert(\vert\psi>_{(A)} + \vert\psi>_{(B)})\vert^2$. It
embodies two different contributions, one stemming from
$\mathcal{E}$$_s$, and the second one from $\mathcal{E}$$_r$. In
other words, we find that

\begin{eqnarray} I = 2 +
2\cos\Bigl(({m\over\hbar})^2glL\tilde{\lambda}\Bigr)\Bigl[\phi\dagger_A(S)\phi_B(S)
+ \phi\dagger_B(S)\phi_A(S)\Bigr]\label{Torsion20}.
\end{eqnarray}

Therefore

\begin{eqnarray} I = 2 +
2\cos\Bigl(({m\over\hbar})^2glL\tilde{\lambda}\Bigr)\Bigl[\cos({\theta_A\over
2})\cos({\theta_B\over 2})
+ [\vec{n}_A\cdot\vec{n}_B]\sin({\theta_A\over 2})\sin({\theta_B\over 2})\Bigr] \nonumber\\
-
2\sin\Bigl(({m\over\hbar})^2glL\tilde{\lambda}\Bigr)\Bigl[\sin({\theta_A\over
2}) \sin({\theta_B\over 2})[\vec{n}_A\times\vec{n}_B] +
\sin({\theta_A\over 2})
\cos({\theta_B\over 2})\vec{n}_A \nonumber\\
-\sin({\theta_B\over 2})\cos({\theta_A\over
2})\vec{n}_B\Bigr]\cdot
\Bigl[2Re(c^{\ast}_{(+)}c_{(-)})\vec{e}_x - 2Im(c^{\ast}_{(-)}c_{(+)})\vec{e}_y  \nonumber\\
+ (\vert c_{(+)}\vert^2 - \vert
c_{(-)}\vert^2\vec{e}_z\Bigr]\label{Torsion21}.
\end{eqnarray}

It is readily seen that
$\cos\Bigl(({m\over\hbar})^2glL\tilde{\lambda}\Bigr)$ corresponds
to the interference term in COW \cite{Colella1}. In other words,
if we discard torsion, then the usual COW result is recovered.
Additionally, $\vec{e}_n$ denotes the unit vector along the
$n$--axis.

Under these conditions we have that

\begin{eqnarray} I = 2 +
2\cos\Bigl(({m\over\hbar})^2glL\tilde{\lambda}\Bigr)\Bigl[\cos({\theta_A\over
2})\cos({\theta_B\over 2}) +
[\vec{n}_A\cdot\vec{n}_B]\sin({\theta_A\over
2})\sin({\theta_B\over 2})\Bigr]\label{Torsion22}.
\end{eqnarray}

Let us now analyze some cases, for instance, $c_{(+)}, c_{(-)}
\in\Re$.

Here we consider $c_{(+)} \not = c_{(-)}$.
\bigskip

\begin{eqnarray} I = 2 +
2\cos\Bigl(({m\over\hbar})^2glL\tilde{\lambda}\Bigr)\Bigl[\cos({\theta_A\over
2})\cos({\theta_B\over 2})
+ [\vec{n}_A\cdot\vec{n}_B]\sin({\theta_A\over 2})\sin({\theta_B\over 2})\Bigr] \nonumber\\
-
2\sin\Bigl(({m\over\hbar})^2glL\tilde{\lambda}\Bigr)\Bigl[\sin({\theta_A\over
2}) \sin({\theta_B\over 2})[\vec{n}_A\times\vec{n}_B] +
\sin({\theta_A\over 2})
\cos({\theta_B\over 2})\vec{n}_A \nonumber\\
-\sin({\theta_B\over 2})\cos({\theta_A\over
2})\vec{n}_B\Bigr]\cdot[\vert c_{(+)}\vert^2 - \vert
c_{(-)}]\vert^2\vec{e}_z\label{Torsion23}.
\end{eqnarray}

Neglecting all derivatives of the axial part of the torsion, a
condition tantamount to $\vec{n}_A = \vec{n}_B$, we obtain

\begin{eqnarray} I = 2 +
2\cos\Bigl(({m\over\hbar})^2glL\tilde{\lambda}\Bigr)
\cos\Bigl(({m\tilde{\lambda}\over\hbar})^3gcl^2K\Bigr)  \nonumber\\
- 2\kappa(0)_z/K\Bigl[\vert c_{(+)}\vert^2 - \vert
c_{(-)}\vert^2\Bigr]
\sin\Bigl(({m\over\hbar})^2glL\tilde{\lambda}\Bigr)
\sin\Bigl(({m\tilde{\lambda}\over\hbar})^3gcl^2K\Bigr)\label{Torsion24}.
\end{eqnarray}

In the foregoing expression the following definition has been
introduced $K = \sqrt{\kappa^2(0)_x+ \kappa^2(0)_y+
\kappa^2(0)_z}$.

Expression (\ref{Torsion21}) provides enough leeway to consider
the possibility of detecting the consequences of torsion, upon the
interference pattern. This can be done modifying the values of
$c_{(+)}$ and $c_{(-)}$. Indeed, choosing $c_{(+)} = 1$ and
$c_{(-)} = 0$,
\bigskip

\begin{eqnarray} I = 2 +
2\cos\Bigl(({m\over\hbar})^2glL\tilde{\lambda}\Bigr)
\cos\Bigl(({m\tilde{\lambda}\over\hbar})^3gcl^2K\Bigr)  \nonumber\\
-
2\kappa(0)_z/K\sin\Bigl(({m\over\hbar})^2glL\tilde{\lambda}\Bigr)
\sin\Bigl(({m\tilde{\lambda}\over\hbar})^3gcl^2K\Bigr)\label{Torsion25}.
\end{eqnarray}

If now $c_{(+)} = 0$ and $c_{(-)} = 1$, then

\begin{eqnarray} I = 2 +
2\cos\Bigl(({m\over\hbar})^2glL\tilde{\lambda}\Bigr)
\cos\Bigl(({m\tilde{\lambda}\over\hbar})^3gcl^2K\Bigr)  \nonumber\\
+
2\kappa(0)_z/K\sin\Bigl(({m\over\hbar})^2glL\tilde{\lambda}\Bigr)
\sin\Bigl(({m\tilde{\lambda}\over\hbar})^3gcl^2K\Bigr)\label{Torsion26}.
\end{eqnarray}

If the parameters are switched from $\{c_{(+)} = 1, c_{(-)} = 0\}$
to $\{c_{(+)} = 0, c_{(-)} = 1\}$ a sign change, in the second
term of the right--hand side, appears. This effect vanishes if the
torsion is null. In other words, this sign change is a direct
consequence of torsion, and appears only if we modify the linear
superposition of the initial spin state vector. As a matter of
fact, considering a series of experiments, in which we begin with
$\{c_{(+)} = 1, c_{(-)} = 0\}$, and gradually we change these two
values (the first parameter diminishes, whereas the second one
increases), then the role, that the absolute value of the second
term plays, peters out. This happens when $c_{(+)} = 1/\sqrt2$.
Afterwards, it starts to appear, once again.

At this point we proceed to estimate the order of magnitude of the
torsion contributions, and then relate it to the current
experimental discrepancy. We have already mentioned that the
theoretical prediction possesses a discrepancy on the order of one
percent in the phase shift \cite{Colella4}. Additionally, a
sufficiently (for our purposes) stringent experimental bound reads
$K\sim 10^{-15}m^{-2}$ \cite{Laemmerzahl2} and, hence (employing
the typical experimental values \cite{Colella1, Colella4}) and
denoting by $\Gamma$ the contribution to this discrepancy, we
deduce

\begin{eqnarray} \Gamma \sim 10^{-16}\label{Torsion27}.
\end{eqnarray}

Let us now analyze this last expression. It is easily seen that
the involved experimental discrepancy cannot be understood,
exclusively, by torsion effects. The reason stems from the fact
that the value of $\Gamma$ is too tiny to provide the necessary
explanation. The introduction of only one new element, Cartan's
torsion tensor, cannot give an explanation to this fact.
Additional work is needed in this direction.

At this point let us remember that the COW experiment has an
additional feature, which has spurred a hot debate about the
validity in the quantum domain of WEP. Indeed, the appearance of
the mass term in the interference expression ($[{m\over\hbar}]^2$)
has been understood by some authors as a possible manifestation of
non--geometricity \cite{Ahluwalia1, Ahluwalia2} in the
gravitational field. Taking a look at (\ref{Torsion24}) it is
readily seen that under the aegis of torsion this trait, not only
does not vanish, but on it an additional term is bestowed, i.e.,
$[\frac{m}{\hbar}]^3$. The only difference between the present
model and the usual COW case has been the introduction of torsion
in the motion equation for the neutrons, see (\ref{Torsion1}).
This implies that the emergence of this extra term in the
interference pattern is a consequence of the coupling between the
torsion tensor and the spin of our neutrons. A fleeting glance at
this term shows us that it involves the mass of the particle, and
we may wonder why the mass appears in the interference pattern,
since the new term involves torsion. Looking at (\ref{Cartan3}) we
find the answer, the Ricci tensor and also its contraction, are a
mixture between the metric and the contorsion tensor, since the
metric is coupled to mass--energy and the contorsion to the spin.
It should not surprise us the appearance of mass in the term
related to the influence of torsion in the interference pattern.

We should be careful in the interpretation of this fact since it
could be conjectured that the term $[\frac{m}{\hbar}]^3$ confirms
that the emergence of the term $[\frac{m}{\hbar}]^2$ in the usual
COW experiment is related to a violation of WEP. Indeed, we could
claim, if torsion entails the violation of WEP, and it appears in
the interference pattern as a parameter involving mass, that
$[\frac{m}{\hbar}]^2$ is a manifestation of a violation of WEP in
the quantum domain. Of course, this last argumentation has to be
supported by further experimental evidence.

\subsubsection{Non--Newtonian Gravity--Induced Interference}

Now we analyze the theoretical predictions, at quantum level, that
a Yukawa term could have. But before doing this let us first
comment some interesting proposals designed to measure the
Newtonian gravitational constant in the realm of atom
interferometry. The pertinency at this point of these proposals
stems from the fact that a good knowledge of the value of $G$ is a
prerequisite in any proposal attempting to determine any
non--Newtonian term associated to gravity, as the expressions
below will clearly show.

These ideas may be divided into two different models. Let us
mention first the so--called MAGIA project \cite{Fattori1,
Fattori2} in which free--falling laser cooled atoms will be used
as probes for the gravitational potential. The acceleration will
be detected resorting to the Raman atom interferometry techniques,
and a value for $G$ will be then deduced. The second proposal
\cite{Fixler1} is based upon a gradiometer which measures the
differential acceleration of two samples of laser--cooled Cs
atoms. Clearly, these two ideas employ independent methods in the
measuring process of $G$.

The idea now in the context of non--Newtonian gravity is to
include in the theoretical background of the gravitational field a
Yukawa term, and calculate the corresponding effects in an
experimental proposal which is very similar to the COW
construction \cite{Colella1}. The goal is to understand the way in
which the parameters, appearing in Fujii's model, impinge upon the
interference pattern of a COW experiment. The parameter $\lambda$
is related to the range of this new force, whereas $\alpha$ has no
dimensions and is connected with the intensity of the interaction
\cite{Fujii1}. We must mention that in this context the analysis
of the effects that a fifth force could have in a COW experiment
has not been done \cite{Fishbach2}.

Consider a Yukawa modification to the Newtonian gravitational
potential \cite{Fujii1}

\begin{eqnarray} V(r) = -G_{\infty}{mM\over r}\left(1 + \alpha
e^{-{r\over \lambda}}\right)\label{Nonnewtonian1}.
\end{eqnarray}

The Lagrangian of a particle with mass $m$, moving in this field,
is

\begin{eqnarray} L = {m\over 2}\dot{\vec{r}}^2 +
G_{\infty}{mM\over r}\left(1 + \alpha e^{-{r\over
\lambda}}\right)\label{Nonnewtonian2}.
\end{eqnarray}

Since we are considering a terrestrial experiment we introduce a
definition in order to have a parameter related to the height
above the surface of the Earth at which the experiment is carried
out. Define $r = R + l$, where $R$ is the Earth's radius and $l$
the height over the Earth's surface. Keeping terms up to second
order in $l$

\begin{eqnarray} L = {m\over 2}\dot{\vec{r}}^2 +
G_{\infty}{mM\over R} {\Bigg(}\left[1 + \alpha + {\alpha
R\over\lambda}\left({R\over 2\lambda} - 1\right)\right] - \left[{1
+ \alpha\over R} - {\alpha R\over 2\lambda^2}\right]l + {1 +
\alpha\over R^2}l^2{\Bigg)}\label{Nonnewtonian3}.
\end{eqnarray}

 Two particles, starting at point $O$, move along two different
trajectories, $C$ and $\tilde{C}$, and afterwards they are
detected at a certain point $S$, see the figure on page 37. Of
course, we require the semiclassical approximation, i.e., the size
of the wavelengths of the packets is much smaller than the size in
which the field changes considerably.

The wave function reads

\begin{eqnarray} \psi(\vec{r}, t) \sim {1\over [E -
V(\vec{r})]^{{1\over 4}}}
\exp\left\{{\pm{i\over\hbar}}\int_{(O)}^{(S)} \sqrt{2m[E -
V(\vec{r})]}d\tilde{L} -
{i\over\hbar}Et\right\}\label{Nonnewtonian4},
\end{eqnarray}

\noindent where $V(\vec{r}) = -G_{\infty}{mM\over R} {\Bigg(}[1 +
\alpha + {\alpha R\over\lambda}({R\over 2\lambda} - 1)] - [{1 +
\alpha\over R} - {\alpha R\over 2\lambda^2}]l + {1 + \alpha\over
R^2}l^2{\Bigg)}$. The line integral in (\ref{Nonnewtonian4}) has
to be calculated along $C$ and $\tilde{C}$, there are two
different trajectories.

The interference term at the detection point is given by

\begin{eqnarray} I
=\cos\left\{-{gm^2Ll_S\Lambda\over\hbar^2}\left[1 - {\alpha
R^2\over 2\lambda^2(1 + \alpha)} - {l_S\over
R}\right]\right\}\label{Nonnewtonian7}.
\end{eqnarray}

$\Lambda$ denotes the initial reduced wavelength of the particles,
and $g_{\infty} = g/(1 + \alpha)$, where $g = {GM\over R^2}$, and
$l_S$ is the non--null coordinate of $S$.

To recover the usual analysis \cite{Sakurai1} we must impose the
condition $\alpha = 0$.

\begin{eqnarray} I_N
=\cos\left\{-{gm^2Ll_S\Lambda\over\hbar^2}\left[1 -{l_S\over
R}\right]\right\}\label{Nonnewtonian8},
\end{eqnarray}

 As a matter of fact, the result of COW does not contain the term that is quadratic in
$l_S$. It appears in our result since we have introduced a less
stringent approximation, to derive the results of COW we need only
a homogeneous Newtonian gravitational field, and expression
(\ref{Nonnewtonian3}) includes the case of an inhomogeneous
gravitational field, i.e., the term ${1 + \alpha\over R^2}l^2$.
This means that expression (\ref{Nonnewtonian8}) is the
interference term when we consider, in a Newtonian field, a
dependence on the height above the surface of the Earth, up to
second order in $l$.

The difference between the Newtonian and non--Newtonian cases
divided by the Newtonian value is (approximately)

\begin{eqnarray} \Delta = {\alpha R^2\over 2\lambda^2(1 +
\alpha)}\left(1 + {l_S\over R}\right)\label{Nonnewtonian9}.
\end{eqnarray}

The Compton wavelength of our new quantum particle (with mass
$m_5$) reads $\lambda = {\hbar\over cm_5}$. Introducing it into
(\ref{Nonnewtonian8})

\begin{eqnarray} \Delta = {\alpha (Rcm_5)^2\over 2\hbar^2(1 +
\alpha)}\left(1 + {l_S\over R}\right)\label{Nonnewtonian10}.
\end{eqnarray}

We may now cast the interference term in the following form

\begin{eqnarray} I = I_N\cos\left\{{gm^2Ll_S\Lambda\alpha
R^2\over 2\hbar^2\lambda^2(1 + \alpha)}\right\} \pm\sqrt{1 -
I_N^2}\sin\left\{{gm^2Ll_S\Lambda\alpha R^2\over
2\hbar^2\lambda^2(1 + \alpha)}\right\}\label{Nonnewtonian11}.
\end{eqnarray}

This last expression contains a deviation from the usual
inverse--square law. If the usual experimental parameters related
to COW \cite{Colella1} are $\alpha\sim 10^{-3}$ and $\lambda\sim
10^{4}$cm \cite{Chan1}, then

\begin{eqnarray} \frac{gm^2Ll_S\Lambda\alpha
R^2}{2\hbar^2\lambda^2(1 + \alpha)}\sim
10^7(cm)^{-1}l_S\label{Nonnewtonian12}.
\end{eqnarray}

The interference pattern becomes now a simple function of $l_S$

\begin{eqnarray} I = I_N\cos\left\{10^7(cm)^{-1}l_S\right\}
\pm\sqrt{1 - I_N^2}\sin\left\{10^7(cm)^{-1}l_S
\right\}\label{Nonnewtonian13}.
\end{eqnarray}

The possibility of testing non--Newtonian gravity appears in
connection with the dependence of (\ref{Nonnewtonian13}) on $l_S$.
The COW experiment can, in principle, be used to detect a Yukawa
type modification to the inverse--square law of gravity. A word of
caution has to be added in this context. From the very beginning
it was clearly stated that we require the validity of the
semiclassical approach.  In this model we have the Newtonian
contribution plus a Yukawa term, this implies that the
semiclassical approximation must take into account these two
fields. The semiclassical limit requires the wavelength of the
particle to be smaller than the distance in which the Newtonian
potential and the Yukawa term have a noticeable change.

\subsection{Non--Demolition Variables and Gravity-Induced Interference}

\subsubsection{Non--Demolition Variables and Restricted Path Integral}

Nowadays one of the fundamental problems in modern physics
comprises the so--called quantum measurement problem
\cite{Omnes1}. Though there are several attempts to solve this old
conundrum (some of them are equivalent \cite{Onofrio1}), here we
will explain the restricted path integral formalism (RPIF)
\cite{Mensky1}. This particular choice has behind it a sound
reason. Indeed, RPIF allows us to test not only the possibility of
a Yukawa term addition to the gravitational interaction
\cite{Fujii1}, but also one of the solutions to the quantum
measurement problem. This last point is quite relevant since we
may find some postures which claim either that there is no
measurement problem in quantum mechanics \cite{Penha1}, or that
quantum mechanics is an incomplete theory \cite{Bohm1}. RPIF
provides us the possibility of testing the existence of a fifth
force and one of the solutions to a fundamental problem (if we
accept that it is a problem) in quantum theory.

Let us now mention briefly the main ideas behind RPIF. This model
explains a continuous quantum measurement with the introduction of
a restriction on the integration domain of the corresponding path
integral \cite{Mensky1}. This last condition can also be
reformulated in terms of a weight functional that has to be
considered in the path integral. Clearly, this weight functional
contains all the information about the interaction between the
measuring device and the measured system.

A measurement in the context of quantum mechanics possesses some
traits which are absent in the classical version. For instance, in
the latter there is no impediment for the simultaneous measuring
of two variables with arbitrary precision. This is, in some cases,
forbidden in the quantum realm \cite{Omnes1, Braginsky1}. The
issue of quantum theory of measurement was considered, for some
decades, as an almost esoteric topic. The main reason behind this
attitude was the irrelevance of the quantum theory of measurement
in the technological aspect. Nevertheless, this started to change
in the 1980s, when technology began to catch up with theory.

The concept of non--demolition variables stems from a
technological requirement. The development of gravitational wave
detectors required methods for measuring macroscopic variables at
levels of precision approaching and even exceeding the standard
quantum limit imposed by Heisenberg's uncertainty relation. The
theoretical analysis of the traditional schemes associated to
measurement proved that the corresponding precisions can never go
beyond the standard quantum limit. This fact spurred the
development of non--traditional measurements procedures, which are
known as quantum non--demolition measurements \cite{Braginsky1}.

The basic idea around the concept of quantum non--demolition (QND)
measurements is to carry out a sequence of measurements of an
observable in such a way that the measuring process does not
diminish the predictability of the results of subsequent
measurements of the same observable . The main idea of a QND
measurement is to deduce a variable such that the unavoidable
disturbance of the conjugate observable does not disturb the
evolution of the chosen variable \cite{Bocko1}.

The dynamical evolution of a system, subject to a measuring
process, limits the class of observables that may be measured
repeatedly with arbitrary precision. Let us explain this a little
bit better and assume that in our case we monitor conti\-nuous\-ly
the observable $A(t)$. Then $A(t)$ is a QND variable if the
commutation relation $[B(t'), B(t'')] = 0$ is fulfilled, for any
$t'$ and $t''$, where $B(t) = \exp\{i{H\over
\hbar}t\}A(t)\exp\{-i{H\over \hbar}t\}$ is the corresponding
operator in Heisenberg picture \cite{Mensky1}.

Let us now suppose that in our case $A(t) = \rho p + \sigma z$,
where $\rho$ and $\sigma$ are functions of time. In this
particular case, the condition that determines when $A(t)$ is a
QND variable may be written as a differential equation
\cite{Braginsky1}

\begin{equation}
{df\over dt} = {f^2\over m} - m\Omega^2,
\end{equation}\label{Nond1}
\bigskip

\noindent where $f(t)=  \sigma/\rho$\label{Nond2}.
\bigskip

It is readily seen that a solution to the differential equation is

\begin{equation}
f(t)= -m\Omega\tanh(\Omega t)\label{Nond3}.
\end{equation}
\bigskip

Choosing $\rho(t) = 1$, we find that in our case a possible QND
variable is provided by

\begin{equation}
A(t) = p - m\Omega z\tanh(\Omega t)\label{Nond4}.
\end{equation}

\subsubsection{Non--Demolition Variables and Non--Newtonian Gravity}

The idea in this part is two--fold: Firstly, the effects of a
Yukawa term upon a quantum system (the one is continuously
monitored) will be calculated and compared against the
corresponding results containing only the usual Newtonian
gravitational potential (in a proposal that has as new element the
fact that it is not a COW type experiment); Secondly, new
theoretical predictions for one of the models in the context of
quantum measurement theory will be found. The first goal offers us
the possibility for testing a fifth force in the form of a Yukawa
term \cite{Fujii1}, whereas the second one provides us a series of
theoretical predictions that embody the premises of one of the
models that claim to be a solution to the quantum measurement
problem, i.e., RPIF.

These two goals will be achieved obtaining a non--demolition
variable for the case of a particle subject to a gravitational
field which contains a Yukawa term such that $\lambda$ (this
parameter is related to the range of this new force) has the same
order of magnitude of the radius of the Earth. This proposal, as
will become clear below, does not involve an interference
experiment in the sense of COW. From this remark we may pose the
following question: Could this idea be considered an interference
experiment? The answer is {\it no}, in the sense of a COW
interference device, but {\it yes} in the context of the meaning
of Feynman path integral version of quantum mechanics, in which
the interference among all the possible trajectories between two
points of the configuration space are fundamental to obtain the
properties of the corresponding quantum system. In this sense it
is an interference experiment and the aspect of the principles of
quantum theory that this idea could allow us to test is the
response of a quantum system to series of continuous measurements,
i.e., the understanding of a quantum system under a continuous
measurement process will be improved.

We have already mentioned that we will assume a certain value for
one of the parameters involved in a fifth force, let us now
explain the reasons behind this choice. The extant experiments
\cite{Fishbach2} set constraints for $\lambda$ for ranges between
10km and 1000km \cite{Smith1}, but the case in which $\lambda
\sim$ Earth's radius remains rather unexplored \cite{Fishbach2}.
Afterwards, we will consider, along the ideas of the restricted
path integral formalism (RPIF) \cite{Mensky1}, the continuous
monitoring of a non--demolition parameter, and then calculate, not
only, the corresponding propagators, but also the probabilities
associated with the different measurement outputs. These
probabilities will provide not only an interesting way to prove
the existence of a Yukawa term as an additional gravitational
element, but also to prove RPIF. The comparison with the purely
Newtonian case will be done, and the effects of the Yukawa term
upon our quantum system will be obtained. In order to do this we
will consider a particular measurement output, and then the
corresponding probabilities, in the Newtonian and non--Newtonian
situations, will be evaluated, and finally the ratio between them
will be obtained.

Suppose that we have a spherical body with mass $M$ and radius
$R$. Let us now consider the case of a Yukawa term \cite{Fujii1},
hence the potential and the Lagrangian are given by expressions
(\ref{Nonnewtonian1}) and (\ref{Nonnewtonian2}) respectively.

We now define $r = R + z$, where $R$ is the body's radius, and $z$
the height over its surface. If $R/\lambda \sim 1$ (which means
that the range of this Yukawa term has the same order of magnitude
as the radius of our spherical body), and if $z<<<R$, then we may
approximate the potential as follows

\begin{eqnarray} V(r)= -G_{\infty}\frac{M}{R} \Bigl([1 + \alpha] -
[\frac{1 + \alpha}{R} + \frac{\alpha}{2\lambda}]z +[\frac{1 +
\alpha}{2R^2} + {\alpha\over 2R\lambda}+ {\alpha\over
2\lambda^2}]z^2\Bigr)\label{Nondemolition3}.
\end{eqnarray}

The Lagrangian of our particle of mass $m$ becomes now

\begin{eqnarray} L = {{\vec{p}}~^2\over 2m} + G_{\infty}{mM\over
R} \Bigl([1 + \alpha] - [{1 + \alpha\over R} + {\alpha\over
2\lambda}]z +[{1 + \alpha\over 2R^2} + {\alpha\over 2R\lambda}+
{\alpha\over 2\lambda^2}]z^2\Bigr)\label{Nondemolition4}.
\end{eqnarray}

If our particle moves from point $N$ to point $W$, then the
propagator becomes

\begin{eqnarray} U(W,\tau'';N, \tau') = \Bigl({m\over 2\pi i\hbar
T}\Bigr)
\exp\{{im\over 2\hbar T}[(x_W - x_N)^2 + (y_W - y_N)^2]\}\nonumber\\
\times\int d[z(t)]\exp\{{i\over\hbar}\int_{\tau'}^{\tau''}
[{m\over 2}\dot{z}^2 + G_{\infty}{mM\over R}
\Bigl([1 + \alpha] \nonumber\\
- [{1 + \alpha\over R} + {\alpha\over 2\lambda}]z +[{1 +
\alpha\over 2R^2} + {\alpha\over 2R\lambda}+ {\alpha\over
2\lambda^2}]z^2\Bigr)]dt\}\label{Nondemolition5},
\end{eqnarray}

\noindent here $T = \tau'' - \tau'$.

Consider now the following two definitions

\begin{eqnarray} F = -G_{\infty}{mM\over R}[{1 + \alpha\over R} +
{\alpha\over 2\lambda}]\label{Nondemolition6},
\end{eqnarray}

\begin{eqnarray} \omega^2 = -2{G_{\infty}M\over R}[{1+
\alpha\over 2R^2}+ {\alpha\over 2\lambda R} + {\alpha\over
2\lambda^2}]\label{Nondemolition7},
\end{eqnarray}

\noindent where $G = G_{\infty}[1 + \alpha]$, and $G$ is the
Newtonian gravitational constant \cite{Fishbach1}. In our case,
$[{1+ \alpha\over 2R^2}+ {\alpha\over 2\lambda R} + {\alpha\over
2\lambda^2}] >0$, hence, $\omega = i\Omega$, where $\Omega\in\Re$.
\bigskip

The propagator becomes now

\begin{eqnarray} U(W,\tau'';N, \tau') = \Bigl({m\over 2\pi i\hbar
T}\Bigr)
\exp\{{im\over 2\hbar T}[(x_W - x_N)^2 + (y_W - y_N)^2]\}\nonumber\\
\times\int_{z_N}^{z_W}d[p]d[z(t)]\exp\{{i\over\hbar}\int_{\tau'}^{\tau''}
[{p^2\over 2m} + (1 + \alpha){G_{\infty}mM\over R} + Fz + {m\over
2}\Omega^2z^2]dt\}\label{Nondemolition8}.
\end{eqnarray}

According to RPIF, if the variable $A(t)$ is continuously
monitored, then we must introduce a particular expression for our
weight functional, i.e., for $w_{[a(t)]}[A(t)]$. As was mentioned
before, the weight functional $w_{[a(t)]}[A(t)]$ contains the
information concerning the measuring process. Here $a(t)$ denotes
the measurement readout. At this point it is noteworthy to comment
that the more probable the ``trajectory'' $A(t)$ is, according to
the output $a$, the bigger that $\omega_{[a(t)]}[A(t)]$ becomes
\cite{Mensky1}. In other words, the value of
$\omega_{[a(t)]}[A(t)]$ is approximately one for all
``trajectories'' $A(t)$ that agree with the measurement output
$a(t)$, and it is almost 0 for those that do not match with the
result of the experiment. Clearly, an issue remains to be
addressed, namely, the choice of our weight functional. In order
to solve this difficulty, let us mention that the results coming
from a Heaveside weight functional \cite{Mensky2} and those coming
from a gaussian one \cite{Mensky1} coincide up to the order of
magnitude. Therefore we may consider a gaussian weight functional
as an approximation to the correct expression.

We choose as our weight functional the following expression

\begin{equation}
\omega_{[a(t)]}[A(t)] = \exp\{-{2\over T\Delta a^2}\int _{\tau
'}^{\tau ''}[A(t) - a(t)]^2dt\}\label{Nondemolition9}.
\end{equation}
\bigskip

$\Delta a$ denotes the resolution of the measuring device.

With our weight functional choice (\ref{Nondemolition9}) the new
propagator may be written as follows

\begin{eqnarray} U_{[a(t)]}(W,\tau'';N, \tau') = \Bigl({m\over
2\pi i\hbar T}\Bigr)
\exp\{{im\over 2\hbar T}[(x_W - x_N)^2 + (y_W - y_N)^2]\}\nonumber\\
\times\int_{z_N}^{z_W}d[p]d[z(t)]\exp\{{i\over\hbar}\int_{\tau'}^{\tau''}
[{p^2\over 2m} + (1 + \alpha){G_{\infty}mM\over R} + Fz + {m\over 2}\Omega^2z^2]dt\}\nonumber\\
\times\exp\{-{1\over T\Delta
a^2}\int_{\tau'}^{\tau''}[A-a]^2dt\}\label{Nondemolition10}.
\end{eqnarray}
\bigskip

This propagator involves two gaussian integrals, and it can be
calculated \cite{Dittrich1}

\begin{eqnarray} U_{[a(t)]}(W,\tau'';N, \tau') = \Bigl({m\over
2\pi i\hbar T}\Bigr)
\exp\{{im\over 2\hbar T}[(x_W - x_N)^2 + (y_W - y_N)^2]\}\nonumber\\
\exp\{{i\over\hbar}(1 + \alpha){G_{\infty}mM\over R}T\}
\exp\{{-T\Delta a^2 + i2m\hbar\over 4m^2\hbar^2 + T^2\Delta a^4}\int_{\tau'}^{\tau''}a^2(t)dt\}\nonumber\\
\times\exp\{{-i\hbar\over
2m\Omega^2}\int_{\tau'}^{\tau''}\Bigl[[{F\over\hbar} +
{4m^2\hbar\Omega a\over 4m^2\hbar^2 + T^2\Delta a^4}\tanh(\Omega
t)
+ i{2ma\Omega T\Delta a^2\over 4m^2\hbar^2 + T^2\Delta a^4}\Bigr]^2\nonumber\\
\times\Bigl[{4m^2\hbar^2[1 + \tanh^2(\Omega t)] + T^2\Delta a^4 -
i2m\hbar T\Delta a^2\tanh^2(\Omega t)\over 4m^2\hbar^2[1 +
\tanh^2(\Omega t)]^2 + T^2\Delta a^4
}\Bigr]dt\}\label{Nondemolition11}.
\end{eqnarray}
\bigskip

The probability, $P_{[a(t)]}$, of obtaining as measurement output
$a(t)$ is given by expression $P_{[a(t)]} = \vert
U_{[a(t)]}\vert^2$ \cite{Mensky1}.  Hence, in this case

\begin{eqnarray} P_{[a(t)]}=
\exp\{{-2T\Delta a^2 \over 4m^2\hbar^2 + T^2\Delta a^4}\int_{\tau'}^{\tau''}a^2(t)dt\}\nonumber\\
\times\exp\{{\hbar\over m\Omega^2}\int_{\tau'}^{\tau''}[2I_1I_2I_3
+ I_4(I^2_2 - I^2_1)]dt\}\label{Nondemolition12}.
\end{eqnarray}
\bigskip

In this last expression we have that

\begin{eqnarray} I _1 = {F\over\hbar} + {4m^2\hbar\Omega a\over
4m^2\hbar^2 + T^2\Delta a^4}\tanh(\Omega
t)\label{Nondemolition13},
\end{eqnarray}

\begin{eqnarray} I _2 = {2ma\Omega T\Delta a^2\over 4m^2\hbar^2 +
T^2\Delta a^4}\tanh(\Omega t)\label{Nondemolition14},
\end{eqnarray}

\begin{eqnarray} I _3 = {4m^2\hbar^2[1 + \tanh^2(\Omega t)] +
T^2\Delta a^4\over 4m^2\hbar^2[1 + \tanh^2(\Omega t)]^2 +
T^2\Delta a^4}\label{Nondemolition15},
\end{eqnarray}

\begin{eqnarray} I _4 = {2m\hbar T\Delta a^2\tanh^2(\Omega
t)\over 4m^2\hbar^2[1 + \tanh^2(\Omega t)]^2 + T^2\Delta
a^4}\label{Nondemolition16}.
\end{eqnarray}

As mentioned before one of the ideas in this proposal involves the
possibility of detecting a Yukawa term in the range of $\lambda
\sim R$. If the probabilities in the non--Newtonian and Newtonian
cases are compared, then we find a manner to evaluate the
feasibility of the proposal. The use of systems with internal
structure complicates the analysis of interference experiments of
the COW type \cite{Borde2, Borde3} (here we mean that it requires
the introduction of additional approximations, for instance, the
so--called rotating wave approximation in the description of the
S-matrices associated to the beam splitters  \cite{Borde3}). In
consequence we assume the motion of thermal neutrons, i.e., $m
\sim 10^{-27}$ kg, and a Compton wavelength of $\Lambda \sim
10^{-10}$m. The experiment lasts $T \sim 10^{-4}$ s., which is the
flight time of the thermal neutrons when the travelled distance is
$l\sim 10^{-2}$m.

Additionally, we require an estimation of the resolution, $\Delta
a$, of a measuring device which could detect $A(t)$. The current
technology cannot detect $A(t)$ \cite{Bocko1}. At this point we
must add that one of the research areas that could, in a near
future, allow the continuous monitoring of the position of a
quantum particle is the use of Paul traps \cite{Thompson1}. Let us
now draw the attention to the fact that $A(t)$ has momentum units
($A(t) = p - m\Omega z\tanh(\Omega t)$). Hence, it will be assumed
that $\Delta a \sim \hbar/\Delta z$, where $\Delta z \sim
10^{-6}$m is the resolution in the case of a Paul trap
\cite{Paul1, Dehmelt1}. This last approximation implies $\Delta a
\sim 10^{-29}$ $kg\cdot m/s$.

The measurement output, $a(t)$, is assumed to be a constant, i.e.,
it equals the initial momentum of the thermal neutrons. In other
words, we suppose that the involved measurement output fulfills
the condition $a(t) = a^{\star} \sim 10^{-24}kg\cdot m/s$.

The current experimental status imposes, already, some
restrictions upon the possible values of $\alpha$, as a function
of $\lambda$. In our case, if $\lambda \sim 10^6$m, then
$\vert\alpha\vert \leq 10^{-7}$ (see page 62 of \cite{Fishbach1}).
Additionally, it is noteworthy to comment that the LAGEOS
satellite, and more generally, the so--called Satellite Laser
Ranging technique provides very accurate measurements for the case
in which $\lambda \approx 1$ A.U. \cite{Iorio1}, and also in the
range of planetary distances, i.e., $10^5m<\lambda<10^7m$
\cite{Bertolami1}.

Let us now denote the prediction in the case in which the Yukawa
term is absent by $P^{N}_{[a(t)]}$, then we have, approximately,
that

\begin{eqnarray} P_{[a^{\star}]}/P^{(N)}_{[a^{\star} ]} =
\exp\{{aT^2\Delta a^2\over 2m^2\hbar^2}\sqrt{{Gm^2M\over
R}}\alpha\}\label{Nondemolition17}.
\end{eqnarray}

Hence

\begin{eqnarray} P_{[a^{\star}]}/P^{(N)}_{[a^{\star}]} = 10^{-2} \label{Nondemolition18}.
\end{eqnarray}

Considering our rough example, the possibility of detecting this
Yukawa term would depend upon the fact that the involved measuring
apparatus could detect probabilities with a precision better than
one per cent.

Let us explain this point, and assume that we have performed this
experiment several times, say $s>>1$, such that the results read
$a_1$, $a_2$,..., $a_s$. Define now a partition of the set of
results \cite{Gallian1}, namely, $a_g\in cl\bigl\{a_t\bigr\}$, if
and only if, $P_{[a_t]} =P_{[a_g]}$. Clearly, we may define the
cardinality of these sets, $C[cl\bigl\{a_t\bigr\}]$. If $s$ is
sufficiently large, then we have that
$C[cl\bigl\{a_t\bigr\}]/C[cl\bigl\{a_r\bigr\}]$ will be a good
approximation to the ratio of the corresponding probabilities. Our
expressions entail that this ratio does depend, in a non--trivial
way, upon the parameters appearing in the Yukawa term. In other
words, with $m$, $a_r$, etc., (\ref{Nondemolition12}) can be
evaluated, and then confronted against
$P_{[a_t^{\star}]}/P_{[a_r^{\star}]}$, and afterwards, compared
with $C[cl\bigl\{a_t\bigr\}]/C[cl\bigl\{a_r\bigr\}]$.

The appearance of the mass term in the COW experiment leads us to
ask about the role that this parameter plays in this new scheme. A
fleeting glance at (\ref{Nondemolition12}) shows us that two
particles with different mass, say $m$ and $\tilde {m}$ have
different probabilities. In other words, the difference in mass
does impinge on a physically detectable function. Nevertheless,
now the dependence upon $m$ is not restricted to functions of
$m/\hbar$, i.e., the inclusion of a measuring process renders a
complication in this function. This means that the way in which
the mass parameter emerges in the COW experiment has not a general
character in the quantum domain.

Concerning (\ref{Nondemolition12}) there is a second point that
draws our attention. Indeed, notice that the probabilities
associated with the different measurement outputs of $a(t)$ do
depend upon the precision of the measuring device. This feature is
a characteristic of RPIF. This model is equivalent to other
formulations of the quantum measurement process \cite{Onofrio1},
and therefore this feature is a property of several approaches in
this context. In other words, those models which deny the
objectivity of this problem cannot predict the dependence upon the
precision of the experimental device embodied in
(\ref{Nondemolition12}), i.e., we have a theoretical prediction
that allows us to test the validity of RPIF, or of any of its
equivalent formulations. As mentioned before, this comprises our
second goal in this part of the work.

\section{Photon Interference}

\subsection{Deformed Dispersion Relations and Coherence Properties of Light}

Optical interferometry has played a fundamental role in some
experimental aspects of gravitational physics \cite{Schleich1}.
For instance, we may mention that there are gravity--waves
detectors which are built following Michelson interferometer
\cite{Thorne1}, or that Sagnac ring interferometer \cite{Scully1}
constitutes the bedrock for the ring laser gyroscopic device, the
one could be used to test the different metric theories of gravity
in the weak--field and slow motion limit \cite{Misner1}.

The use of light as a metric theory probe lies mainly on the
possibility of generating interference patterns, which have to be
related to the properties of the corresponding model. This simple
comment leads us to state, briefly, the conditions that light has
to satisfy in order to show interference. In this short trip we
will encounter the concepts of coherence, order of coherence, etc.
Almost at the beginning of the present work the relation between
interference and the corresponding motion equations was mentioned.
Since the motion of light is governed by a set of equations which
allow the superposition principle, then we may ask when does light
present interference? Everyday experience suggests that
interference must be subject to very special conditions. These
conditions are closely related to the concept of coherence
\cite{Mandel1, Mandel2}. Let us now address briefly the concept of
coherence with an example. Consider two waves whose electric
fields read $\vec{E}_1$ and $\vec{E}_2$. They are said to be
coherent if the following interference term does not vanish in the
region occupied by these electric fields, i.e.,

\begin{eqnarray} <\vec{E}_1\cdot\vec{E}_2> =
\sqrt{I_1I_2}\cos(\delta).\label{Coherence1}
\end{eqnarray}

In this last expression $<\vec{E}^2_i> = I_i$, $i = 1,2$, $<>$
denotes time average, and $\delta$ the phase difference between
the two waves. We may rephrase this stating that two waves are
coherent if the corresponding fields possess a constant phase.
This entails that coherence is associated to the comparison of the
relative phase of two light beams. Of course, this comparison can
be done resorting to the dependence of the waves in terms of
temporal or spatial parameters. The mathematical description of a
comparison between two quantities is done by the so--called
correlation operation \cite{Mandel1}. The temporal correlation of
an electromagnetic wave with itself can be determined by a
Michelson--Morley interferometer, whereas the spatial correlation
is detected by Young's two--slit experiment \cite{Scully1}.

Young's experiment has been mentioned as an example of
first--order coherence. Let us delve deeper into this concept.
Consider a monocromatic light (whose source is depicted in figure
4 by $S$). This beam one split with an opaque screen with two
pinholes. A detecting screen, $D$, is located behind the first
one, and a photo--detector, located at a certain point of the
second screen, measures the intensity, here denoted by $I$.

\setlength{\unitlength}{0.00083300in}%
\begingroup\makeatletter\ifx\SetFigFont\undefined
\def\x#1#2#3#4#5#6#7\relax{\def\x{#1#2#3#4#5#6}}%
\expandafter\x\fmtname xxxxxx\relax \def\y{splain}%
\ifx\x\y   
\gdef\SetFigFont#1#2#3{%
  \ifnum #1<17\tiny\else \ifnum #1<20\small\else
  \ifnum #1<24\normalsize\else \ifnum #1<29\large\else
  \ifnum #1<34\Large\else \ifnum #1<41\LARGE\else
     \huge\fi\fi\fi\fi\fi\fi
  \csname #3\endcsname}%
\else \gdef\SetFigFont#1#2#3{\begingroup
  \count@#1\relax \ifnum 25<\count@\count@25\fi
  \def\x{\endgroup\@setsize\SetFigFont{#2pt}}%
  \expandafter\x
    \csname \romannumeral\the\count@ pt\expandafter\endcsname
    \csname @\romannumeral\the\count@ pt\endcsname
  \csname #3\endcsname}%
\fi \fi\endgroup
\begin{picture}(7449,8460)(2389,-7888)
\thicklines \put(2401,-2161){\line( 1, 0){2000}}
\put(430,-2161){\makebox(6.6667,10.0000){\SetFigFont{10}{12}{rm}}}
\put(4550,-2161){\line( 1, 0){3300}} \put(8026,-2161){\line( 1,
0){1800}}
\put(4801,-4186){\line( 1, 0){2925}} \put(7726,-7186){\line( 0,
1){ 0}}
\put(6050,-4500){\makebox(0,0)[lb]{\smash{\SetFigFont{12}{14.4}{rm}D}}}
\put(6200,-500){\makebox(0,0)[lb]{\smash{\SetFigFont{12}{14.4}{rm}S}}}
\put(6200,-500){\vector(1, -1){1700}}
\put(6200,-500){\vector(-1,-1){1700}}
\put(6200,-500){\vector(-1,-1){1700}}
\put(5900,-5500){\makebox(0,0)[lb]{\smash{\SetFigFont{12}{14.4}{rm}{Figure
4}}}}
\end{picture}

The intensity can be cast in terms of the first--order correlation
function ($G^{(1)}$) in the following form \cite{Scully1}

\begin{eqnarray}
<I(\vec{r},t)> = \vert A\vert^2G^{(1)}(\vec{r}_1, \vec{r}_1;
t-t_1, t-t_1) + \vert B\vert^2G^{(1)}(\vec{r}_2, \vec{r}_2; t-t_2,
t-t_2) \nonumber\\
+ 2Re[ABG^{(1)}(\vec{r}_1, \vec{r}_2
;t-t_1,t-t_2)].\label{Explanation1}
\end{eqnarray}

The first two terms are associated to the average intensities at
the photo--detector stemming from each one of the pinholes, and
the last one is responsible for the interference. In this last
expression $A$ and $B$ are complex factors depending upon the
particular geometry of the pinholes, whose coordinate vectors are
$\vec{r}_1$ and $\vec{r}_2$, respectively, whereas $t_1$ and $t_2$
are the travelling times from these pinholes to the
photo--detector.

The essence of the Hanbury---Brown--Twiss effect involves the
measurement of intensity from two photo--detectors spaced a
distance $l$ apart. The main assumption is that the fluctuations
in the outputs of the photo--detectors must be correlated if the
amplitudes of the two waves are correlated. The difference between
an intensity interferometer and a Michelson--Morley device lies in
the fact that the former measures the square of the modulus of the
complex degree of coherence, whereas the latter detects also the
phase \cite{Mandel1}.

At this point we may wonder why should the Hanbury--Brown--Twiss
effect be considered in the realm of gravitational physics? The
answer comes from the present status of this area. Indeed, the
quantization of the gravitational field defines a long--standing
puzzle in modern physics. Unavoidably, some of the current
attempts in this direction are accompanied by the breakdown of
Lorentz symmetry \cite{Amelino2, Amelino3}. At this point it is
noteworthy to mention that up to now there is no experimental
evidence purporting a possible violation of this symmetry.
Obviously, we have to be more careful with our language because
the phrase {\it violation of Lorentz symmetry} could mean many
things, since this feature embodies several characteristics. For
instance, Local Lorentz Invariance could be violated, i.e., the
results of a local non--gravitational experiment would not be
independent of the velocity of the involved frame \cite{Will1}.
Additionally, Local Position Invariance \cite{Will1} could break
down. In other words, there would be a local non--gravitational
experiment in which the corresponding results do depend upon the
spacetime location of the frame. A violation of Local Lorentz
Invariance does not entail, inexorably, also a violation of Local
Position Invariance, and vice versa, of course.

In the present manuscript the meaning of the phrase Lorentz
violation will take a very precise expression; it will embody a
modification to the so--called dispersion relation.

\begin{equation}
E^2 = p^2c^2\Bigl[1 -
\alpha\Bigl(E\sqrt{G/(c^5\hbar)}\Bigr)^n\Bigr].\label{Viol1}
\end{equation}

In this last expression $\alpha$ is a coefficient, usually of
order 1, and whose precise value depends upon the considered
quantum--gravity model, and $n$, the lowest power in Planck's
length leading to a non--vanishing contribution, is also
model--dependent. This kind of symmetry breakdown is related to
non--critical string theory, non--commutative geometry, and
canonical gravity \cite{Amelino1}. The relevance of this kind of
violations can be easily understood remembering that they involve
a region in which one of the fundamental symmetries of modern
physics becomes only an approximation. In consequence those
experimental proposals that could test this violation acquire
relevance. As will be shown below, a consequence of (\ref{Viol1})
is connected to a modification of the speed of light, i.e., in
this scheme it is not constant \cite{Amelino1}

\begin{equation}
v=c\Bigl[1-\alpha\Bigl(E\sqrt{G/(c^5\hbar)}\Bigr)^n\Bigr]^{3/2}\Bigl[1+
{\alpha (n/2 -1)\Bigl(E\sqrt{G/(c^5\hbar)}\Bigr)^n}\Bigr]^{-1}
\label{Viol2}.
\end{equation}

This last remark opens up the possibility for searching this kind
of effects in the context of photon interferometry
\cite{Amelino1}.

\subsection{First--Order Coherence Experiments and Non--Newtonian Gravity}

In this part of the work we analyze the possibility of detecting a
Yukawa contribution to the gravitational field through the
interference process associated to the modifications upon the
first --order coherence properties in the process of light
emission of a particular quantum system \cite{Camacho5}. In order
to do this consider two identical atoms (located at $P$ and $P'$),
such that each one of them has two levels, and a single photon,
where only one of these two atoms can be excited. The initial
state of our system reads

\begin{equation}
a(\left\vert 0,1'> + \vert 1,0'>\right)\vert \tilde 0> + b\vert 0,
0'>\vert \phi>\label{Coherence1-1}.
\end{equation}

In this last expression some parameters require an explanation.
Here $\vert 0>, \vert1>, \vert 0'>, \vert 1'>$ denote the ground
and excited states of the two atoms, while $\vert \tilde 0>$ is
the vacuum of the electromagnetic field, and finally, $\vert\phi>$
designates the photon. The complex numbers $a$ and $b$ are
normalization constants. It is already known that after a time
larger than the mean decay time, $t_m$, the system decays to

\begin{equation}
\vert \alpha> = {1\over\sqrt{2}}\vert 0,0'>\left[\vert\gamma> +
\vert\gamma'>\right].
\end{equation}

In this last expression $\vert\gamma>$ and $\vert\gamma'>$ denote
the photon states emitted from sites $P$ and $P'$, respectively.

At this point we introduce a Yukawa--type term \cite{Fujii1}

\begin{equation}
V(r) = -{G_{\infty}mM\over r}\left[1 +
\alpha\exp{(-r/\lambda)}\right]\label{Coherence1-2}.
\end{equation}

Once again, we have introduced the following parameters,
$G_{\infty}$ describes the interaction between $M$ and $m$ in the
limit case $r\rightarrow\infty$, i.e., $G_N = G_{\infty}(1 +
\alpha)$, where $G_N$ is the Newtonian constant.

This implies that the gravitational potential generated by $M$
reads

\begin{equation}
U(r) = -{G_{\infty}M\over r}\left[1 +
\alpha\exp{(-r/\lambda)}\right]\label{Coherence1-3}.
\end{equation}

This part of the manuscript analyzes the topic of quantum
interference experiments in the context of tests of fundamental
physics, and at this point we connect with our main goal. Our
interference experiment will detect at point $S$ the light that
results from the decay of the system. There is a red--shift in the
frequency which appears as a consequence of the fact that the
electromagnetic field {\it climbs} in a region where a
non--vanishing gravitational field is present \cite{Misner1}. We
may now wonder what kind of interference experiment is being
considered, i.e., does it measure temporal, or spatial coherence
properties of light? The answer is spatial coherence, it is a
Young's experiment, and this can be seen resorting to the
two--slit analogy in which the present proposal can be
reformulated \cite{Camacho5}.

If the frequency at the emission point is $\nu$, and the radiation
is detected at a point, the one respect to the emission point has
a difference $\Delta U$ in the gravitational potential, then the
frequency at the detection point reads \cite{Misner1}

\begin{equation}
\tilde{\nu} = {\nu\over 1 + \Delta U/c^2}\label{Coherence1-4}.
\end{equation}

The electromagnetic field operator can be separated into two
parts, namely, with positive and negative frequency parts
\cite{Scully1}. Nevertheless, in the case of an experiment which
employs absorptive detectors the measurements are destructive,
and, in consequence, only that part of the field operator
containing annihilation operators, ${\bf E}^{(+)}({\bf r}, t)$,
has to be considered. In order to simplify the model we will
assume that the field is linearly polarized, and that the
radiation emitted from $P$ (or $P'$) is monocromatic. This
approach to our situation embodies, tacitly, the fact that we have
a quantum interference process, i.e., the electromagnetic field is
described by operators.

The possibility of performing this kind of experiments near the
Earth's surface will be at this point introduced, i.e., we have
the condition $R >>\vert z\vert$ ($R$ is the Earth's radius,
whereas $z$ is the height with respect to the Earths's surface).

\begin{equation}
r = R + z \label{Coherence1-5}.
\end{equation}

The field operator containing the annihilation operator reads,
approximately \cite{Scully2}

\begin{equation}
E^{(+)}({\bf r}, t) = \Xi\hat{a}\exp\left\{-i\nu\left[1 -
{g_0\over c^2}h {1 + \alpha e^{(-R/\lambda)}\over 1 +
\alpha}\right]\left[t - \hat{k}\cdot{\bf
r}\right]\right\}\label{Coherence1-6}.
\end{equation}

The parameter $h$ is related to the {\it climbed} distance,
$\hat{k}$ denotes the unitary vector in the direction of
propagation, $\Xi$ is a constant with dimensions of electric
field, $\hat{a}$ is the corresponding annihilation operator, and
$g_0 = g_{\infty}(1 + \alpha)$ is the effective acceleration of
gravity at laboratory distances.

The first--order correlation function is given by \cite{Scully1}
(remembering that $h$ and $h'$ are the {\it climbed} distances
coming from $P$ and $P'$)

\begin{equation}
G^{(1)}({\bf r}, {\bf r}; t, t) = \vert \Xi\vert^2\left\{1 +
\cos\left([{\bf k} - {\bf k'}]\cdot{\bf r} + \tilde{g}[h{\bf k} -
h'{\bf k'}]\cdot{\bf r} + \tilde{g}\nu t\Delta
h\right)\right\}\label{Coherence1-7}.
\end{equation}

In (\ref{Coherence1-7}) we have introduced the following
definition

\begin{equation}
\tilde{g} = {g_0\over c^2}{1 + \alpha e^{(-R/\lambda)}\over 1 +
\alpha}\label{Coherence1-8}.
\end{equation}

Clearly, the condition $g_0 = 0$ allows us to recover the usual
Young's interference pattern \cite{Mandel1}, i.e., the pattern
without gravitational field. Additionally, $\Delta h = h' - h$.
The time dependence of the interference pattern disappears if
$\Delta h = 0$, see (\ref{Coherence1-7}), and under this
restriction the first--order correlation function becomes

\begin{equation}
G^{(1)}({\bf r}, {\bf r}; t, t) = \vert \Xi\vert^2\left\{1 +
\cos\left(A\left[1 + \tilde{g}(h +
h')\right]\right)\right\}\label{Coherence1-9}.
\end{equation}

The behavior of the interference pattern in the absence of
gravitation is encoded in $A$, the one depends upon the geometry
of the interferometer, and also on the wavelength of the emitted
radiation \cite{Scully1}.

The detection of the effects of a fifth force inside the
so--called ``geophysical window" \cite{Gibbons1} imposes
restrictions upon the involved parameters. Here we may consider
the following values $\alpha \in [10^{-3}, 10^{-2}]$ and $\lambda
= 10$m \cite{Talmadge1}.

The expression for the correlation function, together with $g_0 =
G_0M/R^2$, allows us to obtain a condition on $R$ as function of
$h$ and $h'$. For instance, if $\tilde{g}[h + h']\sim 10^{-8}$,
then

\begin{equation}
(h + h')/R^2\sim 10^{-4}m\label{Coherence1-10}.
\end{equation}

From the correlation function \cite{Scully1} we may see that there
are certain time values, $t_n$ ($n\in N$), such that

\begin{equation}
\tilde{g}\Delta h\nu t_n = 2\pi n\label{Coherence1-11}.
\end{equation}

 Hence the interval between $t_{n+1}$ and $t_n$ is

\begin{equation}
\Delta t_n \equiv t_{n+1} - t_{n} = {2\pi\over \tilde{g}\nu\Delta
h}\label{Coherence1-12}.
\end{equation}

The purely Newtonian case implies

\begin{equation}
\Delta t_n^{(N)} = {2\pi c^2\over g_0\nu\Delta
h}\label{Coherence1-13}.
\end{equation}

The detection of a non--Newtonian contribution implies the
comparison between the usual case and the new one. This last
remark implies that the non--Newtonian times are provided by

\begin{equation}
\Delta t_n^{(NN)} = {2\pi c^2\over g_0\nu\Delta h}{1 + \alpha\over
1 + \alpha e^{-R/\lambda}}\label{Coherence1-14}.
\end{equation}

The ratio between these last two parameters entails

\begin{equation}
\Delta t_n^{(NN)}/\Delta t_n^{(N)} = {1 + \alpha\over 1 + \alpha
e^{-R/\lambda}}\label{Coherence1-15}.
\end{equation}

In an approximate way we have

\begin{equation}
\Delta t_n^{(NN)}/\Delta t_n^{(N)} = 1 +
10^{-3}\label{Coherence1-15}.
\end{equation}

The question about the measurability of this term  depends not
only on the order of magnitude of (\ref{Coherence1-15}), but also
on the resolution of the detecting device. Clearly, the
possibility of detecting a {\it fifth force} depends on the
condition that the difference between the non--Newtonian and
Newtonian cases has to be larger than the resolution of the
experimental apparatus, i.e., $\vert\Delta t_n^{(NN)}- \Delta
t_n^{(N)}\vert > \Delta T$, where $\Delta T$ denotes the time
resolution of the measuring device.

Now that we know the condition that the corresponding measuring
device has to satisfy, in order to provide information about the
existence of a Yukawa term, let us now address the conditions that
the use of the optical spectrum entails in this proposal. In other
words, consider, for the emitted field, $\lambda^{(r)}\sim 400$nm.
Finally, we must have an estimation of the order of magnitude of
$\Delta t_n^{(NN)}$. Nevertheless, this variable contains,
implicitly, a condition on $\lambda$ and $\alpha$, a set of data
that we do not know. To solve this impasse we will contemplate
this point from a different perspective, namely, from the
experimental point of view. This phrase means that we will ascribe
to $\Delta t_n^{(NN)}$ an order of magnitude not far from the
current resolutions. The possibility of detecting time intervals
similar to 50fs, based on the interference of two--photon
probability amplitudes \cite {Hong1}, means that time differences
$0.1\sim\mu$s could be within the technological margin. Hence
$\Delta t_n^{(NN)}\sim 0.01\mu$s leads us to a constraint upon
$\Delta h$, as function of $R$

\begin{equation}
\Delta h/R^2 \sim 10^{-8}m^{-1}\label{Coherence1-16}.
\end{equation}

If the experiment were performed near the Earth's surface ($R\sim
10^6$m), then $\Delta h \sim 10^4$m. The difference in climbing
distance has the order of magnitude of tens of kilometers, a
result that implies that this proposal is not very feasible. Let
us now recall the case of an experiment performed some years ago,
in which a test of relativistic gravitation was carried out
comparing the frequencies emitted from hydrogen masers located in
a spacecraft and also at an Earth station \cite{Vessot1, Vessot2}.
The connection with expression (\ref{Coherence1-16}) stems from
the fact that in the experiment mentioned in \cite{Vessot1,
Vessot2} one of the experimental devices was located at an
altitude of 10000km. If we resort to a value of $R=16000$km, then
(\ref{Coherence1-16}) renders a decrease in $\Delta h$ of one
order of magnitude, $\Delta h \sim 10^3$m, a decrease not large
enough to provide a feasible situation.
\bigskip
\bigskip

\subsection{First--Order Coherence Experiments and Deformed Dispersion Relations}

Several quantum--gravity theories are endowed with a modified
dispersion relation \cite{Amelino2, Amelino3}, which can be
characterized, from a phenomenological point of view, by
corrections hinging upon Planck's length, i.e., $l_p$

\begin{equation}
E^2 = p^2c^2\Bigl[1 -
\alpha\Bigl(E\sqrt{G/(c^5\hbar)}\Bigr)^n\Bigr]. \label{Disprel2}
\end{equation}

From these comments it becomes evident that these theories are not
finished. For instance, in loop quantum gravity $\alpha$ and $n$
depend upon the way in which the semiclassical states are
calculated \cite{Montemayor1}. This last argument stresses the
importance of any test that could lead to the detection of this
kind of parameters, or at least to set bounds upon them.

The relation between momentum and wave number, $p =\hbar k$, leads
us to conclude that

\begin{equation}
k =\frac{E/(c\hbar)}{\Bigl[1 -
\alpha\Bigl(E\sqrt{G/(c^5\hbar)}\Bigr)^n\Bigr]^{1/2}}. \label{k1}
\end{equation}

Since, experimentally, these modifications are quite small
(otherwise they would have already been detected) the following
expansion is justified

\begin{equation}
k =\frac{E}{c\hbar}\Bigl[1 +
\frac{\alpha}{2}\Bigl(E\sqrt{G/(c^5\hbar)}\Bigr)^n +
\frac{3}{8}\alpha^2\Bigl(E\sqrt{G/(c^5\hbar)}\Bigr)^{2n}+...\Bigr].
\label{k2}
\end{equation}

A deformed dispersion relation has important physical
implications, among them we may find a non--trivial dependence of
the speed of light upon the energy \cite{Amelino3}. In our case,
the speed of light has already been given, see (\ref{Viol2}).

In this part we resort to a Sagnac interferometer \cite{Mandel1},
which is a first--order coherence experiment, and has already been
used in the context of tests of the gravitomagnetic effect
\cite{Scully3}. The radius and angular velo\-city of this rotating
interferometer are $b$ and $\Omega$, respectively. Since the idea
is to analyze the possibility of carrying out this kind of
experiments near the Earth's surface, then we must consider the
presence of a gravitational potential

\begin{equation}
U(z) = gz.\label{Gravpot1}
\end{equation}

Let us now define our experimental proposal. At the highest point
of the interferometer a light beam, with frequency $\tilde{\nu}$,
enters the device, and it is split up into two parts, one rotating
in the same direction as the interferometer, i.e., clockwise,
whereas the second beam travels in the opposite direction. The
choice for the zero of this potential will be at the lowest point
of the interferometer. The plane where this apparatus is located
has its normal vector perpendicular to the $z$--axis. The presence
of a gravitational potential implies that each beam suffers a
shift in the frequency \cite{Misner1}

\begin{equation}
\nu= \frac{\tilde{\nu}}{1 + \Delta U/v^2}\label{shift1}.
\end{equation}

We have emphasized the fact that this kind of breakdown of Lorentz
symmetry entails that the speed of light is energy--dependent,
i.e., frequency dependent. This last point will be the core factor
in this idea. The speed of each one of the beams will be changing
as they move along the interferometer, and, therefore, an
interference pattern must emerge, the one shall contain
information about the parameters of the deformed dispersion
relation \cite{Camacho4}.

We may approximate the speed of the beams as follows

\begin{equation}
v^3 +
c\Bigl[\alpha\Bigl(\frac{n}{2}+1\Bigr)\Bigl(\hbar\tilde{\nu}/E_p\Bigr)^n
- 1\Bigr]v^2 + n\Delta Uv - cn\Delta U =0\label{speed3}.
\end{equation}

A fleeting glance at (\ref{speed3}) shows that if we set $\alpha
=0$, then $v=c$ is, indeed, a solution. In other words, we recover
the usual situation.

Harking back to our more general case we have that a solution is

\begin{equation}
v =c\Bigl[1 - n\Delta U/c^2-
\frac{\alpha}{3}\Bigl(\frac{n}{2}+1\Bigr)\Bigl(\hbar\tilde{\nu}/E_p\Bigr)^n\Bigr]\label{speed4}.
\end{equation}

Define now the angle $\theta$, the one measures the rotation of
the beams. Since we divided the original beam, there will be two
parameters, $\theta_1$ and $\theta_2$. Here $\theta_i=0$, with $i
= 1, 2$,  coincides with the point at which the beam enters into
the interferometer. If $\beta$ is the angle described by the
interferometer, then any point at the edge of the interferometer
has the following $z$--coordinate

\begin{equation}
z= b\Bigl[1 + \cos(\beta)\Bigr]\label{position1}.
\end{equation}

The beam enters the interferometer at the point

\begin{equation}
z_0= 2b\label{position2},
\end{equation}
then

\begin{equation}
\Delta U = gb\Bigl[\cos(\beta) -1\Bigr].\label{Gravpot2}
\end{equation}

Additionally, we have

\begin{equation}
v = b\frac{d\theta}{dt}.\label{speed5}
\end{equation}

The travelled time, required by the beams, as a function of the
angle is deduced joining expressions (\ref{speed4}) and
(\ref{speed5})

\begin{equation}
t - t_0= \frac{b}{c}\int_0^{\theta}\Bigl[1 -
\frac{\alpha}{3}\Bigl(\frac{n}{2}+1\Bigr)\Bigl(\hbar\tilde{\nu}/E_p\Bigr)^n
- \frac{2gb}{c^2}[\cos(\tilde\theta) -1]
\Bigr]^{-1}d\tilde\theta.\label{time1}
\end{equation}

Therefore,

\begin{equation}
\frac{ct}{2b}\Bigl\{\Bigl[1 -
\frac{\alpha}{3}\Bigl(\frac{n}{2}+1\Bigr)\Bigl(\hbar\tilde{\nu}/E_p\Bigr)^n
+ \frac{2gb}{c^2}\Bigr]^2
-\Bigl(\frac{2gb}{c^2}\Bigr)^2\Bigr\}^{1/2} =
\theta/2.\label{time3}
\end{equation}

The beam moving in the opposite direction, with respect to the
displacement of the interferometer, meets the detection device
after it has moved an angle equal to

\begin{equation}
\beta_1 = t_1\Omega .\label{int1}
\end{equation}

The angle described by the beam reads then

\begin{equation}
\theta_1 = 2\pi -\beta_1.\label{beam1}
\end{equation}

As of the remaining beam, it meets the interferometer after the
measuring device has rotated an angle

\begin{equation}
\beta_2 = t_2\Omega .\label{int2}
\end{equation}

The angle described by the second beam is given by

\begin{equation}
\theta_2 = 2\pi + \beta_2.\label{beam2}
\end{equation}

Introducing these two times, $t_1$ and $t_2$, into (\ref{time3})
renders

\begin{equation}
t_1 = \frac{2\pi b}{c}\Bigl[\gamma +
\frac{b\Omega}{c}\Bigr]^{-1},\label{flight1}
\end{equation}

\begin{equation}
t_2 = \frac{2\pi b}{c}\Bigl[\gamma -
\frac{b\Omega}{c}\Bigr]^{-1},\label{flight2}
\end{equation}

\begin{equation}
\gamma = \Bigl\{\Bigl[1 -
\frac{\alpha}{3}\Bigl(\frac{n}{2}+1\Bigr)\Bigl(\hbar\tilde{\nu}/E_p\Bigr)^n
+ \frac{2gb}{c^2}\Bigr]^2
-\Bigl(\frac{2gb}{c^2}\Bigr)^2\Bigr\}^{1/2}.\label{gamma}
\end{equation}

These two last expressions entail ($\Delta t = t_2 - t_1$)

\begin{equation}
\Delta t = \frac{4\pi b^2\Omega}{c^2}\Bigl\{\Bigl[1 -
\frac{\alpha}{3}\Bigl(\frac{n}{2}+1\Bigr)\Bigl(\hbar\tilde{\nu}/E_p\Bigr)^n
+ \frac{2gb}{c^2}\Bigr]^2 -\Bigl(\frac{2gb}{c^2}\Bigr)^2
-\frac{b^2\Omega^2}{c^2}\Bigr\}^{-1}.\label{timediff1}
\end{equation}

Setting $\alpha =0$ we recover the usual value,
$\Delta\theta^{(0)}$, for the phase shift \cite{Scully1, Mandel1}.
Notice that the presence of a violation to Lorentz symmetry in the
form of a deformed dispersion relation does impinge upon the
difference in time of flight, and in consequence on the
interference pattern. Since the feasibility of the proposal shall
also be addressed, we will consider the case $n =1$. This means
that the time difference becomes

\begin{equation}
\Delta t = \frac{4\pi b^2\Omega}{c^2}\Bigl\{1 +
\alpha\Bigl(\hbar\tilde{\nu}/E_p\Bigr) - 4\frac{gb}{c^2}
 -\frac{\alpha^2}{4}\Bigl(\hbar\tilde{\nu}/E_p\Bigr)^2 + 2\alpha
\Bigl(\frac{gb}{c^2}\Bigr)\Bigl(\hbar\tilde{\nu}/E_p\Bigr) +
\Bigl(\frac{b\Omega}{c}\Bigr)^2\Bigr\}.\label{timediff2}
\end{equation}

In the analysis of the interference pattern we require the optical
path and phase differences associated to (\ref{timediff2}). In
this case they read, respectively

\begin{equation}
\Delta L = c\Delta t,\label{pathdiff1}
\end{equation}

\begin{equation}
\Delta\theta  = \frac{\Delta
L}{{\tilde\lambda}}.\label{phasediff1}
\end{equation}

In consequence

\begin{equation}
\Delta\theta  = \frac{4\pi b^2\Omega}{c\tilde{\lambda}}\Bigl\{1 +
\alpha\Bigl(\hbar\tilde{\nu}/E_p\Bigr)\Bigl(1 -
\frac{\alpha}{4}\Bigl(\hbar\tilde{\nu}/E_p\Bigr)\Bigr) -
4\frac{gb}{c^2}\Bigl(1 -
\frac{\alpha}{2}\Bigl(\hbar\tilde{\nu}/E_p\Bigr)\Bigr) +
\Bigl(\frac{b\Omega}{c}\Bigr)^2\Bigr\}.\label{phasediff2}
\end{equation}

The feasibility of the model hinges upon the difference
$\Delta\theta - \Delta\theta^{(0)}$. Clearly, if $\delta\theta$
denotes the corresponding experimental resolution, then the
requirement, to be fulfilled, reads

\begin{equation}
\delta\theta  < \vert\Delta\theta - \Delta\theta^{(0)}\vert
.\label{phasediff3}
\end{equation}

This may be cast in the following form

\begin{equation}
\delta\theta < \frac{4\pi
b^2\Omega}{c\tilde{\lambda}}\alpha\Bigl(\hbar\tilde{\nu}/E_p\Bigr)\Bigl[1
+ 2\frac{gb}{c^2}\Bigr].\label{phasediff4}
\end{equation}

For any terrestrial proposal of this experiment we have that $g$
is approximately $9.8m/s^2$. We need two experimental parameters
in order to analyze the feasibility of the proposal. One of them
is connected with the experimental resolution of the device, which
will be taken as $\delta\theta\sim 10^{-6}$ \cite{Mandel1}. The
remaining one is associated to the value of $\hbar\tilde{\nu}$.
For this parameter we will choose an optical transition, i.e.,
$\hbar\tilde{\nu}\sim 10^{-3}$ J. At our disposal we have two
variables, namely, the radius of the interferometer, $b$, and its
angular velocity, $\Omega$. Consider the case in which
$b=10^{-1}$m, a value within the current technology. Then

\begin{equation}
\Omega > 10^{25}/s.\label{exp1}
\end{equation}

Obviously, this condition lies outside the current technology.
Demanding a more realistic value for the angular velocity entails
values for the radius that are impossible to achieve in an
experiment. Finally, let us discuss another possibility. For
instance, consider a smoother experimental condition

\begin{equation}
b^2\Omega \sim 10m^2/s.\label{exp2}
\end{equation}

Then we obtain a condition on the energy of the beam being used in
our experiment which does not lie anymore in the realm of optical
transitions

\begin{equation}
\hbar\tilde{\nu} >  10^{9}J.\label{exp3}
\end{equation}

These last two examples show us that Sagnac's device has an
additional parameter, which is absent in the case of a
Michelson--Morley experiment. In Sagnac's idea we have three
experimental parameters at our disposal, $b$, $\Omega$, and
$\hbar\tilde{\nu}$, whereas in a Michelson--Morley experiment, we
have only the frequency (or energy) of the beam and the difference
in optical length. In this sense we may say that Sagnac is richer,
though as this simple model shows the existence of this additional
variable does not imply that the experiment falls within the
current technological possibilities.

\subsection{Second--Order Coherence Experiments and Non--Newtonian Gravity}
\bigskip
\bigskip

Up to now our experimental proposals in the optic realm can be
embodied in the context of first--order coherence \cite{Mandel1,
Mandel2, Scully1}. It has been shown, in connection with Sagnac
interferometry, which is a first--order coherence experiment, that
the detection of a deformed dispersion relation seems to require
either very large angular velocities and radii, or very large
energies for the light beam. This represents a technical problem,
and we may wonder if there is another manner in which this kind of
effects could be detected, without leaving the domain of optics.
The answer stems from the possibility of resorting to
higher--order coherence experiments, for instance, the so--called
Hanbury--Brown --Twiss effect (HBTE), the one falls within the
group of second--order coherence experiments \cite{Hanbury1}.

As mentioned before, HBTE is a technique which involves intensity
correlation between signals collected at two points in space.
Notice that in this last phrase we, tacitly, introduce a new
experimental parameter which is absent in the first--order
coherence models. Indeed, HBTE needs an additional distance
variable related to the fact that in this idea, in contrast to the
situation of any first--order coherence device, two collecting
points are required. The intention in this part of the work is to
use this extra parameter and see if it could lead us to an
experiment closer to the current technology.

The experiment is defined as follow \cite{Camacho5}. We have two
atoms, located at points $P$ and $P'$, but now there are two
detection points, $S_1$ and $S_2$. Initially the atoms are
excited, but there is no electromagnetic field, hence the initial
state vector reads

\begin{equation}
\vert\alpha(t = 0)> = \vert
1,1'>\vert\tilde{0}>\label{Coherence2-1}.
\end{equation}

The system decays, after an interval sufficiently larger, $t_m$.

\begin{equation}
\vert\alpha(t >>t_m)> = \vert 0,0'>\vert\gamma,
\gamma'>\label{Coherence2-2}.
\end{equation}

Here $\vert 0,0'>, \vert1,1'>$ denote the ground and excited
states of the two atoms, while $\vert \tilde 0>$ is the vacuum of
the electromagnetic field, and, finally, $\vert\gamma>$ and
$\vert\gamma'>$ are the photon states emitted from sites $P$ and
$P'$, respectively.

Assuming a plane wave approximation for the emitted radiation, the
definition of second--order correlation function \cite{Hanbury1,
Hanbury2} tells us that the interference pattern is provided by
the following term

\begin{equation}
\cos\left\{\left[{\bf k} - {\bf k'}\right]\cdot\left[{\bf r_2} -
{\bf r_1}\right] + \tilde{g}[h'_2{\bf k'} - h_2{\bf k}]\cdot{\bf
r_2} - \tilde{g}[h'_1{\bf k'} - h_1{\bf k}]\cdot{\bf r_1} +
\tilde{g}\nu t\left[\Delta h - \Delta h'\right]\right
\}\label{Coherence2-3}.
\end{equation}

In this last expression the following parameters have been
introduced; $h_1$ and $h_2$ are the {\it climbed} distances for
the radiation emitted by $P$ and detected at $S_1$ and $S_2$,
respectively (we have the same argument if the light is emitted in
$P'$). Additionally, $\Delta h= h_2 - h_1$, $\Delta h'= h'_2 -
h'_1$.

With these results we may now compare HBTE against the
first--order coherence situation. Indeed, a fleeting glance at
(\ref{Coherence1-7}) and (\ref{Coherence2-3}) allows us to
understand that the order of magnitude of some of the experimental
parameters, for instance, $R$ or $t$, will be the same as in the
analysis of a Young's experiment. In this sense the extra distance
parameter related to HBTE offers no improvement, i.e., the
technical problems are not significantly smoothed.

The time independent terms appearing in (\ref{Coherence2-3}) share
the structure of the corresponding terms of (\ref{Coherence1-7}).
This last statement means that if in the HBTE case we impose the
following condition $\tilde{g}\vert[h'_2{\bf k'} - h_2{\bf
k}]\cdot{\bf r_2}\vert\sim 10^{-8}$, then

\begin{equation}
\vert\Delta h'\vert/R^2\sim 10^{-4}m^{-1}\label{Coherence2-4}.
\end{equation}

In (\ref{Coherence2-4}) appears $\vert\Delta h'\vert$, and not $h
+ h'$, as in the case of first--order coherence case, see
expression (\ref{Coherence1-9}).

\begin{equation}
\Delta t_n = {2\pi\over \tilde{g}\nu\vert\Delta h - \Delta
h'\vert}.
\end{equation}

In other words, if we go from Young's experiment to HBTE, then
$\Delta h$ is replaced by $\vert\Delta h - \Delta h'\vert$. The
additional distance factor that HBTE involves emerges in
$\vert\Delta h -\Delta h'\vert$, which is a parameter that
requires four distance parameters, in lieu of two, that appear in
$h+h'$.

\subsection{Second--Order Coherence and Deformed Dispersion Relations}

Now we consider a deformed dispersion relation and analyze its
effects in a second order coherence device. The conclusions
elicited from the proposal related to the detection of
non--Newtonian gravity with a second--order coherence device do
imply that there is  no significative improvement in the
feasibility of the corresponding experiment, compared with the
case of a first--order coherence device. Therefore, we may wonder
if this scheme could provide an improvement for the case of a
deformed dispersion relation. Fortunately, here, as will be shown
below, the situation is more optimistic, then the extra parameter
associated to HBTE pays off \cite{Camacho1}. As before we take a
deformed dispersion relation, see (\ref{Viol1}).

Since we expect very tiny corrections expression (\ref{k2}) may be
used as a good approximation to the wave number.

The idea now is to consider two photons propagating along the axis
defined by the unit vector $\hat{e}$, though with different
energies.

\begin{equation}
\vec{k} = k\hat{e}, \label{k3}
\end{equation}

\begin{equation}
\vec{k}' = k'\hat{e}. \label{k4}
\end{equation}

Let us now consider the detection of these photons resorting to
HBTE \cite{Hanbury1, Hanbury2}. In other words, we have two
photo--detectors located at points $A_1$ and $A_2$, with position
vectors, $\vec{r}_1$ and $\vec{r}_2$, respectively.

The second--order correlation function reads \cite{Scully1}

\begin{equation}
G^{(2)}(\vec{r}_1, \vec{r}_2; t,t) = \mathcal{E}\Bigl(1 +
\cos\Bigl[(\vec{k}-\vec{k}')\cdot(\vec{r}_2-\vec{r}_1)\Bigr]\Bigr).
\label{scf1}
\end{equation}

$\mathcal{E}$ denotes a constant factor with dimensions of
electric field. If $\Delta\theta^{(n)}$ is the phase difference
for $n$, then the interference pattern (to second--order in
$\Delta E$) is

\begin{eqnarray}
\Delta\theta^{(n)} = \frac{l\Delta E}{c\hbar}\Bigl[\Bigl(1 +
\frac{n+1}{2}\alpha
[E\sqrt{G/(c^5\hbar)}]^n  \nonumber\\
+\frac{3}{8}\alpha^2(2n+ 1)[E\sqrt{G/(c^5\hbar)}]^{2n}\Bigr)\nonumber\\
+ \frac{\Delta E}{E}\Bigl(\frac{n(n+1)}{4}
\alpha[E\sqrt{G/(c^5\hbar)}]^n \nonumber\\
+ \frac{3n(2n+1)}{8}\alpha^2[E\sqrt{G/(c^5\hbar)}]^{2n}
\Bigr)\Bigr]. \label{phase1}
\end{eqnarray}

Here $E = E' + \Delta E$, and in addition, $l=
\hat{e}\cdot(\vec{r}_2-\vec{r}_1)$. The feasibility of the
detection of this kind of corrections depends upon the value of
$n$, at least in the context of a first--order correlation
function. In consequence we will divide our situation in the same
manner, namely, to first order in $\Delta E$ we have
(approximately) that

\begin{eqnarray}
\Delta\theta^{(1)} = \frac{l\Delta E}{c\hbar}[1 +
\alpha[E\sqrt{G/(c^5\hbar)}]\nonumber\\
\times\Bigl(1 + \frac{9}{8}\alpha[E\sqrt{G/(c^5\hbar)}]\Bigr)],
\label{phase2}
\end{eqnarray}

\begin{eqnarray}
\Delta\theta^{(2)} = \frac{l\Delta E}{c\hbar}[1 +
\frac{3}{2}\alpha[E\sqrt{G/(c^5\hbar)}]^2\nonumber\\
\times\Bigl(1 + \frac{5}{4}\alpha[E\sqrt{G/(c^5\hbar)}]^2\Bigr)].
\label{phase3}
\end{eqnarray}

The corrections could be detectable if $\vert\Delta\theta^{(n)}-
\Delta\theta^{(LS)}\vert > \Delta\theta^{(exp)}$. In this last
expression $\Delta\theta^{(LS)}$ and $\Delta\theta^{(exp)}$ denote
the phase difference in the case in which $\alpha =0$, and the
experimental resolution, respectively. Additionally, $\Delta E =
\frac{E}{\gamma}$, with $\gamma >1$.

The last expressions show that the phase difference is a function
of the value of $n$. This in addition entails that if we impose a
condition upon these phase difference, namely, $\Delta\theta^{(n)}
= \Delta\theta^{(exp)}$, then $l$ becomes a function of $n$, see
expression (\ref{phase1}). This dependence will be denoted by
$l^{(n)}$, and now we proceed to find this dependence for some
particular cases. At this point we must mention that this
parameter is a direct consequence of the use of HBTE. Indeed, it
is a function of the distance between the two photo--detectors. To
first order in $E\sqrt{G/(c^5\hbar)}$

\begin{eqnarray}
l^{(1)} \geq\frac{c\hbar\gamma}{\alpha
E^2}\sqrt{c^5\hbar/G}\Delta\theta^{(exp)}, \label{length1}
\end{eqnarray}

\begin{eqnarray}
l^{(2)} \geq\frac{2}{3}\frac{c^6\hbar^2\gamma}{\alpha
GE^3}\Delta\theta^{(exp)}. \label{length2}
\end{eqnarray}

 If we assume the following values for our parameters, $\Delta\theta^{(exp)}\sim 10^{-4}$
 \cite{Guenther1},
$\alpha\sim 1$, $\gamma\sim 10^{2}$, and $E\sim 10^{-6}$J
\cite{Amelino1, Amelino2}, then

\begin{eqnarray}
l^{(1)}\geq 10^{-5} m, \label{length3}
\end{eqnarray}

\begin{eqnarray}
l^{(2)}\geq 10^{11} m. \label{length4}
\end{eqnarray}

The energy that has been considered has the order of magnitude of
the highest energy that nowadays can be produced in a laboratory.
These two last expressions mean that if the corrections to the
dispersion relation entail $n=1$, then a HBTE type--like
experiment with a distance between the photo--detectors greater
than $10^{-5}$m could detect the extra term. In the remaining
case, $n=2$, the required distance implies the impossibility of
detecting the correction, with the assumption introduced for the
involved energy.

In the context of first--order correlation functions
\cite{Camacho4} the possibility of detecting the case $n = 2$ is,
currently, completely an impossible task. Nevertheless, our
approach introduces an additional parameter, and therefore, if we
consider the case of an energy of $E\sim 10$J (which is tantamount
to the energy that could be involved in the observation of
gamma--ray bursts), then

\begin{eqnarray}
l^{(2)}\geq 10^{5}m. \label{length5}
\end{eqnarray}

In the present model the presence of our extra parameter ($l$)
allows us to get closer to its possible detection. Let us now
analyze the feasibility in the context of $n=2$, which is a
tougher situation, experimentally, to handle than $n=1$. The
experimental parameter that should be measured is the normalized
correlation coefficient of the fluctuations in the photoelectric
current outputs \cite{Mandel1}, $C(l)$. The connection with
difference in time of arrival stems from HBTE, namely, the squared
modulus of the degree of coherence function, $\gamma$, is
proportional to the normalized correlation function of the
photocurrent fluctuations

\begin{eqnarray}
C(l) = \delta\vert\gamma(l)\vert^2. \label{HBT1}
\end{eqnarray}

The parameter $\delta$ is the average number of photoelectric
counts due to light of one polarization registered by the detector
in the corresponding correlation time. Experimentally, for thermal
sources of temperature below $10^{5}$ K, $\delta$ is always
smaller than $1$ \cite{Mandel1}. In order to enhance the effect,
i.e., to have a larger value of $\delta$, we may consider the fact
that the number of average photons, as a function of the involved
frequency, $\nu$, reads \cite{Mandel1}

\begin{eqnarray}
\delta = 2\xi(3)\frac{\nu^3}{c^2\pi^2}. \label{MPN1}
\end{eqnarray}

Here $\xi$ is the so--called Riemann zeta function. Clearly,
higher energy implies larger mean number of photons. Hence, for an
energy of $E\sim 10$J we expect a value of $\delta$ not as small
as in the case of $10^{5}$ K. In other words, the higher the
energy of the light beam, the larger the constant between $C(l)$
and $\gamma$ becomes. Of course, this last fact cannot be
considered a shortcoming of the proposal.

In terms of the photocurrents fluctuations at the two
photo--detectors

\begin{eqnarray}
C(l) = \frac{<\Delta I_1(t)\Delta I_2(t)>}{(<[\Delta
I_1(t)]^2>)^{1/2}(<[\Delta I_2(t)]^2>)^{1/2}}. \label{HBT2}
\end{eqnarray}

The feasibility of detecting a deformed dispersion relation in
this context depends upon the aforementioned fluctuations. We may
find already in the extant literature some models that explain the
pulse width of a Gamma Ray Burst (GRB) in terms of the involved
energy \cite{Ping1}, as a power law expression. At least in the
case in which the sources are observed as fireballs. In other
words, we may find non--vanishing sources for $C(l)$. In this case
we expect to have a better experimental situation. In consequence,
we may assert that this proposal is a feasible one.

\subsection{Degree of Coherence Function and Deformed Dispersion Relations}

The Hanbury--Brown--Twiss effect was a watershed in optics since
when it was announced a possible contradiction with quantum theory
was put forward. The situation was understood with the work of
Glauber \cite{Glauber1}, who gave a firm theoretical background to
the quantum theory of coherence. One of the fundamental concepts
in this context is provided by the so--called degree of coherence
function \cite{Mandel1, Scully1}. This function is endowed with a
deep physical meaning. We may state that the interference
properties do depend strongly upon the degree of coherence
function. For instance, in the case of a first--order coherence
function it can be proved \cite{Scully1} that the visibility of
the fringes is a function of the first--order degree of coherence
function. In this part of the work the functional dependence of
the degree of coherence function, when Lorentz symmetry is broken
down in the form of a deformed dispersion relation, will be
analyzed (\ref{Viol1}). This function has a quantum origin, and
therefore the idea here proposed involves a quantum interference
experiment.

In order to deduce the dependence of the degree of coherence
function let us consider two beams with energies $E_1$ and $E_2$,
respectively, such that $E_2 = E_1 + \Delta E$. The experimental
device is a Michelson--Morley apparatus. Each frequency produces
an interference pattern, and at this point it will be supposed
that the corresponding beat frequency is to high too be detected
\cite{Scully1}, i.e., the output intensity is obtained adding the
intensities associated with each frequency contained in the input.
Under these conditions the measured intensity reads
\cite{Camacho2}

\begin{equation}
I = I_1\Bigl[1 + \cos(\omega_1\tau_1)\Bigr] + I_2\Bigl[1 +
\cos(\omega_2\tau_2)\Bigr]. \label{Intensity1}
\end{equation}

In this last expression $I_1$ and $I_2$ denote the intensities of
the two beams, $\omega_1$, $\omega_2$ the corresponding
frequencies, and

\begin{equation}
\tau_1 =2d/c_1, \label{Time1}
\end{equation}

\begin{equation}
\tau_2 =2d/c_2. \label{Time2}
\end{equation}

The difference in length in the two interferometer arms is denoted
by $d$, and $c_1$, $c_2$, the corresponding velocities. As
mentioned before the velocity has a non--trivial energy dependence
\cite{Amelino3}, i.e., $c_1 \not=c_2$.

From now on we will assume that $I_1 = I_2$ (this is no
restriction at all), such that $I_0=I_1+ I_2$. The detected
intensity can be cast in the following form

\begin{equation}
I = I_0\Bigl[1 + \gamma(d)\Bigr]. \label{Intensity2}
\end{equation}

In this last equation the so--called degree of coherence function
has been introduced, the one for our situation reads ($k_1$ and
$k_2$ are the corresponding wave numbers)

\begin{equation}
\gamma(d)= \cos\Bigl([k_1 + k_2]d/2\Bigr)\cos\Bigl([k_1 -
k_2]d/2\Bigr). \label{Degree1}
\end{equation}

A fleeting glance at (\ref{k1}) clearly shows that (\ref{Degree1})
does depend upon $\alpha$ and $n$, and, in consequence, the roots
of the degree of coherence function will be modified by the
presence of a deformed dispersion relation.

The expression providing us the roots  of the degree of coherence
function is

\begin{equation}
\Bigl(k_1 - k_2\Bigr)d/2 = \pi/2. \label{Root1}
\end{equation}

We may cast (\ref{Root1}) as

\begin{eqnarray}
d = c\hbar\pi\Bigl\{\Delta E +
\frac{\alpha}{2}E_1\Bigl(E_1\sqrt{G/(c^5\hbar)}\Bigr)^n\nonumber\\
\times\Bigl[(n+1)\frac{\Delta E}{E_1}
+\frac{n(n+1)}{2}\Bigl(\frac{\Delta E}{E_1}\Bigr)^2 +...
\Bigr]\Bigr\}^{-1}. \label{Root12}
\end{eqnarray}

Introduce now the following definition $\beta = \Delta E/E_1$, a
real number smaller than 1. In the present proposal we will
consider two possible values for $n$, namely, $n= 1, 2$.

For the case $n=1$  we have that the roots of the degree of
coherence function become, approximately

\begin{eqnarray}
d = \frac{c\hbar\pi}{E_1}\Bigl\{\beta
-\frac{\alpha}{2}\Bigl(E_1\sqrt{G/(c^5\hbar)}\Bigr) \Bigl[2 +
\beta\Bigr]\Bigr\}. \label{Root3}
\end{eqnarray}

Assume that $\alpha\sim 1$. The possibility of detecting this
deformed dispersion relation will hinge upon the fulfillment of
the condition

\begin{eqnarray}
\vert D - d\vert > \Delta d. \label{Exp1}
\end{eqnarray}

In this last equation $D$ denotes the usual value in the
difference of the interferometer arms at which the degree of
coherence function vanishes (that is when $\alpha =0$), whereas
$\Delta d$ is the corresponding experimental resolution. Then

\begin{eqnarray}
\frac{\Delta E}{E_1}> \frac{2\Delta d}{\pi l_p} - 1. \label{Exp2}
\end{eqnarray}

Since it was assumed that our device cannot detect the beat
frequencies, i.e., if $T$ denotes the time resolution of the
measuring device, then

\begin{eqnarray}
\vert\omega_2 - \omega_1\vert T/2>>1. \label{Exp3}
\end{eqnarray}

This last condition may be rewritten as

\begin{eqnarray}
T\Delta E>\hbar. \label{Exp4}
\end{eqnarray}

In other words, (\ref{Exp2}) and (\ref{Exp4}) are the two
conditions to be fulfilled, if the case $n=1$ and $\alpha \sim 1$
is to be detected.

Similarly, for $n = 2$ and $\alpha \sim 1$ the roots of the degree
of coherence function read, approximately

\begin{eqnarray}
d = \frac{c\hbar\pi}{E_1}\Bigl\{\beta
-\frac{\alpha}{2}\Bigl(E_1\sqrt{G/(c^5\hbar)}\Bigr)^2\Bigl[3 +
3\beta+ \beta^2\Bigr]\Bigr\}. \label{Root4}
\end{eqnarray}

The expression tantamount to (\ref{Exp2}) is

\begin{eqnarray}
\Bigl\{3 + 3\Bigl(\frac{\Delta E}{E_1}\Bigr) + \Bigl(\frac{\Delta
E}{E_1}\Bigr)^2\Bigr\}E_1> 2\frac{c\hbar\Delta d}{\pi l^2_p}.
\label{Exp5}
\end{eqnarray}

The impossibility of detecting beat frequencies translates, once
again, as

\begin{eqnarray}
T\Delta E>\hbar. \label{Exp6}
\end{eqnarray}

The experimental feasibility of this idea is related to
(\ref{Exp2}), the one implies that (together with $\Delta E/E_1
<1$) the resolution of the measuring device has to be close to
Planck's length, i.e., $\Delta d\sim l_p$, a condition far away
from the current technology.

\section{Conclusions}

\subsection{Neutron Interference: Limitations and Possibilities}

\subsubsection{Metric Theories and Semiclassical Approximation}

The use of neutron interference to test the structure of
space--time has been a fundamental tool, and provided a deeper
insight into the connection between gravity and quantum theory.
From the very beginning it has been mentioned that there are
several experimental devices explicitly designed for the detection
of the phenomenon of interference \cite{Mandel1}.

At this point we may wonder what kind of coherence, either
spatial, or temporal, is involved in a COW experiment? Notice that
the experimental proposal divides a neutron beam several times
\cite{Colella1}, and, afterwards, the interference pattern is
measured. Clearly, this situation falls within the case of
temporal coherence.

We have been considering the concept of coherence from diverse
perspectives, i.e., temporal coherence, spatial coherence, etc.
Let us at this point delve a little bit deeper in these ideas.

We may state that optical coherence addresses the statistical
description of fluctuations and that the phenomena of optical
coherence may be regarded as manifestations of the correlations
between them \cite{Mandel1, Mandel2}. We may also ask what is the
relation between coherence and interference. The answer to this
question leads us to accept that interference is intimately
associated to optical coherence, and that it may be considered the
simplest example revealing correlation between light beams. These
last remarks could be enlightening but do not provide a physical
insight to the origin of coherence.

Searching for a more profound understanding of the concept of
coherence let us consider a point--like source of light, and for
the sake of clarity, we assume that the emitted beam is
quasimonocromatic. The first question that we pose concerns the
conditions of our source that give rise to the concept of temporal
coherence. At any point subjected to the influence of our light
beam a very swiftly oscillating field exists. For a very small
time interval, about $10$ns,  this field behaves as a sinusoid,
and then its phase shows a {\it random jump}, and once again,
behaves as a sinusoid, and so on. Two aspects must be mentioned
about this last remark. Firstly, this {\it random jump} emerges as
a consequence of the emitting process of the source, i.e., it is a
direct consequence of the laws that govern emitting processes.
Secondly, the time interval between two successive {\it random
jumps} of the phase is known as temporal coherence. If we denote
by $\Delta \nu$ the bandwidth in frequency of our
quasimonocromatic source, then the time during which the relation
$\Delta t= 1/\Delta \nu$ is satisfied is called temporal
coherence, $\Delta t_c$. The concept of coherence length is
defined in terms of coherence time, namely, $\Delta l_c = c\Delta
t_c$. We may endow this last parameter with a physical meaning, it
is the distance in which the wave has a sinusoidal behavior.

We may say that temporal coherence is a manifestation of spectral
purity, i.e., if the source were ideally monocromatic, then the
beam would be a perfect sinusoid, but, of course, this is an
idealization, and the concept of temporal coherence is a parameter
which determines how far the situation is from the ideal case.

What about the concept of spatial coherence? In order to address
this issue we must recognize that in the previous example we,
tacitly, accepted an additional assumption, the emitting source is
point--like. Spatial coherence is related to this supposition.
Indeed, let us now consider a extended object, the one emits
electromagnetic radiation. Consider now two points on this object,
say $A$ and $B$. Concerning this particular situation we may
wonder if the phase of the light emitted by $A$ is correlated with
the phase of the light emitted by $B$. The concept of spatial
coherence measures the correlation shown by two different points
on the emitting source.

Now that we have a deeper comprehension of the concepts of
temporal and spatial coherence let us proceed to explain their
relation with the different experimental proposals that here have
been considered. Concerning the wave front division
interferometers we may state that the prototype of them is Young's
proposal. Clearly, a single beam is diffracted by two holes which
can be considered as two sources emitting the same frequency
\cite{Chartier}. This last comment leads us to conclude that this
device measures spatial coherence, i.e., it compares the light
emitted from {\it two different points of an extended body}. A
Michelson--Morely apparatus belongs to the family of amplitude
splitting interferometers. Just for the sake of completeness let
us mention that Mach--Zehnder and Sagnac devices are also elements
of this family. In this kind of interferometers if the difference
in optical path is larger than the coherence length, then the
interference pattern will not be detectable \cite{Chartier}. In
other words, amplitude splitting devices allow us to measure
coherence length, and in consequence, temporal coherence.

Harking back to the COW experiment, we have stressed that it
requires the use of the semiclassical approach. This restriction
is imposed by two facts with a very different origin. On one hand
this origin has an experimental core. Indeed, the breakdown of the
semiclassical approximation, for a COW experiment performed on the
surface of the Earth, would require a spreading of the wave packet
larger than the radius of the Earth, as shown below. On the other
hand, even if this experimental condition were somehow solved a
conceptual difficulty would emerge.

We now proceed to explain this conceptual problem. We may
understand the semiclassical approximation as a situation in which
the wave packet becomes a point. This statement needs an
explanation. {\it The wave packet becomes a point} means that the
wavelength is much smaller (almost a point) than any other
physically significative length parameter of the situation, i.e.,
the distance in which the potential has a noticeable change
\cite{Sakurai1}. Here is where the problem with the possible
generalizations of COW experiments lies. In metric theories the
motion of point particles and null rays are endowed with a special
status. In contrast, wave properties do not pertain to intrinsic
geometric properties of space--time \cite{Mashhoon2}. Obviously,
if we loosen the restrictions inherent to the semiclassical
approximation, then the motion of the neutrons will share all the
conceptual difficulties that wave motion has in general
relativity, and, in consequence, the interpretation of the
measurement readouts will face difficulties.

These last remarks do impinge upon those COW--like experiments
that attempt to test fundamental physics in the context of
gravitation. The theoretical framework embodied in the postulates
of metric theories cannot analyze, consistently, those proposals
in which the semiclassical approximation breaks down. We may
fathom this last statement noting that these cases would involve
the description of wave phenomena, the one represents a conceptual
problem in metric theories. This fact sets a limit to the
information about gravitational theories that can be elicited from
COW experiments. In connection with this remark there is an
additional aspect of this experiment to be discussed. Indeed, we
know that the Newtonian gravitational potential (here denoted by
$\phi$) enters as an element of the components of the metric
tensor, i.e., it is proportional to the deviations of the metric
from the Minkowskian situation, $\phi \sim h_{\mu\nu}$,
($g_{\mu\nu} = \eta_{\mu\nu} + h_{\mu\nu}$). Here we assume a
weak--field limit of the theory \cite{Misner1}, and, therefore,
the gravitational acceleration, $g$, is proportional to the
Christoffel symbols, $g \sim \Gamma^{\alpha}_{\mu\nu}$. Since the
postulates of metric theories of gravity imply that locally
gravitational effects can be {\it gauged away}, then the COW
experiment involves degrees of freedom that do not have an
invariant meaning under coordinate transformations (if we believe
in metric theories). In this last phrase we bear in mind an
experiment in which the whole experimental device falls freely.
Clearly, this is not the case in the usual experiment. Indeed, the
detecting screen is at rest with respect to the Earth, a fact that
implies that the screen is not falling freely.

To consider effects in this kind of experiments involving second
derivatives of the metric tensor (which cannot in general be {\it
gauged away} \cite{Misner1}) the spreading of the wave packet has
to be larger than the region in which physics behaves according to
special relativity. This means that the spreading of the wave
packet $\Delta x$, must fulfill the condition $\Delta x \geq
\frac{1}{\vert\phi_{,\mu\nu}\vert}$, i.e., larger than the inverse
of the absolute value of the second--order derivatives of the
gravitational potential. What means this last condition? For the
case of the Earth, whose radius and mass will be denoted by $R$
and $M$, respectively, the aforementioned restriction entails
$\Delta x \geq R\sqrt{\frac{R}{M}}$ (here we use geometrized
units, in which length and mass have the same units
\cite{Misner1}). In the weak--field limit $\frac{M}{R}<1$, and in
consequence $\Delta x \geq R$. The introduction of this condition
would imply the breakdown of the semiclassical approximation, the
one is always present as a condition in COW \cite{Colella1,
Colella4}. In other words, the use of COW experiments to detect
elements of the gravitational field with an invariant geometrical
meaning requires two initial premises: (i) The validity of the
semiclassical approximation, this fact means that the size of the
wave packet has to be smaller than the region in which the
gravitational potential has a noticeable change \cite{Sakurai1};
(ii) A wave packet larger than the region in which the flatness
theorem is valid \cite{Misner1}, this means that the wave packet
has to be larger than the region in which the gravitational field
has a considerable change. Clearly, these two conditions cannot be
fulfilled simultaneously, i.e., the lesson to be elicited from
here is that we cannot resort to COW experiments and test those
degrees of freedom of the gravitational field which are coordinate
invariant. This is, perhaps, the most important limitation of COW
experiments in connection with the principles of metric theories.

To finish this part let us address the measuring process
associated to the Aharonov--Bohm effect. This issue is an
interesting one, since the experimental verification of this
effect created a hot debate. Indeed. for some time it was claimed
that no effect of inaccessible electromagnetic fields could be
present in quantum mechanics, namely, continuity conditions for
the vector potential would be responsible for the disappearance of
the Aharonov--Bohm effect \cite{Roy1}. The detection of this
effect was done in the context of a spatial coherence device
\cite{Tonomura2}, i.e., a Young type experiment. This fact becomes
clear if we notice that two electron beams were used as part of
the experimental idea \cite{Tonomura2}. Of course, this last
remark does not imply a spatial coherence experiment. This
conclusion appears after noting that these two electron beams
where not obtained dividing a {\it primeval} electron beam. In
other words, COW experiments resort to temporal coherence, whereas
the experimental verification of the Aharonov--Bohm effect is
related to spatial coherence.

\subsubsection{Torsion, Fifth Force, and Semiclassical Approximation}

Finally, the discrepancy between theory an experiment
\cite{Colella4, Werner1} that the COW presents nowadays cannot be
solved by the inclusion of torsion, as we have shown. Other
effects, for instance, the consequence of the rotation of the
Earth upon the interference pattern \cite{Page1, Mashhoon1}, the
influence of gravity on the beam splitter \cite{Laemmerzahl4}, or
the possible dependence of the dynamical diffraction on the
bending and strains in the interferometer, do not provide a
complete answer to this discrepancy \cite{Colella4}. This is an
issue that up to now remains an open problem.

As shown in the corresponding section, COW experiments can also be
used to test the possibility of a fifth force, though once again,
the semiclassical approximation has to be used. Since in this case
an additional interaction appears, then the semiclassical limit
will involve a second condition, i.e., the wavelength of the
thermal neutrons has to be smaller than $\lambda$, the Compton
wavelength of the field, this means smaller than the range of the
force. In other words, since thermal neutrons imply a wavelength
of about $ 10^{-8}$m, then this model will work if $\lambda >
10^{-8}$m. When this condition is not fulfilled, then the
measurement readouts will not match the theoretical background.

Now that we understand better the limitation of COW experiments,
let us address the issue that the case of non--demolition
variables offers us. In general, mass not only does not disappear
from the interference pattern, but it acquires a more complicated
dependence upon mass than in COW experiments. In the context of
the principles of quantum theory it predicts a very particular
dependence of the probabilities upon the precision of the
measuring device, and, therefore, it does provide a way to
confront the RPIF model \cite{Mensky1}, or any of its equivalent
formulations \cite{Onofrio1}, against measurement readouts. It has
to be underlined that the current technology is far from being
capable of carrying out the required experiments.

Usually, the experiments in the quantum domain are divided into
two groups \cite{Dittus2, Laemmerzahl3, Fishbach3}, either
experiments which involve the evolution of free particles (neutron
interferometry falls within this group), or spectroscopy of bound
states, the Hughes--Drever experiment is a typical case of this
situation \cite{Hughes1, Drever1}. The aforementioned limitations
concerning neutron interferometry, due to the need of the
semiclassical approximation, demands  a new type of experiments in
the quantum realm. One possibility in this direction could be
provided by experiments joining gravitation and quantum
measurement theory, since they could lead to a conceptual
development in gravitation \cite{Onofrio2}.

Summing up, the void that wave phenomena has in the principles of
metric theories is the main hindrance to test the principles of
metric theories in the quantum domain, at least by means of
interference experiments. Clearly, this implies that, for
instance, if we are looking for the existence of a fifth force
then the wavelength of the particle has to be smaller than the
range of the sought force. If the fifth force had a Compton
wavelength smaller than the wavelength associated to that of
thermal neutrons then this idea cannot be employed. This comment
clearly defines a stringent limitation for this technique.

\subsection{Photon Interference: Limitations and Possibilities}

\subsubsection{First Coherence Experiments and Gravity}

It has been stressed the fact that the description of wave
phenomena has conceptual difficulties in the context, not only of
general relativity, but of metric theories. The limitations that
we have mentioned in the domain of neutron interference will be
shared by the technique of photon interference. In other words, we
may test the principles of metric theories as long as the eikonal
limit is a good approximation to the motion of light. If this
limit is abandoned, then the results obtained by the corresponding
experiment will imply premises falling outside the assumptions
behind metric theories. The consequences of this conceptual
drawback have to be contemplated in the realm of the present
proposal, which involves, unavoidably, a wave phenomenon. The
analysis of the results of an experiment of this sort shall be
done in the domain in which the aforementioned shortcoming can be
circumvented, at least partially. Otherwise the corresponding
results would be {\it weaved} with effects that are not
contemplated by our metric theory. For instance, since visible
light has a wavelength between 400nm and 700nm, then a fifth force
with a Compton length smaller than 400nm cannot be detected with
an optical experiment, this, clearly, represents a shortcoming.

Let us now make a brief summary of the applications of
first--order effects as tools to test gravitational physics. The
Michelson--Morley idea has been an important experimental device
in gravitational physics, but this is not the only optical
possibility. Sagnac's proposal has an additional parameter, which
is absent in the case of a Michelson--Morley experiment. A Sagnac
interferometer involves $b$, $\Omega$, and $\hbar\tilde{\nu}$, the
radius, the angular velocity of the interferometer, and the energy
of the beam, respectively. In this sense we may say that Sagnac is
richer, though as shown above, the detection of a violation to
Lorentz symmetry, in the form of a deformed dispersion relation,
lies outside the technological possibilities.

Another experiment which falls within the group of the
first--order coherence type is Young's, the one has been
considered in the present work as an idea for testing
non--Newtonian contributions to the gravitational force.
Unfortunately, the proposal requires travelling distances, for the
involved light beams, that avert (if the experiment is to be
carried out near the Earth's surface) the feasibility of the
proposal. The detection of a fifth force by means of first--order
experiments seems to be impossible.

\subsubsection{Second Coherence Experiments and Gravity}

The just mentioned drawbacks concerning first--order experiments,
in the context of test of fundamental physics, lead us to seek for
additional options. The realm of Hanbury--Brown--Twiss effect,
which is a second--order coherence effect, has been partially
explored. For non--Newtonian contributions it has been proved that
it offers no significant improvement with respect to Young's
situation, i.e., very large travelling distances have to be
considered.

Nevertheless, the breakdown of Lorentz symmetry can be tested
resorting to HBTE. The additional distance parameter, that this
model introduces, pays off in the quest of these violations. It
has an additional advantage. Up to now the use of interference,
either neutron, or optical, has always required the fulfillment of
the semiclassical limit (for quantum systems), or of the eikonal
limit (for light). Nevertheless, the measurement output in HBTE is
the normalized correlation coefficient of the fluctuations in the
photoelectric current obtained with the photo--detectors located
at the two detection points \cite{Mandel1, Scully1}. Obviously, it
depends upon the distance, $l$, between the photo--detectors. This
correlation function also hinges upon the properties of the
gravitational field, but $l$ is not the distance that a wave
travels. It can be much smaller, or large, than the wavelength of
the corresponding system and no conceptual difficulty would
emerge. In this sense HBTE circumvents the aforementioned
restriction.

We may sum up the limitations and possibilities of photon
interference to test the principles of metric theories stating
that in the context of first--order coherence experiments the
possibility of setting bounds to some gravitational effects,
either fifth force, or deformed dispersion relations, is
practically null. The quest for more optimistic scenarios leads us
to consider higher--order effects, as HBTE, which, for some
possible gravitational features could provide interesting results
\cite{Camacho8}.

\subsection{Neutron and Photon Interference: Coincidences}

In the present work the topic of interference has been divided
into two realms, i.e., neutron and photon interference. This could
lead us to conclude that, though sharing a set of common
properties (stemming from a linear motion equation), they do not
have too much in common. The truth is that there is a profound
physical relation between these two phenomena. Indeed, the analogy
between the case of light propagating in a moving dielectric and
the motion of a charged particle (described by quantum mechanics),
under the presence of a magnetic field has already been underlined
\cite{Cook1}. This result implies, among other things, that: (i)
Interference is a phenomenon which does not, necessarily, depend
upon the nature of the wave, rather, it hinges on the propagation
of wave motion in a medium; (ii) Some of the phenomena found in
neutron motion will have its similar in photon motion, and vice
versa. For instance, an optical Aharonov--Bohm effect has already
been put forward \cite{Leonhardt1}. Furthermore, there is an
interesting analogy between the propagation of light in a moving
dielectric and the motion of light in a gravitational field, where
the metric of the moving medium is defined by its dielectric
properties \cite{Leonhardt2}. This last remark poses interesting
questions. For instance, the motion of light in a medium (under
the condition that polarization does not impinge on the motion of
light) resembles the motion of light in general relativity
\cite{Leonhardt2}. If we now take into account the effects of
polarization of light, what is the corresponding analogy in the
context of a gravitational theory? A metric theory, or a more
general model?

\subsection{Additional Alternatives}

Of course, the principles of metric theories can be tested
resorting to experiments based on astrophysical sources, an
already known possibility, as the analysis of the motion of
Mercury \cite{Misner1} proves. In our case we may mention, for
instance, the Auger project \cite{Alfaro1}, a proposal that
possesses undeniable advantages, but that has a drawback, i.e., in
some of its aspects it is not controlled by the experimentalist,
as all proposals hinging upon astrophysical sources. Another
possibility is the analysis of the effects that, for instance, a
violation to Lorentz symmetry has on white dwarfs \cite{Camacho9},
though this kind of tests fall outside the realm of interference
models.

Finally, the use of atom interferometry opens up an interesting
spectrum of possibilities. Indeed, it as already been mentioned
that atom interferometry provides several ways in which the
Newtonian gravitational constant can be measured \cite{Fattori1,
Fattori2, Fixler1}. Fortunately, the options that this technique
offers do not end at this topic. For instance, it could be
feasible to obtain a very good accuracy, up to 1 part in
$10^{15}$, in tests of general relativity resorting to atom
interferometry \cite{Dimo1}. The detection of rotation or gravity
may also be contemplated from the perspective of atom
interferometry \cite{Dubetsky1}.

These cases illustrate, quite vividly, the possibilities that atom
interferometry could offer in the context of gravitational
physics, namely, tests of some of the postulates associated to
metric theories, etc.

\subsection{Mathematical Notation}

In this part a short list containing the main mathematical symbols
is provided.

\begin{enumerate}
\item Gravitational symbols.
\begin{enumerate}
\item Newton's gravitational constant, $G$ \item Einstein tensor,
$G_{\mu\nu}$ \item Ricci tensor, $R_{\mu\nu}$ \item Ricci scalar
of curvature, $R$ \item Christoffel symbols,
$\tilde{\Gamma}^k_{ij}$ \item Contorsion tensor, $K^k_{ij}$ \item
Planck's energy, $E_p$ \item Metric tensor, $g_{\mu\nu}$ \item
Acceleration of gravity, $g$ \item Affine connection,
$\Gamma^k_{ij}$ \item Torsion tensor, $S^k_{ij}$ \item
Gravitomagnetism--related PPN parameters, $\Delta_2$, $\Delta_1$
\end{enumerate}
\item Quantal symbols.
\begin{enumerate}
\item Pauli matrices, $\sigma^l$ \item Planck's constant, $\hbar$
\item Berry's phase, $\gamma$
\end{enumerate}
\end{enumerate}

\begin{acknowledgments}
 This research was  supported by CONACYT Grant 47000--F and by the Mexico--Germany collaboration grant
CONACyT--DFG J110.491/2006. A. Camacho would like to thank A. A.
Cuevas--Sosa and M. Fern\'andez for useful discussions and
literature hints.
\end{acknowledgments}

{}

\begin{thebibliography}{}

\bibitem{OED1} C. Soanes and A. Stevenson, Eds., {\em Oxford Dictionary of English}, (Oxford University Press,
2005).

\bibitem{Gaga1}G. Galilei, {\em Schriften, Briefe, Dokumente}, (VMA-Vertriebsgesellschaft, 2005).

\bibitem{Servet1} M. Hillar and C. S. Allen, {\em Michael Servetus:
Intellectual Giant, Humanist, and Martyr}, (University Press of
America, Lexington, 2002).

\bibitem{Newton1} I. Newton, {\em Mathematical Principles of Natural Phylosophy and His System
of the World}, (University of California Press, Berkeley and Los
Angeles, 1962).

\bibitem{Jose1} J. V. Jos\'e and E. J. Saletan, {\em Classical Dynamics. A Contemporary Approach},
(Cambridge University Press, Cambridge, 2002).

\bibitem{Mandel1} L. Mandel and E. Wolf, {\em Optical Coherence and Quantum Optics},
(Cambridge University Press, Cambridge, 1995).

\bibitem{Trouton1} F. T. Trouton and H. R. Noble, {\it The mechanical forces acting on a charged electric condenser
moving through space}, Phi. Trans. Royal Society London {\bf 202},
165-181 (1903).

\bibitem{Chase1} C. T. Chase, {\it The Trouton--Noble ether drift experiment}, Phys. Rev. {\bf 30}, 516--519 (1927).

\bibitem{Teukolsky1} S. A. Teukolsky, {\it The explanation of the Trouton--Noble experiment revisited},
Am. J. Phys. {\bf 64}, 1104--1109 (1996).

\bibitem{Turyshev1} S. G. Turyshev et al, {\it Fundamental Physics with the Laser
Astrometric Test Of Relativity}, ESA Spec. Publ. {\bf 588}, 11-18
(2005).

\bibitem{Bohm1} D. Bohm and J. Bub, {\it A proposed solution of the measurement problem
in quantum mechanics by a hidden variables theory}, Rev. Mod.
Phys. {\bf 38}, 453--469 (1966).

\bibitem{Nelson1} E. Nelson, {\em Dynamical Theories of Brownian Motion},
(Princeton University Press, Princeton, 1967).

\bibitem{Amelino3} G. Amelino--Camelia, C.
L\"ammerzah, and A. Mac\'{\i}as, {\it The Search for Quantum
Gravity Signals}, in {\em Gravitation and Cosmology. Second
Mexican Meeting on Mathematical and Experimental Physics}, A.
Mac\'{\i}as, C. L\"ammerzahl, and D. Nu\~nez, editors, American
Institute of Physics, New York (2005).

\bibitem{Kiefer1} C. Kiefer, {\em Quantum Gravity},
(Oxford Science Publications, Oxford, 2004).

\bibitem{Rovelli1} C. Rovelli, {\em Quantum Gravity},
(Cambridge University Press, Cambridge, 2004).

\bibitem{Kudoh1} H. Kudoh, A. Taruya, T. Hiramatsu,and Y.
Himemoto, {\it Detecting a gravitational-wave background with
next-generation space interferometers}, Phys.Rev. {\bf D73},
064006 (2006).

\bibitem{Guenther1} R. D. Guenther, {\em Modern Optics},
(John Wiley and Sons, New York, 1990).

\bibitem{Acustica1} E. Bavu, J. Smith, and J. Wolfe, {\it Torsional waves in a
bowed string}, Acustica {\bf 91}, 241-246 (2005).

\bibitem{Colin1} L. Garc\'{\i}a-Colin, {\em Introducci\'on a la Termodin\'amica Cl\'asica},
(Trillas, M\'exico D.F., 2002).

\bibitem{Scully1} M. O. Scully and M. S. Zubairy, {\em Quantum Optics},
(Cambridge University Press, Cambridge, 1996).

\bibitem{Mandel2} L. Mandel and E. Wolf, {\it Optical coherence properties of optical fields},
Rev. Mod. Phys. {\bf 37}, 231--287 (1965).

\bibitem{Paul1} H. Paul, {\it Interference between independent photons},
Rev. Mod. Phys. {\bf 58}, 209--231 (1986).

\bibitem{Einstein1} A. Einstein, Zur Elektrodynamik bewegter K\"orper,
Ann. d. Physik {\bf 17}, 891--896 (1905).

\bibitem{Whittaker1} E. Whittaker, {\em A History of the Theories of Aether},
(Harper, New York, 1960).

\bibitem{Ohanian1} H. Ohanian and R. Ruffini, {\em Gravitation and Spacetime},
(W.W. Norton and Company, New York, 1994).

\bibitem{Camacho1} A. Camacho, {\it Deformed dispersion relations and the
Hanbury--Brown--Twiss effect}, Gen. Rel. Grav. {\bf 37},
1405--1410 (2005).

\bibitem{Camacho2} A. Camacho and A. Mac\'{\i}as, {\it Deformed dispersion relations and
the degree of coherence function}, Gen. Rel. Grav. {\bf 38},
547--551 (2006).

\bibitem{Amelino1} G. Amelino--Camelia, J. Ellis, N. E. Mavromatos, D. V.
Nanopoulos, and S. Sarkar, {\it Tests of quantum gravity from
observations of -ray bursts}, Nature {\bf 393}, 763--765 (1998).

\bibitem{Amelino2} G. Amelino--Camelia, {\it Gravity-wave interferometers as quantum-gravity detectors}, Nature {\bf 398}, 216--218 (1999).

\bibitem{Omnes1} R. Omn\'es, {\em The Interpretation of Quantum Mechanics},
(Princeton University Press, Princeton, 1994).

\bibitem{Lamb1} W. E. Lamb Jr., {\em The Interpretation of Quantum Mechanics},
(Rinton Press, Princeton, 2001).

\bibitem{Penha1} Luis de la Pe\~na, {\em Introducci\'on a la Mec\'anica Cu\'antica},
(CECSA, M\'exico D.F., 1979).

\bibitem{Sakurai1} J. J. Sakurai, {\em Modern Quantum Mechanics},
(Addison--Wesley Publishing Company, New York, 1994).

\bibitem{Bub1} J. Bub, {\em Interpreting the Quantum World},
(Cambrdige University Press, Cambridge, 1999).

\bibitem{Giulini1} D. Giulini, E. Joos, C. Kiefer, J. Kupsch, I. -O Stamatescu, and H. D. Zeh, {\em Decoherence and the Appearance of a Classical World in
Quantum Theory}, (Springer--Verlag, Heidelberg, 1996).

\bibitem{Zeh1} H. D. Zeh, {\it On the interpretation of measurement in quantum theory},
Found. Phys. {\bf 1}, 69--76 (1970).

\bibitem{Joos1} E. Joos and H. D. Zeh, {\it The emergence of classical properties
through interaction with the environment}, Z. Phys. {\bf B59},
223--243 (1985).

\bibitem{Colella1} A. W. Overhauser and R. Colella, {\it Experimental test of
gravitationally induced quantum interference}, Phys. Rev. Lett.
{\bf 33}, 1237--1239 (1974).

\bibitem{Ahluwalia1} D. V. Ahluwalia, {\it On a new
non-geometric element in gravity}, Gen. Rel. Grav. {\bf 29},
1491-1501 (1997).

\bibitem{Ahluwalia2} D. V. Ahluwalia, {\it Can
general-relativistic description of gravitation be considered
complete?}, Mod. Phys. Lett. {\bf A13}, 1393--1400 (1998).

\bibitem{Will1}  C. M. Will, {\em Theory and Experiment in Gravitational
Physics}, (Cambridge University Press, Cambridge, 1993).

\bibitem{Simon1} C. Simon and W. T. E. Irvine, {\it Robust long-distance entanglement and a
loophole-free bell test with ions and photons}, Phys. Rev. Lett.
{\bf 91}, 110405 (2003).

\bibitem{Haeffner1} H. H\"affner et al, {\it Scalable multi-particle
entanglement of trapped ions}, Nature {\bf 438}, 643-646 (2005) .

\bibitem{Hannay1} J. H. Hannay, {\it Angle variable holonomy in adiabatic
excursion of an integrable Hamiltonian}, J. Phys. A: Math. Gen.
{\bf 18}, 221--230 (1985).

\bibitem{Berry1} M. V. Berry, {\it Quantal phases factors accompanying adiabatic changes},
Proc. R. Lond. {\bf A392}, 47--57 (1984).

\bibitem{Griffiths1} D. J. Griffiths, {\em Introduction to Quantum Mechanics},
(Prentice Hall, New Jersey, 1995).

\bibitem{Aharonov1} Y. Aharonov and D. Bohm, {\it Significance of electromagnetic
potentials in quantum theory}, Phys. Rev. {\bf 115}, 485--491
(1959).

\bibitem{Bayer1} M. Bayer et al, {\it Optical Detection of the Aharonov--Bohm effect on a
Charged Particle in a Nonoscale Quantum Ring}, Phys. Rev. Lett.
{\bf 90}, 186801 (2003).

\bibitem{Tonomura1} A. Tonomura et al, {\it Observation of Aharonov-Bohm effect by electron
holography}, Phys. Rev. Lett. {\bf 48}, 1443-1446 (1982).

\bibitem{Tomita1} A. Tomita and R. Chiao, {\it Observations of Berry's topological phase by use of
an optical fiber}, Phys. Rev. Lett. {\bf 57}, 937--940 (1986).

\bibitem{Jackson1} J. D. Jackson, {\em Classical Electrodynamics},
(John Wiley and Sons, Inc., New York, 1999).

\bibitem{ACasher1} Y. Aharonov and A. Casher, {\it Topological Quantum Effects for Neutral Particles},
Phys. Rev. Lett. {\bf 53}, 319-321 (1984).

\bibitem{Spavieri1} G. Spavieri, {\it Quantum Effect of the Aharonov--Bohm Type for Particles
with an Electric Dipole Moment}, Phys. Rev. Lett. {\bf 82},
3932-3935 (1999).

\bibitem{Boyer1} T. H. Boyer, {\it Proposed Aharonov--Casher Effect: Another example of an Aharonov--Bohm
effect arising from classical lag}, Phys. Rev. {\bf A36},
5083-5086 (1987).

\bibitem{APearle1} Y. Aharonov, P. Pearle, and L. Vaidman, {\it Comment on: Proposed Aharonov--Casher Effect: Another
example of an Aharonov--Bohm effect arising from classical lag},
Phys. Rev. {\bf A37}, 4052-4055 (1988).

\bibitem{Sjoqvist1} E. Sj\"oqvist, {\it Locality and Topology in the Molecular Aharonov--Bohm Effect},
Phys. Rev. Lett.{\bf 89}, 210401 (2002).

\bibitem{Marques1} G. de A. Marques and V. B. Bezerra, {\it On a gravitational analogue of the
Aharonov--Bohm effect}, Phys. Lett. {\bf A318}, 1-5 (2003).

\bibitem{Misner1} Ch. W.  Misner, K. S. Thorne, and J. A. Wheeler, {\em Gravitation},
(W. H. Freeman and Company, San Francisco, 1973).

\bibitem{Lockerbie1} N. Lockerbie, J. C. Mester, R. Torii, S. Vitale, and P. W. Worden, {\it STEP: A Status Report},
in {\em Gyros, Clocks, Interferometers...; Testing Relativistic
Gravity in Space}, C. L\"ammerzahl, C. W. F. Everitt, and F. W.
Hehl, editors, (Springer--Verlag, Heidelberg, 2001).

\bibitem{Dittus1} W. Vodel, et al, {\it High Sensitive DC SQUID Based Position Detectors for Application
in Gravitational Experiments at the Drop Tower Bremen}, in {\em
Gyros, Clocks, Interferometers...; Testing Relativistic Gravity in
Space}, C. L\"ammerzahl, C. W. F. Everitt, and F. W. Hehl,
editors, (Springer--Verlag, Heidelberg, 2001).

\bibitem{Haugan1} M. P. Haugan and C. L\"ammerzahl, {\it Principles of Equivalence:
Their Role in Gravitation Physics and Experiments That Test Them},
in {\em Gyros, Clocks, Interferometers...; Testing Relativistic
Gravity in Space}, C. L\"ammerzahl, C. W. F. Everitt, and F. W.
Hehl, editors, (Springer--Verlag, Heidelberg, 2001).

\bibitem{Adel1} E. G. Adelberger, {\it Modern tests of the universality of free fall},
Class. Quant. Grav. {\bf 11}, A9--A21 (1994).

\bibitem{Adler1} R. Adler, M. Bazin, and M. Schiffer, {\em Introduction to General Relativity},
(McGraw--Hill Book Company, New York, 1975).

\bibitem{Richard1} J. P. Richard, {\it Tests of Theories of Gravity in the Solar System},
in {\em General Relativity and Gravitation}, G. Shaviv and J.
Rosen, editors, (John Wiley and Sons, New York, 1975).

\bibitem{Ritter1} R. C. Ritter and J. W. Beams, {\it A Laboratory Measurement of the Constancy of G},
in {\em On the Measurement of Cosmological Variations of the
Gravitational Constant}, L. Hapern, editor, (University Press of
Florida, Gainsville, 1978).

\bibitem{Brans1} C. Brans and R. H. Dicke, {\it Mach's principle and a relativistic
theory of gravitation}, Phys. Rev. {\bf 124}, 925--935 (1961).

\bibitem{Rosen1} N. Rosen, {\it Bimetric gravitation theory on a cosmological basis},
Gen. Rel. Grav. {\bf 9}, 339--351 (1978).

\bibitem{Ciufolini1} I. Ciufolini and J. A. Wheeler, {\em Gravitation and Inertia},
(Princeton University Press, Princeton, 1995).

\bibitem{Ciufolini2} I. Ciufolini, E. Pavlis, F. Chieppa, E. Fernandes--Vieira, and J. P\'erez--Mercader,
{\it Test of general relativity and measurement of the
Lense--Thirring effect with two Earth satellites}, Science {\bf
279}, 2100--2103 (1998).

\bibitem{Mashhoon2} B. Mashhoon, {\it On the spin-rotation-gravity coupling},
Gen. Rel. Grav. {\bf 31}, 681--690 (1999).

\bibitem{Krisher1} T. P. Krisher, J. D. Anderson, and A. H.
Taylor, {\it Voyager 2 test of the radar time delay effect},
Astrophys. J. {\bf 373}, 665--670 (1991).

\bibitem{Dickey1} J. O. Dickey et al, {\it Lunar laser ranging: A continuing legacy of the Apollo program},
Science {\bf 265}, 482--490 (1994).

\bibitem{Fluegge1} S. Fl\"ugge, {\em Rechenmethoden der Quantentheorie},
(Springer--Verlag, Heidelberg, 1993).

\bibitem{Rebka1} R. V. Pound and G. A. Rebka, {\it Apparent weight of photons},
Phys. Rev. Lett. {\bf 4}, 337--341 (1960).

\bibitem{Colella2} J.-L. Straudemann, S. A. Werner, R. Colella,
and A. W. Overhauser, {\it Gravity and inertia in quantum
mechanics, coherence effects in neutron diffraction and gravity
experiments}, Phys. Rev. {\bf A21}, 1419--1438 (1980).

\bibitem{Colella3} D. M. Greenberg and A. W. Overhauser, {\it Coherence effects in
neutron diffraction and gravity experiments}, Rev. Mod. Phys. {\bf
51}, 43--78 (1979).

\bibitem{Colella4} K. C. Littrell, B. E. Allman, and S. A. Werner,
{\it Two-wavelength-difference measurement of gravitationally
induced quantum interference phases}, Phys. Rev. {\bf A56},
1767--1780 (1997).

\bibitem{Borde2} Ch. J. Bord\'e, {\it Atomic with internal states labelling},
Phys. Lett. {\bf A140}, 10--12 (1989).

\bibitem{Borde3} Ch. J. Bord\'e, N. Courtier, F. du Burck, A. N. Goncharov, and M.
Gorlicki, {\it Molecular interferometry experiments}, Phys. Lett.
{\bf A188}, 187--197 (1994).

\bibitem{Bonse1} U. Bonse and T. Wroblevski, {\it Measurement of Neutron Quantum Interference in Noninertial Frames},
Phys. Rev. Lett. {\bf 51}, 1401--1410 (1983).

\bibitem{Laemmerzahl10} C. L\"ammerzahl, {\it On the equivalence principle in quantum theory},
Gen. Rel. Grav. {\bf 28}, 1043--1070 (1996).

\bibitem{Raum1} K. Raum et al, {\it Effective--Mass Enhanced Deflection of Neutrons in Noninertial Frames},
 Phys. Rev. Lett. {\bf 74}, 2859--2862 (1995).

\bibitem{Varju1} K. Varj\'u and L. H. Ryder {\it General Relativistic treatment of the Colella --Overhauser--Werner
experiment on neutron interference in a gravitational field}, Am.
J. Phys. {\bf 68}, 404--409 (2000).

\bibitem{Mannheim1} P. D. Mannheim, {\it Classical underpinnings of gravitationally induced quantum interference},
Phys. Rev. {\bf A57}, 1260--1264 (1998).

\bibitem{Onofrio2} L. Viola and R. Onofrio, {\it Testing the equivalence principle through freely falling
quantum objects}, Phys. Rev. {\bf D55}, 455--462 (1997).

\bibitem{Camacho10} A. Camacho, {\it On a quantum equivalence principle},
Mod. Phys. Lett. {\bf A14},  275--288 (1999).

\bibitem{Borde1} C. J. Bord\'e, J. C. Houard, and A. Karasiewicz, {\it Relativistic Phase Shifts
for Dirac Particles Interacting with Weak Gravitational Fields in
Matter--Wave Interferometers}, in {\em Gyros, Clocks,
Interferometers...; Testing Relativistic Gravity in Space}, C.
L\"ammerzahl, C. W. F. Everitt, and F. W. Hehl, editors,
Springer--Verlag, Heidelberg (2001).


\bibitem{Will2} C. M. Will, {\it The confrontation between general relativity and
 experiment}, Living Rev. Relativity {\bf 9}, 3--92 (2006).

 \bibitem{Hughes1} V. W. Hughes, H. G. Robinson, and V. Beltr\'an--L\'opez, {\it Upper limit for
 the anisotropy of inertial mass from nuclear resonance experiments}, Phys. Rev. Lett.
{\bf 4}, 342--344 (1960).

\bibitem{Drever1} R. W. P. Drevers, {\it A search for anisotropy of inertial mass using
a free precession technique}, Phyl. Mag. {\bf 6}, 683--687 (1961).

\bibitem{Anandan1} J. Anandan, {\it Gravitational and rotational effects in
quantum interference}, Phys. Rev. {\bf D15}, 1448--1457 (1977).

\bibitem{Mashhoon1} B. Mashhoon, {\it Neutron interferometry in a rotating frame of reference},
Phys. Rev. Lett. {\bf 61}, 2639--2642 (1988).

\bibitem{Mashhoon3} B. Mashhoon, {\it Wave propagation in a gravitational field},
Phys. Lett. {\bf A122}, 299--304 (1986).

\bibitem{Truehaft1} C. M. Will, {\it Violation of the Weak Equivalence Principle
in theories of gravity with nonsymmetric metric}, Phys. Rev. Lett.
{\bf 62}, 369--372 (1989).

\bibitem{Neugebauer1} C. L\"ammerzahl and G. Neugebauer, {\it The Lense--Thirring Effect: From the
Basic Notions to the Observed Effects}, in {\em Gyros, Clocks,
Interferometers...; Testing Relativistic Gravity in Space}, C.
L\"ammerzahl, C. W. F. Everitt, and F. W. Hehl, editors,
(Springer--Verlag, Heidelberg, 2001).

\bibitem{Ni1} W.-T. Ni, {\it A new theory of gravity}, Phys. Rev. {\bf D7}, 2880--2883 (1973).

\bibitem{Carmeli1} M. Carmeli, {\em Group Theory and General Relativity},
(World Scientific Company, Singapore, 2000).

\bibitem{Hehl1} F. W. Hehl, P. von der Heyde, G. D. Kerlick, and J. M. Nester,
{\it General relativity with spin and torsion: Foundations and
prospects},  Rev. Mod. Phys. {\bf 48}, 393--416 (1976).

\bibitem{Yaskin1} P. B. Yasskin and W. R. Stoeger, {\it Propagation equations for test bodies with spin
and rotation in theories of gravity with torsion}, Phys. Rev. {\bf
D21}, 2081--2094 (1980).

\bibitem{Stoeger1} R. William, S. J. Stoeger, and P. B. Yaskin, {\it Can a macroscopic gyroscope feel
torsion?}, Gen. Rel. Grav. {\bf 11}, 427--431 (1979).

\bibitem{Vessot1} R. F. C. Vessot and L. W. Levien, {\it A test of the equivalence
principle using a space-borne clock}, Gen. Rel. Grav. {\bf 10},
181--204 (1979).

\bibitem{Vessot2} R. F. C. Vessot, et al, {\it Test of relativistic gravitation
with a space-borne Hydrogen maser}, Phys. Rev. Lett. {\bf 45},
2081--2084 (1980).

\bibitem{Taylor1} J. H. Taylor, {\it Pulsar timing and relativistic gravity}, Class. Quant. Grav. {\bf 10} S167--S174
(1993).

\bibitem{Damour1} T. Damour and J. H. Taylor, {\it Strong-field tests of relativistic gravity and binary
pulsars}, Phys. Rev. {\bf D45}, 1840--1868 (1992).

\bibitem{Damour2} T. Damour and G. Esposio--Far\'ese, {\it Tensor-scalar gravity and binary-pulsar experiments},
Phys. Rev. {\bf D54}, 1474--1491 (1996).

\bibitem{Long1} D. R. Long, {\it Why do we believe Newtonian
gravitation at laboratory dimensions?}, Phys. Rev. {\bf D9},
850--852 (1974).

\bibitem{Gibbons1} G. W. Gibbons and B. F. Whiting, {\it Newtonian gravity measurments impose
constraints on unification theories}, Nature {\bf 291}, 636--638
(1981).

\bibitem{Smith1} G. L. Smith C. D. Hoyle, J. H. Gundlach, E. G. Adelberger,
B. R. Heckel, and H. E. Swanson, {\it Short-range tests of the
equivalence principle}, Phys. Rev. {\bf D61}, 022001 (2000).

\bibitem{Carugno1} G. Carugno, Z. Fontana, R. Onofrio, and C.
Rizzo, {\it Limits on the existence of scalar interactions in the
submillimeter range}, Phys. Rev. {\bf D55}, 6591--6595 (1997).

\bibitem{Fujii1} F. Fujii1, {\it Dilaton and possible non--Newtonian gravity}, Nature {\bf 234}, 5--7 (1971).

\bibitem{Fishbach2} E. Fishbach and C. L. Talmadge, {\em The
Search for Non--Newtonian Gravity}, (Springer--Verlag, New York,
1999).

\bibitem{Yu1} H.-T. Yu, W.-T. Ni, C. Hu, F. Liu, C. Yang, and W.
Liu, {\it Experimental determination of the gravitational forces
at separations around 10 meters}, Phys. Rev. {\bf D20}, 1813--1815
(1979).

\bibitem{Zumberge1} M. A. Zumberge et al, {\it Submarine measurement of the Newtonian
gravitational constant}, Phys. Rev. Lett. {\bf 67}, 3051--3054
(1991).

\bibitem{Milonni1} P. W. Milonni, {\em
The Quantum Vacuum}, (Academic Press, San Diego, Cal., 1994).

\bibitem{Mostepanenko1} V. M. Mostepanenko and I. Yu. Sokolov, {\it Hypothetical
long-range interactions and restrictions on their parameters from
force measurements}, Phys. Rev. {\bf D47}, 2882--2891 (1993).

\bibitem{Laemmerzahl3} C. L\"ammerzahl, {\it Quantum tests of the foundations of
general relativity}, Class. Quant. Grav. {\bf 15}, 13--27 (1998).

\bibitem{Damour3} T. Damour, {\it Experimental Tests of Relativistic
Gravity}, in {\em Proc. Supplement of the 1998 Texas Symp.}, Nucl.
Phys. B.

\bibitem{Fishbach1} E. Fishbach, D. Sudarsky, A. Szafer, C. L.
Talmadge, and S. H. Aronson, {\it Reanalysis of the E\"otv\"os
experiment}, Phys. Rev. Lett. {\bf 56}, 3--6 (1986).

\bibitem{Werner1} S. A. Werner, Gravitational, {\it Rotational
and Topological Quantum Phase Shifts in Neutron Interferometry},
Class. Quant. Grav. {\bf 11}, A207--A226 (1994).

\bibitem{Laemmerzahl1} C. L\"ammerzahl, {\it Constraints on spacetime torsion from
Hughes--Drever experiments}, Phys. Lett. {\bf A228}, 223---231
(1997).

\bibitem{Camacho3} A. Camacho and A. Mac\'{\i}as, {\it Spacetime torsion
contribution to quantum interference phases}, Phys. Lett. {\bf
B617} 118--123 (2005).

\bibitem{Laemmerzahl2} J. Audretsch and C. L\"ammerzahl, {\it Neutron interference: general theory
of the influence of gravity, inertia and space--time torsion}, J.
Phys. A: Math. Gen. {\bf 16}, 2457--2477 (1983).

\bibitem{Audretsch1} J. Audretsch, {\it Dirac electron in space--times with torsion: Spinor propagation,
spin precesion, and nongeodesic orbits}, Phys. Rev. {\bf D24},
1470--1477 (1981).

\bibitem{Cohen1} C. Cohen--Tannoudji, B. Diu, and F. Lalo\"e, {\em
Quantum Mechanics, Volume II}, (John Wiley and Sons, New York,
1977).

\bibitem{Fattori1} M. Fattori et al,
{\it Towards an atom interferometric determination of the
Newtonian gravitational constant}, Phys. Lett. {\bf A318}, 184-191
(2003).

\bibitem{Fattori2} J. Stuhler et al,
{\it MAGIA--using atom interferometry to determine the Newtonian
gravitational constant}, J. Opt. B {\bf 5}, S75--S81 (2003).

\bibitem{Fixler1} J. B. Fixler et al,
{\it Atom Interferometer Measurment of the Newtonian Constant of
Gravity}, Science {\bf 315}, 74--77 (2007).

\bibitem{Chan1} H. A. Chan, M. V. Moody, and H. J. Paik,
{\it Null Test of the Gravitational Inverse Square Law}, Phys.
Rev. Lett. {\bf 49}, 1745-1748 (1982).

\bibitem{Onofrio1} C. Presilla, R. Onofrio, and U. Tambini, {\it Measurement quantum mechanics and
experiments on quantum Zeno effect}, Annals Phys. {\bf 248},
95-121 (1996).

\bibitem{Mensky1} M. B. Mensky, {\em Continuous Quantum Measurements and Path Integrals},
(IOP, Bristol, 1993).

\bibitem{Braginsky1} V. B. Braginsky and F. Ya. Khalili, {\em Quantum Measurement},
(Cambridge University Press, Cambridge, 1995).

\bibitem{Bocko1} M. F. Bocko and R. Onofrio, {\it On the measurement of a weak classical force
coupled to a harmonic oscillator: experimental progress}, Rev.
Mod. Phys. {\bf 68}, 755-799 (1996).

\bibitem{Mensky2} M. B. Mensky, {\it Quantum restrictions for continuous observation of an oscillator},
Phys. Rev. {\bf D20}, 384--387 (1979).

\bibitem{Dittrich1} W. Dittrich and M. Reuter, {\em Classical and
Quantum Dynamics}, (Springer--Verlag, Berlin, 1996).

\bibitem{Thompson1} R. Thompson, {\it Quantum optics with trapped
ions}, in {\em Latin--American School of Physics XXXI ELAF}, S.
Hacyan, R. J\'auregui, and R. L\'opez--Pe\~na, eds., American
Institut of Physics, Woodbury, (1999).

\bibitem{Dehmelt1} H. Dehmelt, {\it Experiments with an isolated subatomic particle at rest},
Rev. Mod. Phys. {\bf 62}, 525---530 (1990).

\bibitem{Iorio1} L. Iorio, {\it First preliminary tests of the general relativistic
gravitomagnetic field of the Sun and new constraints on a
Yukawa-like fifth force from planetary data}, gr--qc/0507041.

\bibitem{Bertolami1} O. Bertolami and J. Paramos, {\it Astrophysical constraints on
saclar field models}, Phys. Rev. {\bf D71}, 023521 (2005).

\bibitem{Gallian1} J. A. Gallian, {\em Contemporary Abstract Algebra}, (Houghton Mifflin Company, Boston, 1998).

\bibitem{Schleich1} W. Schleich and M. O. Scully, {\it General Relativity and Modern
Optic}, in {\em Mo\-dern Trends in Atomic and Molecular Physics},
Proceedings of Les Houches Su\-mmer School, Session XXXVIII, eds.
R. Stora and G. Grynberg, North--Holland, Amsterdam (1984).

\bibitem{Thorne1} K. S. Thorne, {\it Multipole expansions of gravitational
 radiation}, Rev. Mod. Phys. {\bf 52}, 299--339 (1980).

\bibitem{Camacho5} A. Camacho, {\it Non--Newtonian gravity and coherence properties of light},
Phys. Lett. {\bf A287}, 339--343 (2001).

\bibitem{Scully2} M. O. Scully and K. Dr\"uhl, {\it Quantum eraser: A proposed photon correlation
 experiment concerning observation and "delayed choice" in quantum mechanics},
 Phys. Rev. {\bf A25}, 2208--2213 (1982).

\bibitem{Talmadge1} C. Talmadge, J. -P. Berthias, R. W. Hellings,
and E. M. Standish, {\it Model-independent constraints on possible
modifications of Newtonian gravity}, Phys. Rev. Lett. {\bf 61},
1159-1162 (1988).

\bibitem{Hong1}  C. K. Hong, Z. Y. Ou, and L. Mandel, {\it Measurement of subpicosecond
time intervals between two photons by interference}, Phys. Rev.
Lett. {\bf 59}, 2044--2046 (1987).

\bibitem{Montemayor1} R. Montemayor, L. F. Urrutia, {\it Synchrotron Radiation in Lorentz-violating
Electrodynamics: the Myers-Pospelov},  Phys. Rev. {\bf D72},
045018 (2005).

\bibitem{Scully3} M. O. Scully, M. S. Zubairy, and M. P. Haugan, {\it Proposed optical
test of metric gravitation theories}, Phys. Rev. {\bf A24},
2009-2016 (1981).

\bibitem{Camacho4} A. Camacho and E. Castellanos, {\it Deformed dispersion relations and Sagnac interferometer}, in {\em
Gravitation and Cosmology. Second Mexican Meeting on Mathematical
and Experimental Physics}, A. Mac\'{\i}as, C. L\"ammerzahl, and D.
Nu\~nez, editors, American Institute of Physics, New York (2005).

\bibitem{Hanbury1} R. Hanbury Brown and R. Q. Twiss, {\it Correlation between
Photons in two Coherent Beams of Light}, Nature {\bf 177}, 27-29
(1956).

\bibitem{Hanbury2} R. Hanbury--Brown, {\em The Intensity Interferometer}, (Taylor and Frances, London, 1974).

\bibitem{Ping1} Y.-P. Qin, Y.-M. Dong, R.-J. Lu, B.-B. Zhang, and L.-W. Jia,
{\it Relationship between the gamma-ray burst pulse width and
energy due to the Doppler effect of fireballs}, Astrophys. J. {\bf
632}, 1008--1015 (2005).

\bibitem{Glauber1} R. J. Glauber, in {\em Quantum Optics}, eds. C. DeWitt, A. Blandin, and C. Cohen--Tannoudji,
(Gordon and Breach, New York, 1970).

\bibitem{Chartier} G. Chartier, {\em Introduction to Optics}, (Springer--Verlag, New York, 2005).

\bibitem{Roy1} S. M. Roy, {\it Condition for the Nonexistence of Aharonov--Bohm Effect},
Phys. Rev. Lett. {\bf 44}, 111--114 (1980).

\bibitem{Tonomura2} A. Tonomura, et al, {\it Evidence for Aharonov--Bohm effect with Magnetic Field Completely
Shielded from Electron Wave }, Phys. Rev. Lett. {\bf 56}, 792--795
(1986).

\bibitem{Page1} L. A. Page, {\it Effect of Earth's rotation in
neutron interferometry}, Phys. Rev. Lett. {\bf 35}, 543 (1975).

\bibitem{Laemmerzahl4} C. L\"ammerzahl and Ch. J. Bord\'e, {\it Atomic interferometry in
gravitational fields: Influence of gravitation on the beam
splitter}, Gen. Rel. Grav. {\bf 31}, 635--652 (1999).

\bibitem{Dittus2} H. Dittus and C.
L\"ammerzahl, {\it Experimental tests of the Equivalence Principle
and Newton's Law in Space}, in {\em Gravitation and Cosmology.
Second Mexican Meeting on Mathematical and Experimental Physics},
A. Mac\'{\i}as, C. L\"ammerzahl, and D. Nu\~nez, editors, American
Institute of Physics, New York (2005).

\bibitem{Fishbach3} E. Fishbach and B.-S Freeman, {\it Testing general relativity at the quantum level},
Gen. Rel. Grav. {\bf 11}, 377--381 (1979).

\bibitem{Amelino4} G. Amelino--Camelia, {\it Proposal of a second generation
of quantum-gravity-motivated Lorentz-symmetry tests: sensitivity
to effects suppressed quadratically by the Planck scale}, Int. J.
Mod. Phys. {\bf D12} 1633--1640 (2003).

\bibitem{Amelino5} G. Amelino-Camelia and C. Lammerzahl, {\it Quantum-gravity-motivated
Lorentz-symmetry tests with laser interferometers}, Class. Quant.
Grav. {\bf 21}, 899-916 (2004).

\bibitem{Camacho8} A. Camacho, {\it Continuous distribution of frequencies and deformed dispersion
relations}, Class. Quant.Grav. {\bf 22}, 2101-2106 (2005).

\bibitem{Cook1} R. J. Cook, H. Fearn, and P. W. Milonni,
{\it Fizeau's experiment and the Aharonov--Bohm effect}, Am. J.
Phys. {\bf 63}, 705--710 (1995).

\bibitem{Leonhardt1} U. Leonhardt and P. Piwnicki, {\it Optics of nonumiformly moving media},
Phys. Rev. {\bf A60}, 4301--4312 (1999)

\bibitem{Leonhardt2} U. Leonhardt and P. Piwnicki, {\it Relativistic Effects of Light in Moving Media with
Extremely Low Group Velocity}, Phys. Rev. Lett. {\bf 84}, 822--825
(2000).

\bibitem{Alfaro1} J. Alfaro, {\it Quantum gravity induced Lorentz invariance violation in the
standard model: Hadrons}, Phys. Rev. {\bf D72}, 024027 (2005).

\bibitem{Camacho9} A. Camacho, {\it White dwarfs as test objects of Lorentz
violations}, Class. Quant. Grav. {\bf 23}, 7355--7368 (2006).

\bibitem{Dimo1} S. Dimopoulos et al, {\it Testing General
Relativity with Atom Interferometry}, Phys. Rev. Lett.{\bf 98},
111102 (2007).

\bibitem{Dubetsky1} B. Dubetsky et al, {\it Atom
interferometer as a selective sensor of rotation or gravity},
Phys. Rev.{\bf A74}, 023615 (2006).

\end{thebibliography}
\end{document}